\documentclass[iop]{emulateapj}

\usepackage{graphicx}
\usepackage{txfonts}
\usepackage{color}
\usepackage{url}

\newcommand{\OI}{O\,{\sc i}}

\newcommand{\CII}{C\,{\sc ii}}
\newcommand{\HII}{H\,{\sc ii}}
\newcommand{\HI}{H\,{\sc i}}
\newcommand{\NII}{N\,{\sc ii}}
\newcommand{\SII}{S\,{\sc ii}}

\newcommand{\OIII}{O\,{\sc iii}}
\newcommand{\ArIII}{Ar\,{\sc iii}}
\newcommand{\PIII}{P\,{\sc iii}}
\newcommand{\SIII}{S\,{\sc iii}}

\slugcomment{Accepted for publication in the Astrophysical Journal, 2015 August 12}

\shorttitle{Velocity-resolved  [\CII] emission and [\CII]/FIR mapping along OMC\,1}
\shortauthors{Goicoechea et al.}

\begin{document} 

\title{Velocity-resolved [\CII] emission and [\CII]/FIR Mapping along Orion  with \textit{Herschel} \footnotemark[*,**]}

\footnotetext[*]{\textit{Herschel} is an ESA space observatory with science instruments provided by European-led
Principal Investigator consortia and with important participation from NASA.}

\footnotetext[**]{Uses observations obtained with the IRAM-30m telescope. 
IRAM is supported by INSU/CNRS (France), MPG (Germany), and IGN (Spain).\\}

\author{Javier R. Goicoechea$^1$,  D. Teyssier$^2$, M. Etxaluze$^{1,3}$, 
P.F. Goldsmith$^4$,
V. Ossenkopf$^{5}$,
M. Gerin$^{6,7}$,
E.A. Bergin$^8$,
J.H. Black$^9$,\\
J. Cernicharo$^1$,
S. Cuadrado$^1$,
P. Encrenaz$^{6}$,
E. Falgarone$^{6,7}$,
A. Fuente$^{10}$,
A. Hacar$^{11}$,
D.C. Lis$^{6}$,
N. Marcelino$^{12}$,\\
G.J. Melnick$^{13}$,
H.S.P. M\"uller$^{5}$,
C. Persson$^{9}$,
J. Pety$^{14}$,
M. R\"ollig$^{5}$,
P. Schilke$^{5}$,
R. Simon$^{5}$,
R.L. Snell$^{15}$,
J. Stutzki$^{5}$
}


\affil{$^1$Instituto de Ciencia de Materiales de Madrid (CSIC).
Calle Sor Juana Ines de la Cruz 3, E-28049 Cantoblanco, Madrid, Spain \email{jr.goicoechea@icmm.csic.es}}

\affil{$^2$Herschel Science Centre, ESA/ESAC, P.O. Box 78, Villanueva de la Ca\~nada, E-28691 Madrid, Spain}

\affil{$^3$RAL Space, Rutherford Appleton Laboratory, Didcot OX11 0QX, UK}

\affil{$^4$Jet Propulsion Laboratory, California Institute of Technology, 4800 Oak Grove Drive, Pasadena, CA 91109-8099, USA}

\affil{$^5$I. Physikalisches Institut der Universit\"at zu K\"oln, Z\"ülpicher Str. 77, 50937 K\"oln, Germany}

\affil{$^6$LERMA, Observatoire de Paris, PSL Research University, CNRS, Sorbonne Universit\'es, UPMC Univ. Paris 06, F-75014, Paris, France}

\affil{$^7$\'Ecole Normale Sup\'erieure,  24 rue Lhomond, F-75005, Paris, France}

\affil{$^8$Department of Astronomy, University of Michigan, 500 Church Street, Ann Arbor, MI 48109, USA}

\affil{$^9$Department of Earth and Space Sciences, Chalmers University of Technology, Onsala Space Observatory, SE-43992 Onsala, Sweden}

\affil{$^{10}$Observatorio Astron\'omico Nacional (OAN IGN),  Apdo. 112, 28803, Alcal\'a de Henares, Spain}

\affil{$^{11}$Institute for Astrophysics, University of Vienna, T\"urkenschanzstrasse 17, 1180, Vienna, Austria}

\affil{$^{12}$INAF, Istituto di Radioastronomia, via P. Gobetti 101, 40129, Bologna, Italy}

\affil{$^{13}$Harvard-Smithsonian Center for Astrophysics, 60 Garden Street, MS 66, Cambridge, MA 02138, USA}

\affil{$^{14}$Institut de Radioastronomie Millim\'etrique, 300 rue de la Piscine, 
38406 Saint-Martin d'H\'{e}eres, France,}

\affil{$^{15}$Department of Astronomy, University of Massachusetts, LGRT-B 619E, 710 North Pleasant Street, Amherst, 
MA 01003, USA}

\begin{abstract}

We present  the first $\sim$7.5$'$$\times$11.5$'$ velocity-resolved ($\sim$0.2~km\,s$^{-1}$) map of the 
[\CII]\,158\,$\mu$m line 
toward  the Orion molecular cloud\,1~(OMC\,1)
taken with the \mbox{\textit{Herschel}/HIFI} instrument.
In combination with \mbox{far-infrared} (FIR) photometric images and velocity-resolved maps of 
the H41$\alpha$ hydrogen recombination and CO~$J$=2-1 lines,
this data set provides an unprecedented view of the intricate small-scale kinematics of the \mbox{ionized/PDR/molecular} 
gas interfaces and of the radiative feedback from massive stars. 
The main contribution  to the [\CII] luminosity  ($\sim$85~\%) is from the extended, FUV-illuminated
face of the cloud \mbox{($G_0$$>$500, $n_{\rm H}$$>$5$\times$10$^3$\,cm$^{-3}$)} and from  dense PDRs 
($G_0$$\gtrsim$10$^4$, $n_{\rm H}$$\gtrsim$10$^5$\,cm$^{-3}$)  at the interface
between OMC\,1 and the \HII~region surrounding the \mbox{Trapezium} cluster. Around
$\sim$15\,\%~of the [\CII] emission arises from a different gas component without CO counterpart.
The [\CII] excitation, PDR gas turbulence, line opacity
(from  [$^{13}$\CII]) and role of the geometry of the illuminating stars with respect to
the  cloud are investigated. 
We construct maps of the $L$[\CII]/$L_{\rm FIR}$
and  $L_{\rm FIR}$/$M_{\rm Gas}$ ratios  and show that  $L$[\CII]/$L_{\rm FIR}$ decreases 
from the extended cloud component
(\mbox{$\sim$10$^{-2}$$-$10$^{-3}$}) to the more opaque star-forming cores 
(\mbox{$\sim$10$^{-3}$$-$10$^{-4}$}). The lowest values are reminiscent
of the ``[\CII] deficit'' seen in local ultra-luminous IR galaxies hosting vigorous star formation.
Spatial correlation analysis shows that the decreasing
 $L$[\CII]/$L_{\rm FIR}$ ratio correlates better with the column density of dust through
  the molecular cloud
 than with $L_{\rm FIR}$/$M_{\rm Gas}$. 
We conclude that the   [\CII] emitting column relative to the total dust
column along each line of sight is responsible for  the observed  $L$[\CII]/$L_{\rm FIR}$ variations
through the cloud. 

\end{abstract}


\section{Introduction}

The \mbox{$^{2}P_{3/2}$−-$^{2}P_{1/2}$} fine structure line  ($\Delta E$/$k$$\simeq$91 K) of ionized carbon, 
[\CII]~158\,$\mu$m, is one of the most important
cooling lines of the cold neutral medium \mbox{\citep{Dal72}}
and among the brightest lines in  photodissociation regions \citep[PDRs, e.g.,][]{Hol99}.
Early   observations    showed that the line is very luminous \mbox{\citep[][]{Rus80}}, 
carrying  
\mbox{from $\sim$0.1 to 1~\%} of the total far-IR (FIR) luminosity of  galaxies \citep{Cra85}.
The neutral carbon atom has an ionization potential of 11.3~eV, so that the ion C$^+$ traces the H$^+$/H/H$_2$ transition,
the critical conversion from atomic to molecular ISM. 
As a consequence, [\CII] emission/absorption from different ISM phases is expected.
Recent velocity-resolved line
surveys show that the average contribution to the [\CII] emission in the Milky Way is the H$_2$ gas illuminated by 
\mbox{far-UV  \mbox{(FUV; $E$$\simeq$6-13.6~eV) }} photons \mbox{(55-75~\%)}, 
the cold  atomic gas  \mbox{(20-25~\%)} and the \HII~ionized gas 
\mbox{(5-20~\%)} \citep[][]{Pin13,Pin14}. 

Star-forming complexes  hosting massive OB stars are irradiated by strong FUV fields and most of their [\CII] emission
is likely to arise from dense PDRs at the interface between 
 \HII~regions and their parental molecular cloud \citep[e.g.,][]{Sta93}.
Therefore, the [\CII]~158\,$\mu$m  line traces the FUV radiation field from massive stars and indirectly, 
 the star formation rate (SFR)  \citep[e.g.,][]{Sta10,Pin14,Her15}.
As in the Milky Way, the [\CII] emission in nearby  galaxies is widespread and ultimately related  to the 
star formation activity 
\citep[][]{Rod06,Mal01,Kap15}.  In addition, an undetermined fraction of the [\CII] 
arises from the diffuse neutral gas \citep{Mad93}  and from the \HII~ionized gas, 
the latter varying 
from $\approx$5 to 50\%, depending on the ionizing radiation  strength and 
electron density \citep[e.g.,][]{Abe06a}.

In normal and starburst galaxies, the [\CII]-to-FIR luminosity\footnote{In this work we refer to the \textit{luminosity} ($L$)
computed in power units (erg\,s$^{-1}$ or $L_{\odot}$). The conversion from velocity-integrated  
line intensities $\int \Delta T_{\rm b}\, d$v = $W$\,(K\,km\,s$^{-1}$), with $\Delta T_{\rm b}$ the 
continuum-subtracted brightness temperature (see Section~\ref{sec-observations}),
to line surface brightness ($I$), is $I$=2$k$\,$W$\,$\nu^3/c^3$. For the \mbox{[\CII]~158\,$\mu$m} and
\mbox{CO $J$=2-1} lines respectively,
the conversion is $I$\,(erg\,s$^{-1}$\,cm$^{-2}$\,sr$^{-1}$)=7.0$\times$10$^{-6}$\,$W$\,(K\,km\,s$^{-1}$)
and 1.3$\times$10$^{-8}$\,$W$\,(K\,km\,s$^{-1}$).} 
ratio ($L$[\CII]/$L_{FIR}$) is  in the  10$^{-2}$ to  10$^{-3}$ range. 
Low metallicity dwarf galaxies show the highest ratios \citep[$\gtrsim$10$^{-2}$, e.g.,][]{Mad97},
whereas  observations with \textit{ISO} showed that
local ultra-luminous galaxies 
(ULIRGs, $L_{\rm FIR}$$>$10$^{12}$\,$L_{\odot}$) display a 
 deficiency, $L$[\CII]/$L_{FIR}$$\simeq$10$^{-4}$, 
that is not straightforward  to interpret \citep[][]{Mal97,Luh98}.
This fractional luminosity deficiency has been confirmed with new observations of 
ULIRGs, hosting intense star formation and  often associated with galaxy mergers \citep[e.g.,][]{Gra11,Gon15}.
 This deficit, however, does not necessarily hold in the early Universe at high-$z$
\mbox{\citep[e.g.,][]{Sta10,Bri15}}.

Interestingly, while it is difficult from the ground to detect the [\CII] emission in the local Universe 
(balloon, airborne, or space telescopes are needed  to overcome the low atmospheric transmission
at 158\,$\mu$m), the line does become accessible to ground-based submm observatories  such as ALMA at redshifts \mbox{$z>$1}.
Indeed, the [\CII] line has been detected toward young $z>5$ galaxies \citep[][]{Rie14,Cap15}.

Most of the  [\CII] emission arising from  PDR gas in the Milky Way is associated with modest average FUV radiation
fields, between  \mbox{$G_0$$\simeq$1-20} \citep{Pin13} and
$G_0$$\simeq$100  \citep[][]{Cub08}, with  $G_0$  the mean
interstellar FUV field in Habing units \citep[$G_0$=1 is equal to 1.6$\times$10$^{-3}$\,erg\,cm$^{-2}$\,s$^{-1}$,][]{Hab68}.
These radiation fields are likely not representative of the efficient massive star forming modes expected
in ULIRGs showing  $L$[\CII]/$L_{FIR}$ deficits.
Galactic high-mass star-forming regions, however, offer a more interesting template  in which 
to investigate the [\CII] emission 
in the more extreme irradiation conditions  prevailing these active galaxies.
The Orion molecular cloud~1 (OMC\,1), in the Orion~A complex, lies
behind the Orion Nebula cluster  \mbox{\citep[e.g.,][]{Ode01}} and is the nearest
($\sim$414\,pc), and probably most studied high-mass star-forming region of the Milky Way \mbox{\citep[e.g.,][]{Gen89}}. 

OMC\,1 is directly exposed to the UV radiation emitted by young massive stars in the 
Trapezium cluster, located $\sim$0.3~pc in front of the molecular cloud  \mbox{\citep[e.g.,][]{Bal08}}. 
A region of  several parsecs in size is  exposed to  FUV  fluxes much higher than the average
 $G_0$ in the Milky Way \citep[e.g.,][]{Sta93,Luh94,Rod98}.
The Trapezium stars have created a blister \HII~region that  on the far side is bounded by the molecular cloud.
The most famous \HII/OMC\,1 interface  is the Orion Bar~PDR (the ``Bright~Bar'' seen in the visible-light
images),
a protrusion of OMC\,1 
in which the neutral cloud acquires a nearly edge-on geometry and thus the optically thin 
PDR emission is limb-brightened
 \citep[][and references therein]{Hog95,Cua15}. 
 On the  near side, toward the observer, several components of neutral atomic gas revealed by \HI~absorption 
cover the nebula,   the \mbox{Orion's Veil} \citep[e.g.,][]{vdW13}.

Intermediate- and  high-mass star formation is currently taking place in two locations:  the 
Becklin-Neugebauer/Kleinmann-Low (BN/KL) region within OMC\,1 \citep[][]{Bla87,Ter10},
 and  Orion~South~(S), an isolated molecular core located within the ionized gas cavity in front of OMC\,1 \citep[e.g.,][]{Ode09}.
Therefore, Orion is a unique region  to study the spatial distribution of [\CII] and related 
 quantities ($L$[\CII]/$L_{FIR}$, $L$[\CII]/$L_{\rm CO}$, $L_{\rm FIR}$/$M_{\rm Gas}$, etc.) in well characterized environments
(dense PDRs, shocks, \HII~regions, and lower density quiescent gas).

The [\CII]~158\,$\mu$m line has been previously mapped toward OMC\,1 at 
low angular ($\sim$55$''$) and spectral ($\sim$67~km\,s$^{-1}$) resolution
with the \textit{Kuiper Airborne Observatory} \citep[KAO,][]{Sta93,Her97}
and also with balloon-borne FIR telescopes \citep[e.g.,][]{Moo03}.
Pointed observations toward the Trapezium cluster region were also carried out  by \citet{Bor96}
with a heterodyne receiver on board KAO. These observations employed a velocity resolution of 0.5~\,km\,s$^{-1}$ 
and allowed the detection of two of the three [$^{13}$\CII] hyperfine  line components.
The HIFI instrument \citep{dG10} on board  \mbox{\textit{Herschel}} \citep{Pil10} allowed us
to spectrally resolve ($\sim$0.2~\,km\,s$^{-1}$ resolution) and to 
map the [\CII]~158\,$\mu$m line with  unprecedented sensitivity and angular 
resolution ($\sim$11.4$''$). Velocity-resolved spectroscopy offers the best way to characterise the origin of all possible
[\CII] emission components and their kinematics. In addition, it allows one to map [$^{13}$\CII] 
and estimate the [\CII] line opacity, a critical parameter to determine C$^+$ column densities
 \citep[][]{Sta91b,Bor96,Gol12,Oss13}.

In this work, we present initial results from a  \textit{Herschel} Open-Time program (OT1\_jgoicoec\_4) devoted to
the spatial and velocity structure of the ionized and the warm molecular gas in OMC\,1.
In particular we present a large-scale velocity-resolved map of the [\CII]~158\,$\mu$m line, complemented by a 
map of the H41$\alpha$ recombination line, and existing \mbox{CO $J$=2-1} and FIR photometric observations. 
The paper is organized as follows. In Section~\ref{sec-observations} we present the data set.
 The  [\CII] spatial distribution and gas kinematics is studied in Section~\ref{sec-results}.
In \mbox{Section~\ref{sec-anal}} we analyze the [\CII]~158\,$\mu$m \mbox{non-LTE} excitation 
and derive C$^+$ column densities.  A spatial correlation analysis of different   quantities
often used in the extragalactic discussion is also carried out.
Finally, our results are discussed in the context of the  [\CII], CO and FIR diagnostic power
and their link with the extragalactic  emission.


\begin{figure*}[t]
\centering
\includegraphics[scale=0.48,angle=0]{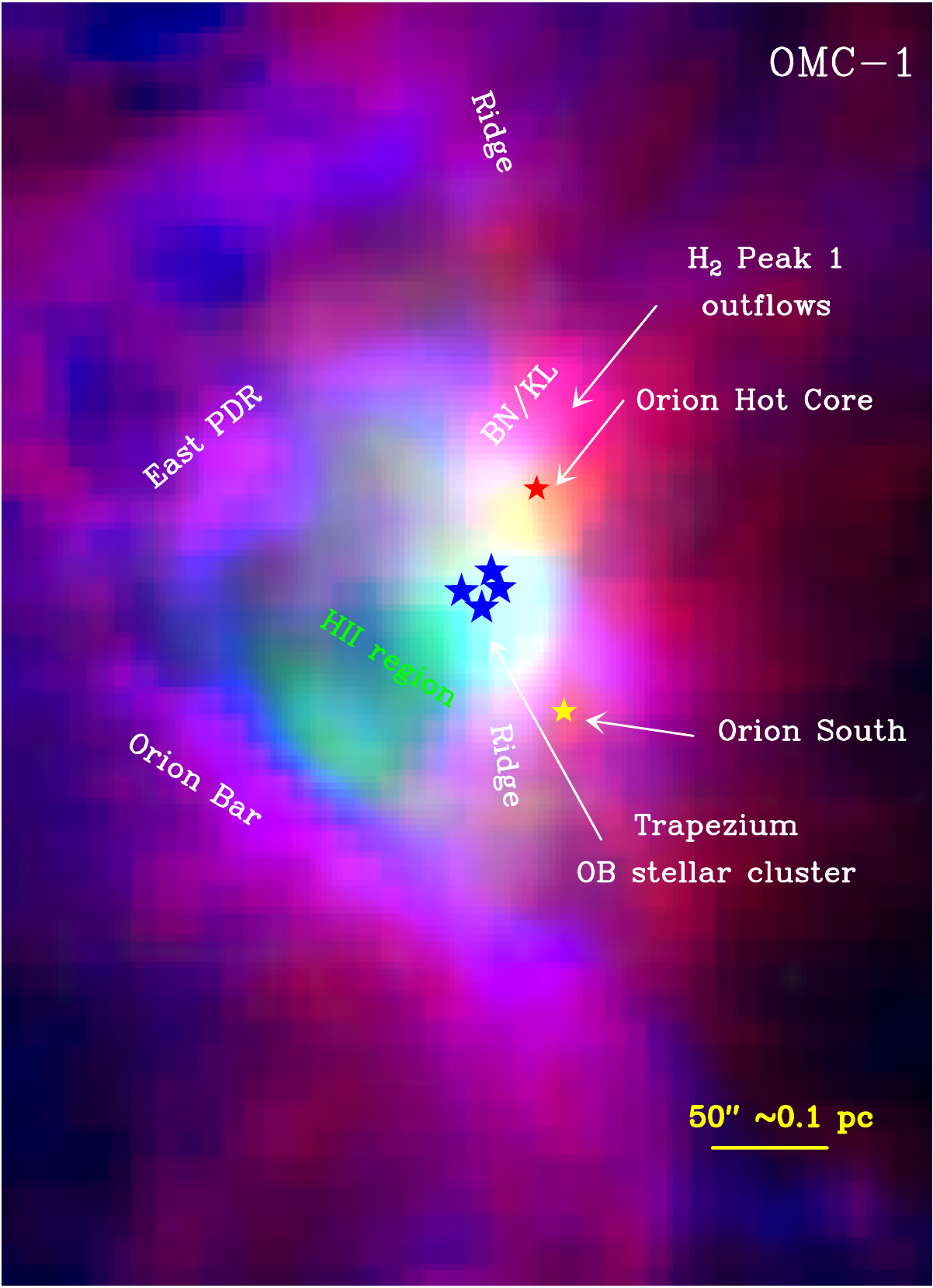} 
\hspace{2.5cm}
\includegraphics[scale=0.38,angle=0]{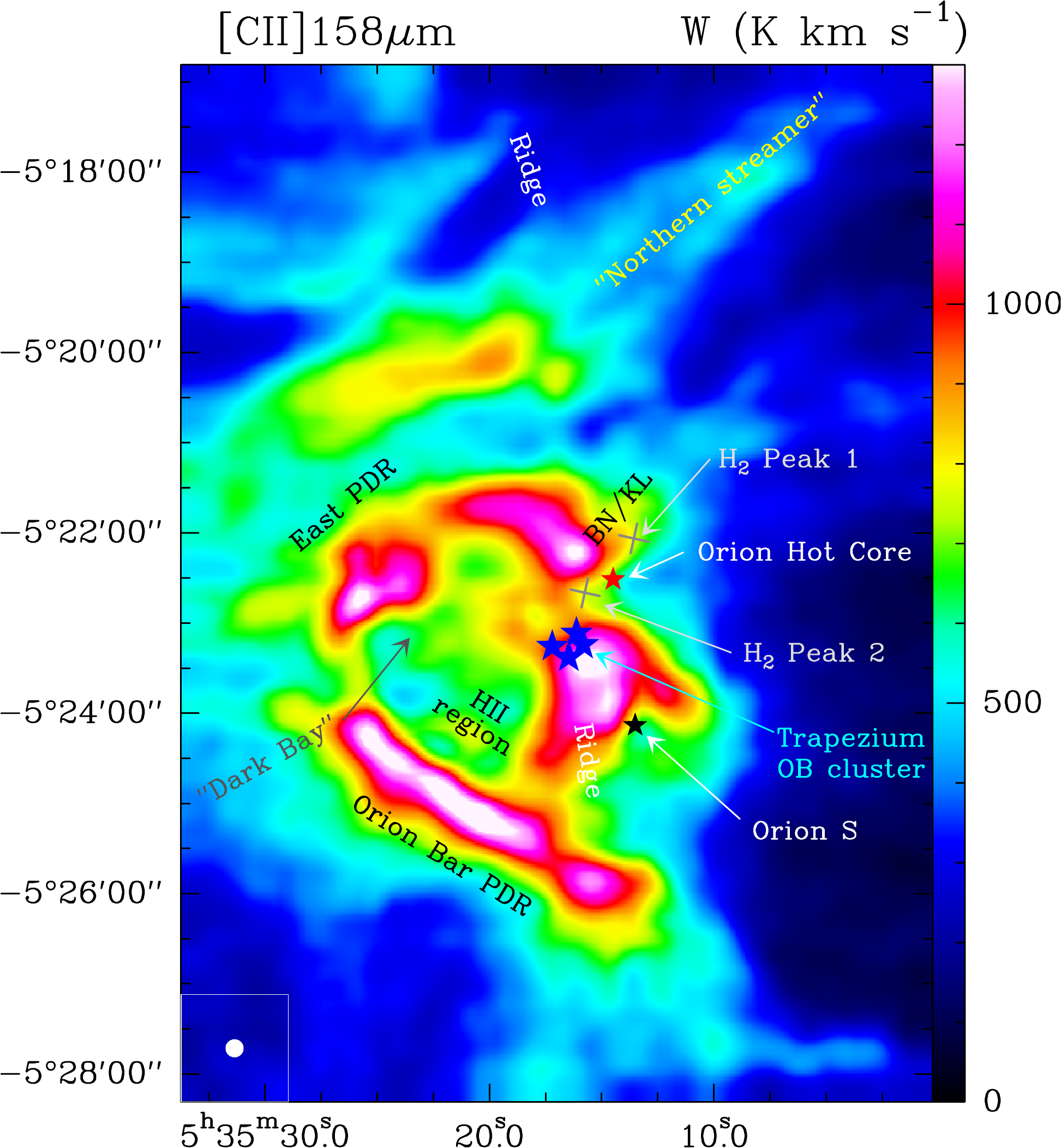} 
\caption{(Left): Composite image with the 
H41$\alpha$ (green), [\CII]\,158\,$\mu$m (blue) 
and CO~2-1 (red) integrated line intensities. 
The positions of the main sources  in OMC\,1 are shown.
(Right): \textit{Herschel}/HIFI map of the   total [\CII]\,158\,$\mu$m  
integrated line intensity$^{3}$  (in K\,km\,s$^{-1}$) from v$_{\rm LSR}$=$-30$ to $+30$~km\,s$^{-1}$.
} 
\label{fig_rgb}
\end{figure*}

\section{Observations and Data Reduction}
\label{sec-observations}
 
\subsection{{\rm[\CII]}, H41$\alpha$ and CO 2-1 Heterodyne Observations}

The [\CII] line at 1900.537~GHz \citep{Coo86} was mapped over a total area of $\sim$7.5$'$$\times$11.5$'$
($\sim$0.9\,pc\,$\times$1.4\,pc)  in 2012 August, using the HIFI instrument on board \mbox{\textit{Herschel}}. 
A pair of Hot Electron Bolometer mixers was employed to observe the [\CII] line in HIFI~band 7b in two orthogonal 
polarizations. We used the Wide Band Acousto-Optical Spectrometer providing a spectral resolution of 1.1\,MHz
and a bandwidth of 2.4~GHz ($\simeq$0.2\,km\,s$^{-1}$ and $\simeq$380\,km\,s$^{-1}$ at 1900.5\,GHz, respectively).

\begin{table}[t]
\caption{\label{table_freqs}  Spectroscopic parameters of the lines discussed in this work} 
\centering
\begin{tabular}{rcrcc}
\hline\hline
        &                              &  Frequency  & $E_{\rm u}/k$     & $A_{\rm ij}$    \\
Species & Transition                   &    (GHz)    &   (K)             & (s$^{-1}$)  \\
\hline
$^{12}$C$^+$           & $^{2}P_{3/2}$-$^{2}P_{1/2}$ & 1900.5369$^a$     &    91.21       &   2.32($-$6)   \\
$^{13}$C$^+$    & $^{2}P_{3/2}$-$^{2}P_{1/2}$        &                   &                &                \\
                &              $F$=1-0               & 1900.9500$^{a,b}$ &    91.23       &   1.55($-$6)   \\
                &              $F$=2-1               & 1900.4661$^{a,b}$ &    91.25       &   2.32($-$6)   \\
                &              $F$=1-1               & 1900.1360$^{a,b}$ &    91.23       &   7.73($-$7)    \\
CO              & $J$=2-1                            &  230.5379         &    16.60       &   6.92($-$7)   \\
H41$\alpha$     & $n$=42-41                          &   92.0344$^c$     &    89.50       &   4.85($+$1)   \\
\hline
\end{tabular}
\tablecomments{$^a$ From \citet{Coo86}. $^b$See \citet{Oss13} for corrected relative line-strengths.
$^c$ From \citet{Lil68}.}
\end{table}

Three $\sim$7.5$'$$\times$4$'$ On-The-Fly (OTF) submaps of 3\,h each were needed to cover the region with
a 10$''$ sampling (\mbox{ObsIDs:~1342250415}, 1342250414 and 1342250412). The half-power beam width (HPBW) of HIFI at these
  frequencies is $\sim$11.4$''$, i.e., the map is not fully sampled.
\mbox{Because} of the large scale [\CII]\ emission \citep{Sta93} and risk of contaminating emission in the
reference position, the OTF maps were performed using the Load-Chop referencing scheme \citep{Roe12}. In this mode,
an internal chopper alternates between the sky and an internal cold load as the telescope slews over a scan leg. Although
a nearby reference position (at a radial offset of $\sim$ 9$'$ from the map centre) was also observed, we
reprocessed the data without subtracting this reference emission from the on-source data, therefore being 
insensitive to line contamination from
self-chopping. The resulting data are mostly affected by so-called Electrical Standing Waves, that were removed in
 HIPE\footnote{HIPE is a joint development by the Herschel Science Ground Segment Consortium, consisting of ESA, the NASA Herschel Science Center, and
the HIFI, PACS, and SPIRE consortia.}
using the \mbox{{\it doHebCorrection}} task \citep{Kes14}. 
The residual optical standing waves resulting from
resonances against the internal hot load ($\sim$100\,MHz
period) and the diplexer optics ($\sim$625\,MHz) were treated 
using the \mbox{{\it FitHifiFringe}} task in HIPE.
The three sub-maps were then compared pair by pair in a region of $\sim$7.5$'$$\times$0.6$'$ where 
they overlap on the sky. 
Antenna temperatures ($T_{\rm A}^{*}$) were extracted for each overlapping pixel and  [\CII] velocity channel. 
Scatter plots were formed to derive the inter-calibration factor between the respective coverage in each channel. We derived linear slopes deviating from unity by $\sim$5\% between pairs of sub-maps.
Since we cannot know a priori which map corresponds to the true intensity, we re-adjusted the
fluxes by $\pm$2.5\% depending on the sub-map, i.e. re-calibrated with respect to the mean intensities in the overlapping regions.

Through this work we use the main beam temperature scale ($T_{\rm mb}$ in K) as opposed to $T_{\rm A}^{*}$.
Given the latest HIFI efficiencies, both 
scales\footnote{See: \path{http://herschel.esac.esa.int/twiki/pub/Public/HifiCalibrationWeb/HifiBeamReleaseNote_Sep2014.pdf}} are related by  
\mbox{$T_{\rm mb}$$\simeq$1.64\,$T_{\rm A}^{*}$}.
 For semi-extended   (not infinite but larger than the main beam width of \textit{Herschel}/HIFI) emission sources of uniform brightness
temperature ($T_{\rm b}$), the main beam temperature 
is the most appropriate intensity scale ($T_{\rm mb}$$\simeq$$T_{\rm b}$).
The achieved rms noise was typically  $\sim$1.5\,K (1$\sigma$) per 0.2\,km\,s$^{-1}$ resolution channel and position.
 This rms is only a factor of $\sim$2.5 worse than that achieved by \citet{Bor96} in their deep pointed observations after
an integration of $\sim$3\,h.
\mbox{Figure~\ref{fig_rgb}~(right)} shows the baseline-subtracted [\CII] 
integrated line intensity map.
    
The H41$\alpha$ millimeter recombination line at 92.034~GHz was mapped with the IRAM-30m 
telescope at Pico Veleta (Spain) using the E090 receiver and the 200\,kHz FFTS backend in 2013 November.
The HPBW at this frequency is $\sim$27$''$ and the  typical rms noise in the map is low, $\sim$0.05\,K per 0.65\,km\,s$^{-1}$
 resolution channel. 
In this work we also make use of part of the high-sensitivity \mbox{CO 2-1} large-scale map (1$\times$0.8$^{\circ}$) at 230.538~GHz
obtained by \citet{Ber14} with the multi-beam receiver HERA, also at the \mbox{IRAM-30m} telescope. 
The spectral and angular resolutions are
 320\,kHz and $\sim$11$''$, respectively. The achieved rms noise  is $\simeq$0.2\,K per 0.4\,km\,s$^{-1}$ channel
 \citep[see][]{Ber14}. Both H41$\alpha$  and \mbox{CO 2-1} OTF maps are fully sampled.
 
\textit{Herschel}/HIFI and \mbox{IRAM-30m} data were processed with the GILDAS/CLASS software. 
\mbox{Figures~\ref{fig_rgb}, \ref{fig_channels_H41}, and \ref{fig_channels_CO}} show the [\CII]~158\,$\mu$m, H41$\alpha$ 
and CO~2-1 emission at their native angular resolutions but resampled to a common velocity grid of
\mbox{0.65\,km\,s$^{-1}$}.  Offsets of selected positions are given with respect
to the center of the [\CII] map: \mbox{$\alpha_{2000}$:~5$^h$35$^m$17.0$^s$}, \mbox{$\delta_{2000}$:~$-5^o$22$'$33.7$''$}.
 Table~\ref{table_freqs} summarizes the main spectroscopic parameters of the
lines discussed in this work.
To match the FIR photometric observations and carry out a combined analysis, 
the line maps were also gridded and convolved to an uniform angular resolution of 25$''$ ($\sim$0.05\,pc).

\begin{figure*}[t]
\centering
\includegraphics[scale=0.352, angle=-0]{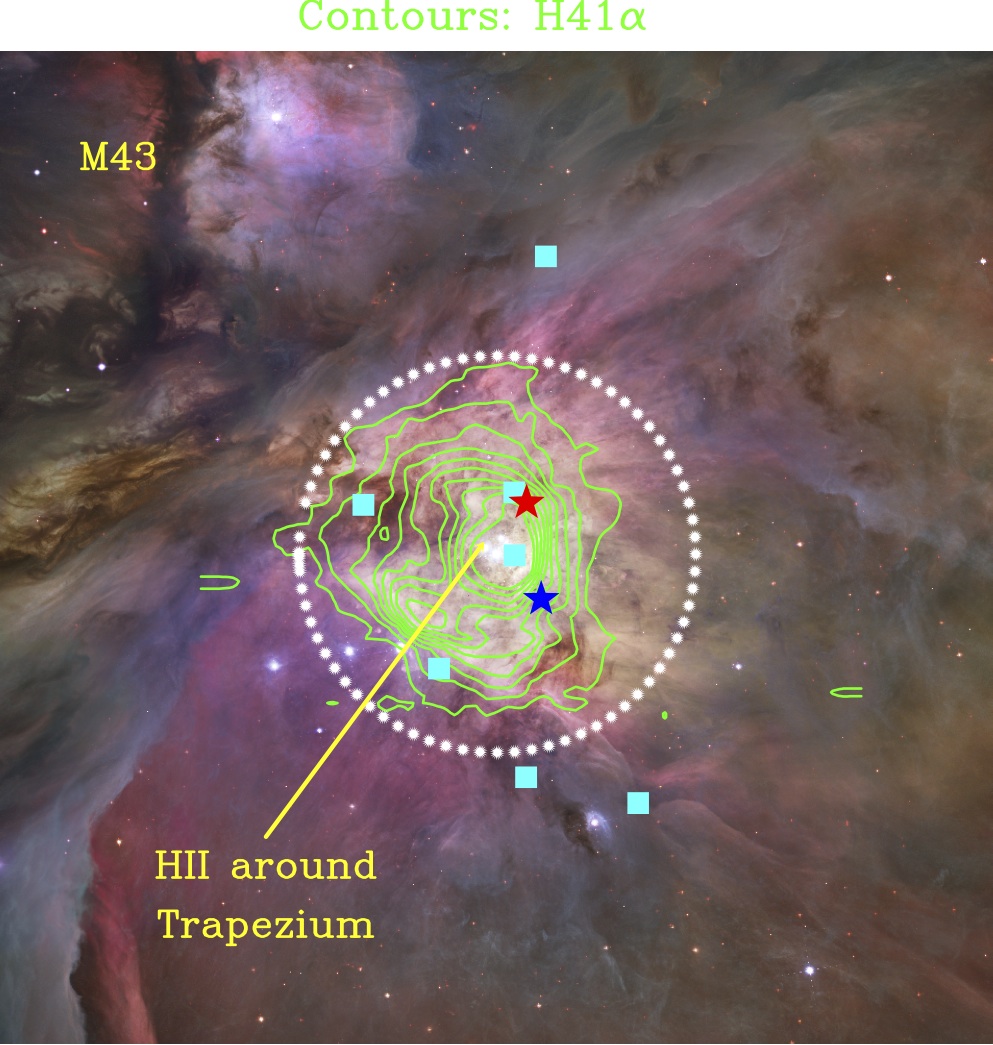}   
\includegraphics[scale=0.352, angle=-0]{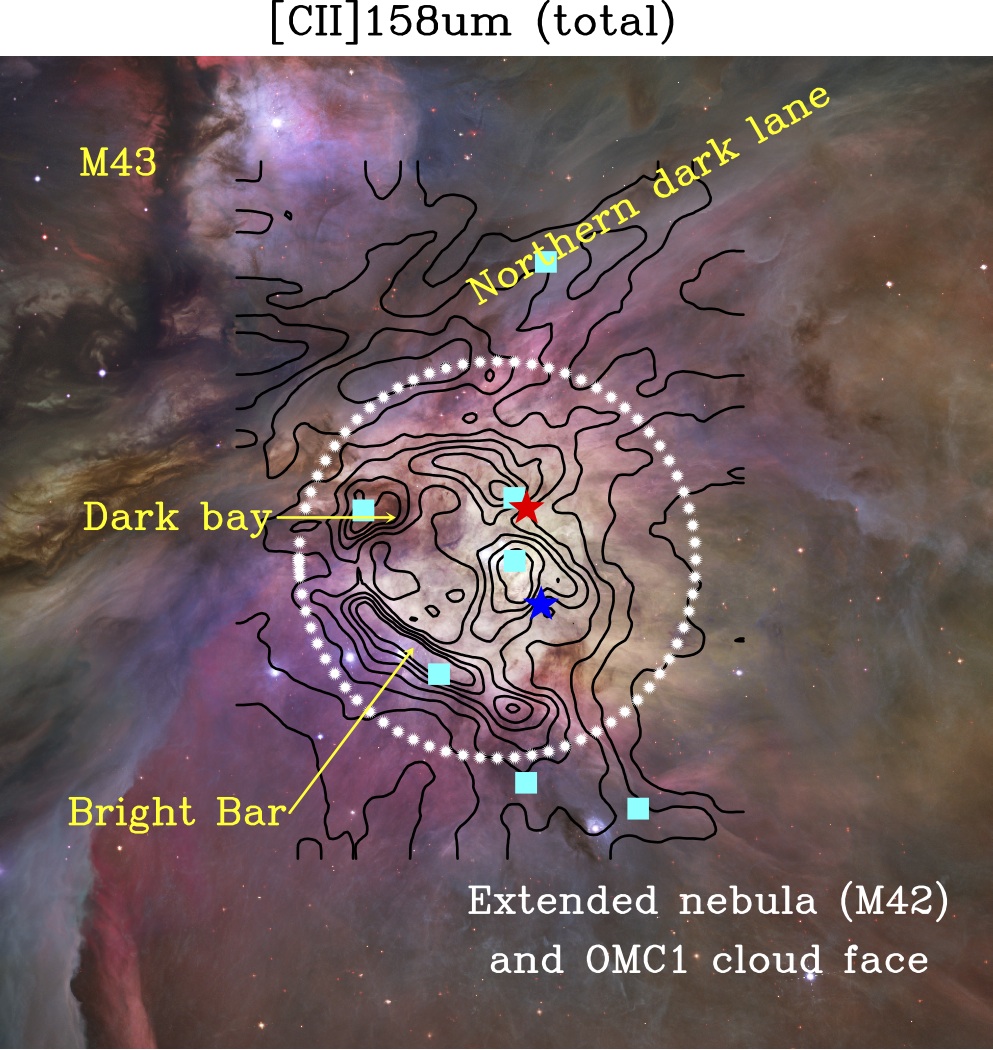} 
\includegraphics[scale=0.352, angle=-0]{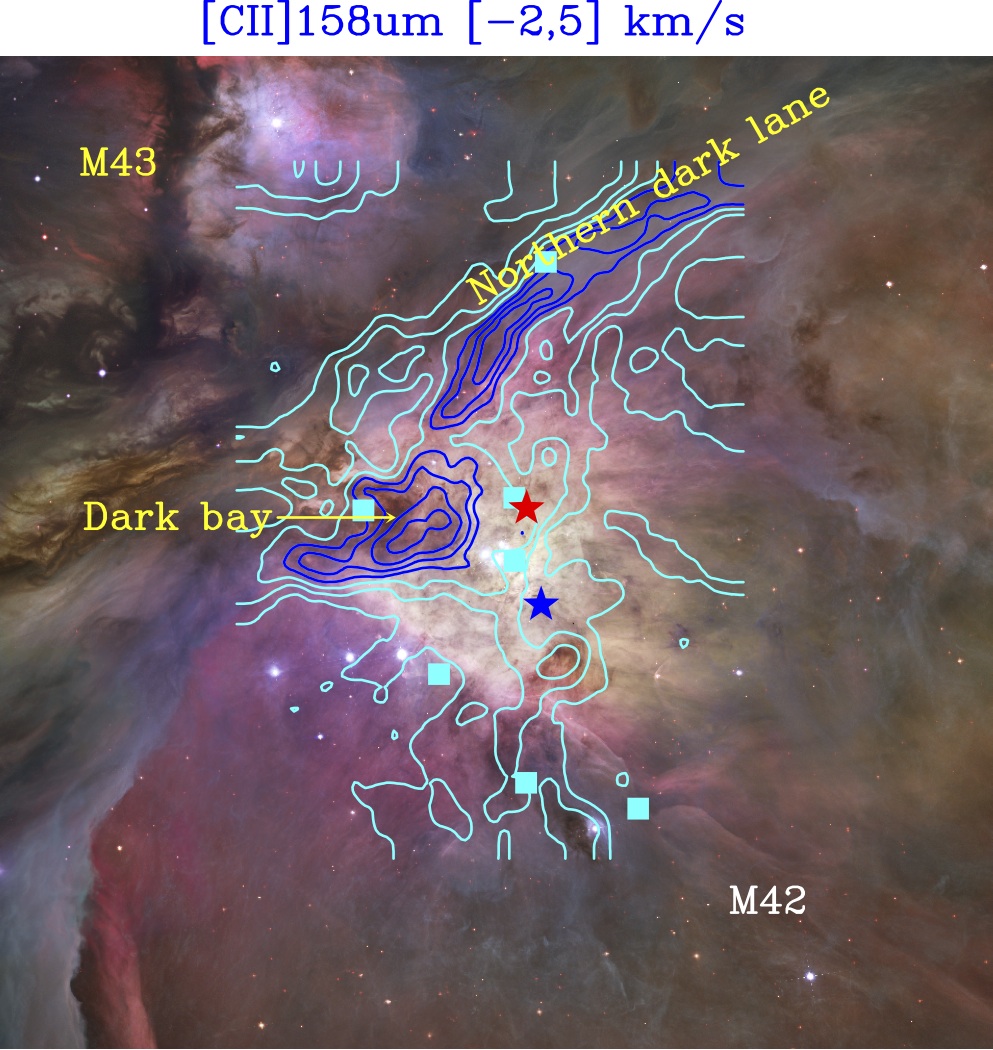}  
\caption{Color-composite visible image of the Orion Nebula (M42) taken with the ACS instrument
on board the \textit{Hubble Space Telescope (HST)}. The assigned colors are: 
red and orange for [\SII] and H$\alpha$, blue for $B$, and green for $V$ filters respectively \citep[][]{Rob13}.
\textit{(Left)}: Green contours show the integrated intensity ($W$) of the H41$\alpha$ recombination line
(from 3 to 21~K\,km\,s$^{-1}$ in steps of 2~K\,km\,s$^{-1}$).
The white dotted 200$''$ (0.4\,pc) radius circle  is centered on $\theta^1$\,Ori\,C, the brightest star of the Trapezium. 
We define the ``extended cloud face'' component to lie outside this circle. 
\textit{Middle}: Black contours show the total [\CII]\,158\,$\mu$m integrated intensity  
(v$_{\rm LSR}$=$-30$ to $+30$~km\,s$^{-1}$ with 
$W$ from 0 to 1300 K\,km\,s$^{-1}$  in steps of 150~K\,km\,s$^{-1}$).
\textit{Right}:  [\CII] ``blue-shifted streamer''   
(emission from \mbox{v$_{\rm LSR}$=$-$2 to $+$5~km\,$^{-1}$}) with $W$ from 0 to 60 K\,km\,s$^{−1}$ (cyan) and from 80 to 140 K\,km\,s$^{−1}$
 (blue), all in steps of 20 K\,km\,s$^{-1}$. 
 This emission coincides with the \textit{Northern dark lane} and with the \textit{Dark bay} seen in
visible-light images of the Nebula.  The Orion hot core in the BN/KL region and Orion~S  positions are indicated by red and blue  stars respectively.
The positions where line spectra have been extracted (shown in Figure~\ref{fig_spectra}) 
are indicated by cyan squares.} 
\vspace{0.25cm}
\label{fig_nebula}
\end{figure*}

\subsection{Photometric Observations and SED Fits}\label{photometry}

In order to determine the FIR luminosity\footnote{$L_{FIR}$ from 40 to 500\,$\mu$m \citep{San96} 
and using $D$=414\,pc.} and the dust continuum opacity toward each position of the [\CII] map,
we also make use of fully sampled calibrated images of OMC\,1 taken with the \textit{Herschel}/SPIRE 
and PACS\footnote{The central 
region of the photometric image (BN/KL outflows and IRc  sources) 
contain saturated data points that are not used in our analysis.}
cameras \citep{Gri10,Pog10} at 70, 160, 250, 350\,$\mu$m photometric bands
(from programs \mbox{GT2$\_$pandre}  and \mbox{OT1$\_$nbillot}), and
 with JCMT/SCUBA2 at 850\,$\mu$m \citep[project JCMTCAL;][]{Hol2013}.

We convolved all the photometric images to a 25$''$ resolution (that of the 350\,$\mu$m image) 
and fitted the resulting spectral energy distribution (SED) per 12.5$''$-pixel with  
a modified black body at an \textit{effective} dust temperature ($T_{\rm d}$); 
\begin{equation}
\label{eq-flux}  
F_{\lambda}=B_{\lambda}(T_{\rm d})\cdot(1-e^{-\tau_{\rm d,\lambda}})\cdot \Omega
\end{equation}
where $\Omega$ is the solid angle subtended by each pixel. Fit examples are shown in the Appendix. 
The dust continuum opacity is parametrized as 
\mbox{$\tau_{\rm d,\lambda}=\tau_{\rm d,160}(160/\lambda)^\beta$}, with $\tau_{\rm d,160}$ the 
dust opacity close to the [\CII]~158\,$\mu$m line, and $\beta$ a grain emissivity index.
The resulting $L_{FIR}$ and $\tau_{\rm d,160}$ maps, where we have allowed $\beta$ to be a free parameter
 of the fits, are shown in \mbox{Figures~\ref{fig_lums}(a)} and (f). The mean  $\beta$ value is 2.4$\pm$0.5~(1$\sigma$) but we note that
the derived FIR luminosities and dust opacities, the relevant quantities for our analysis,
 do not change significantly by fixing $\beta$ to
a constant value of 2. 
We also estimated the \mbox{pixel-averaged} gas column density and mass. In particular, we
used the SPIRE\,250\,$\mu$m image assuming that it corresponds to optically thin dust emission and
computed  the equivalent gas mass per pixel as: 
\begin{equation} 
\label{eq-mass}
M_{\rm Gas}=\frac{R_{\rm gd} S_{250} D^2}{\kappa_{250}  B_{250}(T_{\rm d})}
\end{equation}
\citep[][]{Hil83}, where $S_{250}$ is the 250\,$\mu$m flux density within $\Omega$, $R_{\rm gd}$ is the gas-to-dust mass 
ratio  and $\kappa_{250}$ is the dust absorption cross-section divided by the dust mass. 
We  adopt  \mbox{$R_{\rm gd}$/$\kappa_{250}$=25\,gr\,cm$^{-2}$},   
computed from grain properties taken from Draine's tabulations \citep[][and references therein]{Li01}
for a grain size distribution that leads to a $A_V$/$N_{\rm H}$ ratio compatible with observations of Orion
 (see Section~\ref{sect-circles}).
\section{Results}
\label{sec-results}

\subsection{Global Morphology of the [\CII]\,158$\mu$m Emission}

Figure~\ref{fig_rgb} (left) shows a composite image of OMC\,1 in the  integrated line emission
of \mbox{CO~2-1} (red), [\CII]\,158\,$\mu$m (blue) 
and H41$\alpha$ (green). This image shows the \HII/PDR/molecular gas stratification
expected in molecular clouds irradiated by strong UV fields from nearby massive stars.

The  \mbox{H41$\alpha$ recombination line} traces dense ionized gas  from
the blister \HII~region surrounding the Trapezium. The H41$\alpha$ line
peaks near the OB stellar cluster and is also bright adjacent to the Orion Bar PDR. 
Despite the different angular resolutions, the H41$\alpha$ emission morphology is similar to the brightest regions seen in the
 visible-light H$\alpha$ images of the Orion Nebula (M42) 
\citep[\mbox{Figure \ref{fig_nebula}~$\textit{left}$};][]{Rob13}.

\begin{figure}[t]
\centering
\includegraphics[scale=0.55, angle=0]{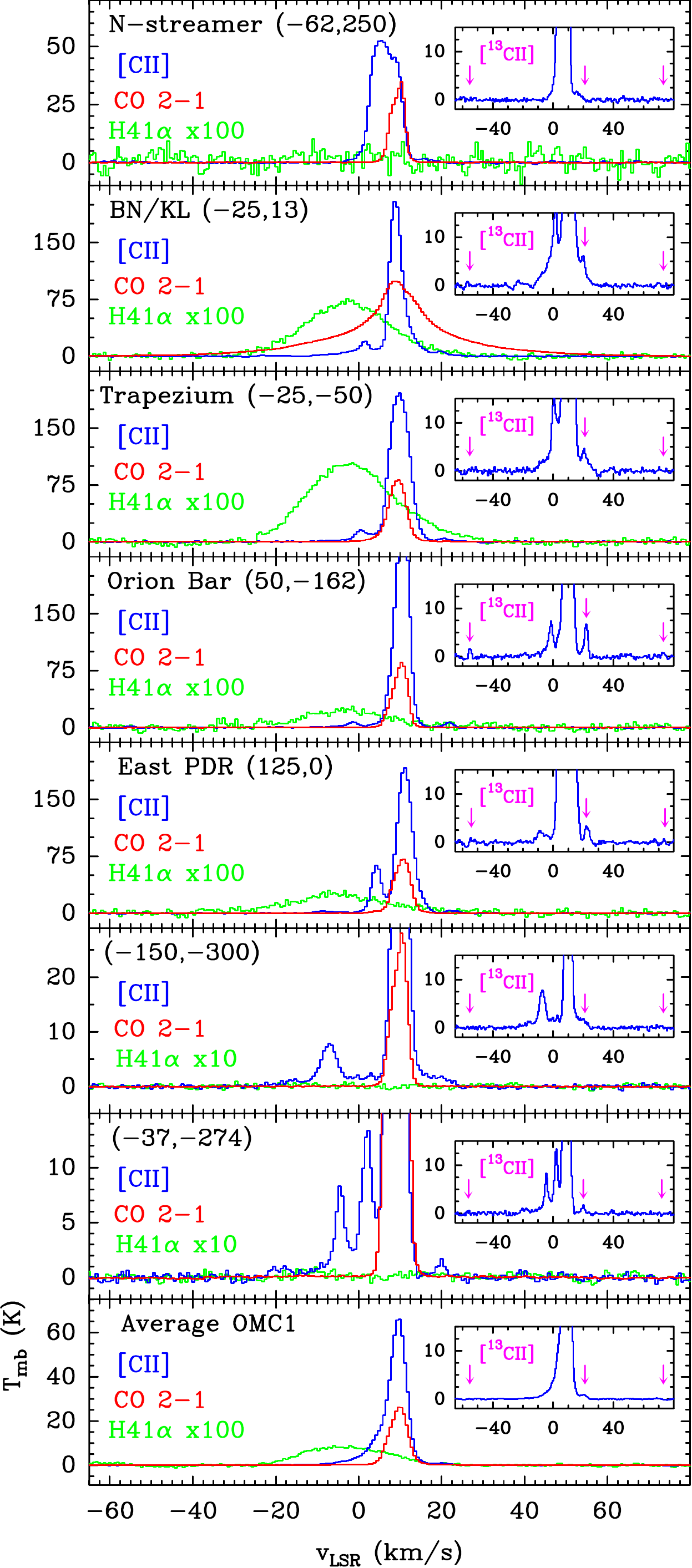}
\hspace{1cm}
\caption{[\CII]\,158\,$\mu$m, CO 2-1 and H41$\alpha$ line profiles toward 
 representative positions in Orion. Offsets in arcsec are given with respect
to the HIFI map center at \mbox{$\alpha_{2000}$:~5$^h$35$^m$17.0$^s$}, \mbox{$\delta_{2000}$:~$-5^o$22$'$33.7$''$}. 
The last panel shows the average spectrum over the region.
The inset panels show a zoom into the spectra.
The  magenta arrows mark the expected positions of the [$^{13}$\CII] \mbox{$F$=1-0}, \mbox{2-1} and \mbox{1-1} 
hyperfine components (from left to right).
The [\CII] and CO spectra were extracted from maps convolved to 25$''$ resolution.}
\label{fig_spectra}
\end{figure}

\begin{figure}[t]
\centering
\includegraphics[scale=0.38, angle=-0]{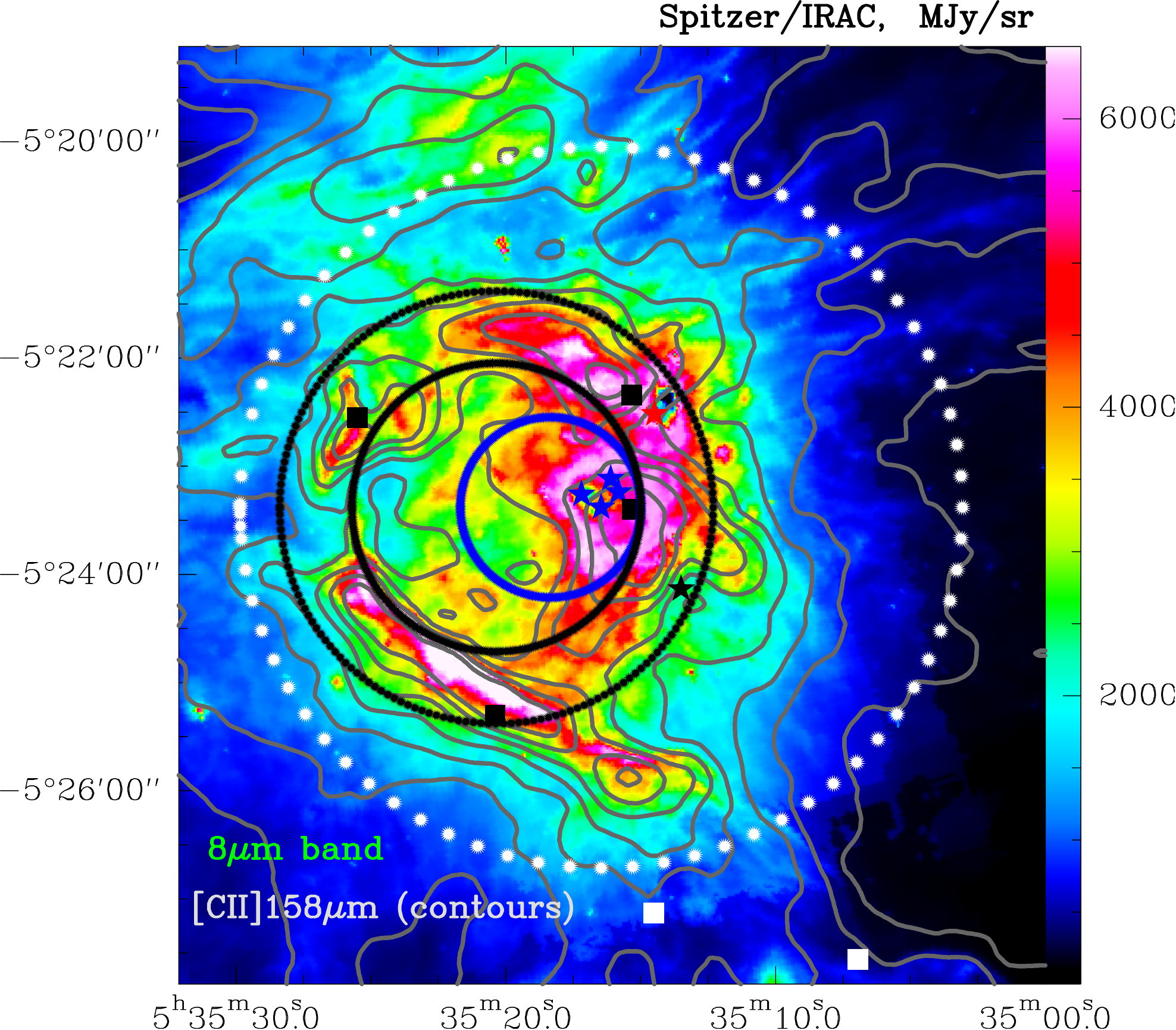}
\caption{ \textit{Spitzer}/IRAC~8\,$\mu$m filter image (dominated by PAH emission)
toward OMC\,1. Contours show the 
 total [\CII]~158\,$\mu$m integrated intensity 
 (from 0 to 1300~K\,km\,s$^{-1}$ in steps of 150~K\,km\,s$^{-1}$).
The different circles and regions are defined in Section~\ref{sect-circles}.
The positions where line spectra have been extracted (shown in Figure~\ref{fig_spectra}) 
are indicated by squares.} 
\label{fig_PAHs}
\end{figure}

Figure~\ref{fig_spectra} shows [\CII], CO~2-1 and H41$\alpha$  spectra extracted from representative lines of sight toward OMC\,1, namely 
the northern streamer, BN/KL, the Trapezium, the Orion Bar, the East PDR, and the extended cloud face in the south.
OMC\,1 has an overall face-on geometry, with the
molecular cloud located \mbox{$\sim$0.3-0.6\,pc} behind  the ionizing stars \citep[e.g.,][]{Hog95,vdW13}.
Therefore, the observed  \mbox{$\sim$10-15\,km\,s$^{-1}$}  blue-shift of the  H41$\alpha$ line relative to the [\CII] and CO lines 
(Figure~\ref{fig_spectra}) is consistent
with streamers of ionized gas escaping from the high-pressure \HII/OMC\,1 interface and moving toward the observer 
\citep[e.g.,][]{Gen89}.

The regions of brightest [\CII]~158\,$\mu$m emission delineate the  far edges of the \HII~blister and follow 
a roughly spherical shell distribution (as projected in the plane of the sky) with emission peaks near the Trapezium, at the east of BN/KL (that we
refer to as the \mbox{``East PDR''}) and along
the Orion Bar PDR (\mbox{Figure~\ref{fig_rgb}~right}). These \mbox{``interface''} regions 
 show very bright [\CII] emission, with 
 main beam temperature peaks of $\gtrsim$150\,K (or a surface brightness of
$\gtrsim$5$\times$10$^{-3}$\,erg\,s$^{-1}$\,cm$^{-2}$\,sr$^{-1}$ for the average
\mbox{$\Delta$v$\simeq$5\,km\,s$^{-1}$}
 line-width, see  Figure~\ref{fig_spectra}). In addition,
a more extended component of fainter [\CII]~158\,$\mu$m emission exists.
In particular, the average line peak  temperature in the map is $\sim$70\,K. 
This reflects the presence of a lower density and  cooler component, the extended FUV-illuminated ``face'' of OMC\,1
previously detected in [\CII] at lower angular and spectral resolutions \citep{Sta93,Her97}. 
 This extended component is also seen in the wide-field images of M42 taken in the emission from
other low ionization potential ($<$13.6\,eV) atoms (see the reddish nebular emission in
Figure~\ref{fig_nebula} taken with the [\SII] filter in the visible).
It is also seen in the near-IR vibrationally excited H$_2$ emission
attributed to large-scale FUV fluorescence \citep{Luh94}.

Most of the  [\CII] (and obviously CO) emission is expected to arise from the neutral cloud. This can be readily seen
from the similar [\CII]~158\,$\mu$m  and \mbox{CO 2-1} line profiles and also similar line peak velocities
(Figure~\ref{fig_spectra}). 
At the small spatial scales
revealed by our observations ($\sim$5000\,AU), however, both the integrated  intensity map (Figure~\ref{fig_rgb} left)
and the velocity channel maps (Figure~\ref{fig_channels_CO})  show 
that the [\CII] and \mbox{CO\,2-1} spatial distribution does not exactly match. 
This is more obvious toward (nearly) edge-on PDRs, like the Orion Bar,  because  the irradiated cloud edge  (emitting [\CII]) can be resolved from the  more shielded molecular gas, where CO reaches its abundance peak
 ($A_V$$\gtrsim$3\,mag for dense gas and strong  FUV fields,  see  the representative model in Fig.~\ref{fig_PDR}).

The integrated intensity  maps also reveal regions of moderately bright 
[\CII]~158\,$\mu$m emission, but very weak  CO emission counterpart (northeastern areas). In addition,
they show regions,  such as the shocked  gas toward \mbox{H$_2$~Peaks 1 and 2}
in \mbox{Orion~BN/KL outflows}, where the
\mbox{CO} line luminosities are exceptionally high \citep[see also][]{Goi15}. 

Finally, Figure~\ref{fig_PAHs} shows the
good spatial correspondence between the [\CII]~158\,$\mu$m integrated intensity (contours) and the 
\textit{Spitzer}/IRAC~8\,$\mu$m  emission 
\citep[e.g.,][]{Meg11}.
This image is dominated by fluorescent IR emission from PAHs 
\citep[after absorption of UV photons, e.g.,][]{All89} and by very small grain emission, 
 with minor contribution from
[\ArIII] toward the \HII~region  and from 
H$_2$ lines toward particular positions of enhanced shocked gas emission \citep{Ros00}.

\subsection{Template Regions and Representative PDR Model}
\label{sect-circles}

Based on our knowledge of Orion \citep[e.g.,][]{Gen89,Sta93,Bal08}, and in order to ease the interpretation of 
the [\CII] and related maps,  we  introduce three different \textit{template} regions (or sightlines).  
We first consider a 200$''$ (0.4\,pc) radius circle  centered on $\theta^1$\,Ori\,C, the brightest star of the Trapezium 
(e.g., the white dotted circle shown in \mbox{Figure~\ref{fig_PAHs}}). 
We define the \mbox{``extended cloud face''} component for positions outside this circle. 
Secondly,  a small $R$$<$0.1\,pc circle around the Trapezium defines lines of sight that go through
 the \mbox{``\HII~region''} component around the H41$\alpha$ emission peak.
Finally, we consider a spherical shell, or 
annulus, delineated by the \mbox{$R$=0.16\,pc} and \mbox{$R$=0.24\,pc} circles. This region represents \mbox{``PDR interfaces"}
between OMC\,1 and the \HII~region. These PDRs (specially the Bar) show a more
edge-on configuration and are illuminated  by the strong, almost unattenuated, UV field  from the Trapezium stars.

\begin{figure}[t]
\centering
\includegraphics[scale=0.48, angle=-0]{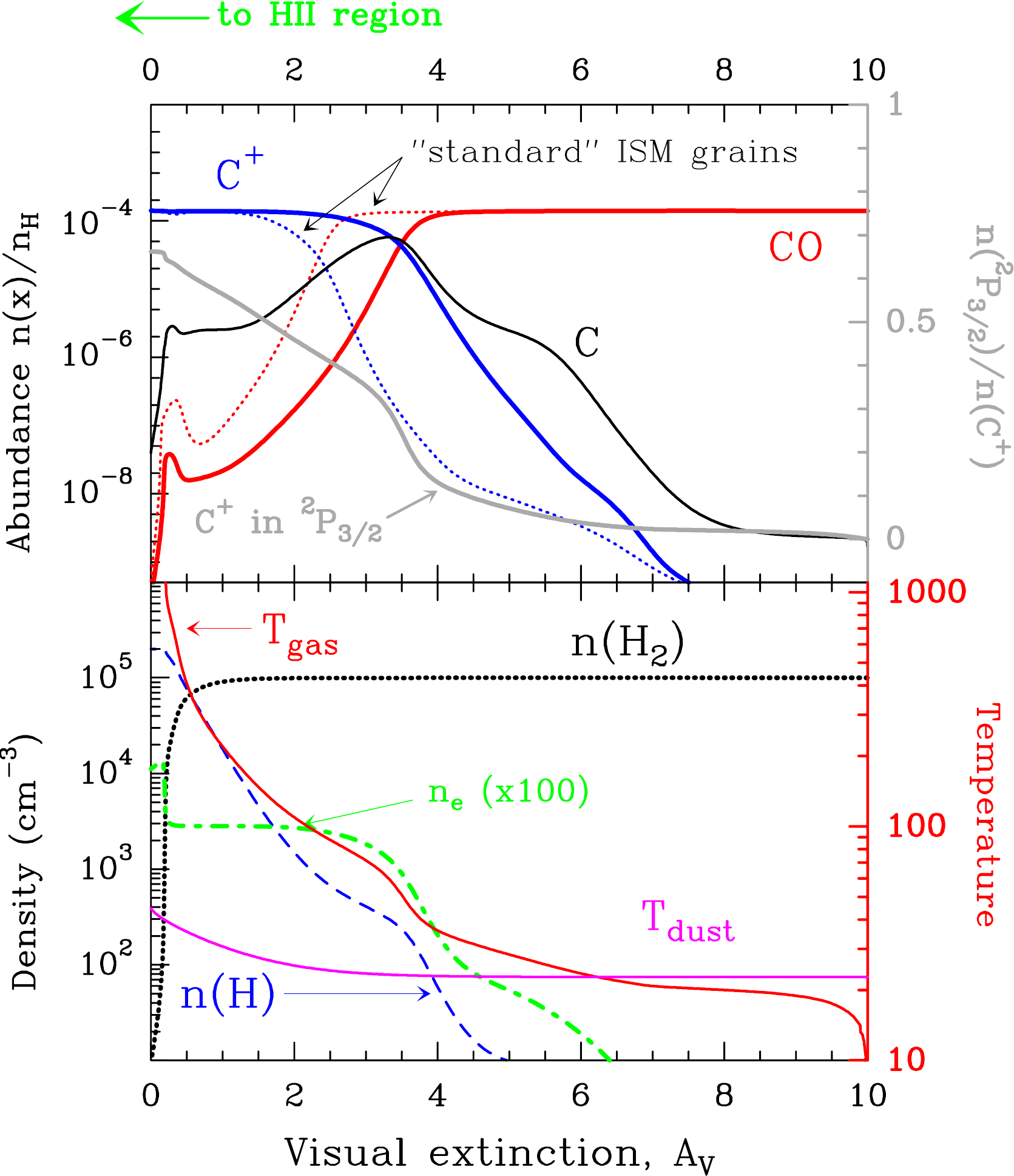}
\caption{H/H$_2$ and C$^+$/C/CO transitions predicted by the \textit{Meudon}~PDR code
for physical conditions and grain properties representative of OMC\,1 near the Trapezium cluster 
($G_0$=2$\times$10$^4$ and \mbox{$n_{\rm H}$=$n$(H)+2\,$n$(H$_2$)=2$\times$10$^5$~cm$^{-3}$)}. 
The equivalent length of the $A_V$=10\,mag slab is $\simeq$10,000\,AU ($\simeq$24$''$ at the distance 
to OMC\,1). The dotted C$^+$ and CO abundance profiles in the upper panel are obtained using standard, 
diffuse ISM, grain properties (see Section~\ref{sect-circles}). }
\label{fig_PDR}
\end{figure}

 To guide our interpretation, Figure~\ref{fig_PDR} shows the predicted
 structure of the H/H$_2$ and C$^+$/C/CO transitions as a function of visual extinction ($A_V$)
for physical conditions ($G_0$=2$\times$10$^4$ and \mbox{$n_{\rm H}$=2$\times$10$^5$~cm$^{-3}$)} 
representative of the \HII/OMC\,1 interface near the Trapezium (see Section~\ref{sec-anal}).
This homogeneous, stationary slab cloud model was computed with the \textit{Meudon} PDR code 
\citep[][]{LP06}. We adopted [C/H]=1.4$\times$10$^{-4}$ \citep[from \textit{HST} absorption observations
towards the star HD~37021 in the Trapezium,][]{Sof04}  and a flat extinction curve compatible with observations of Orion \citep[][]{Lee68,Car89,All05,Abe06}. 
In particular, an extinction to color index ratio {\mbox{($R_V=A_V/E_{B-V}$}) of  5.5, 
and \mbox{$A_V/N_{\rm H}$=3.5$\times$10$^{-22}$\,mag\,cm$^{-2}$} were adopted.
These values are different than the standard grain properties used in diffuse cloud models ({\mbox{$R_V = 3.1$}} and 
$A_V/N_{\rm H}$=5.3$\times$10$^{-22}$\,mag\,cm$^{-2}$) and are consistent with larger than standard size grains
in  Orion. Larger grains  lead to an increased penetration
of FUV photons, larger size of the C$^+$ zone (\mbox{$\Delta A_V$(C$^+$)$\gtrsim$1\,mag}), 
higher C$^+$ column densities, and shift the peak CO formation layer to larger $A_V$
deeper inside the cloud  \mbox{\citep[see also][]{Goi07}}.

 More in general, an ionization front can move into the molecular cloud. 
Ionized gas will rapidly photoevaporate from the cloud surface and the PDR structure will not be in equilibrium 
{\mbox{\citep[e.g.,][]{Ber96}}. Thus, dynamical effects might alter the picture shown in Figure~\ref{fig_PDR}. 
\mbox{Non-stationary} PDR models
predict that a well defined C$^+$/CO transition layer always exists, although the layer is slightly shifted closer
to the cloud surface (by only $\Delta A_V$$\simeq$0.5 for $G_0$/$n_{\rm H}$ conditions appropriate to OMC\,1)
due to advection of CO from  inside the  cloud \citep{Sto98}.
Since the predicted C$^+$ column density in these models does not vary much compared to that in stationary models,
the general predictions  shown in Figure~\ref{fig_PDR} for C$^+$  should remain valid.

\begin{figure*}[t]
\centering
\includegraphics[scale=0.605, angle=-0]{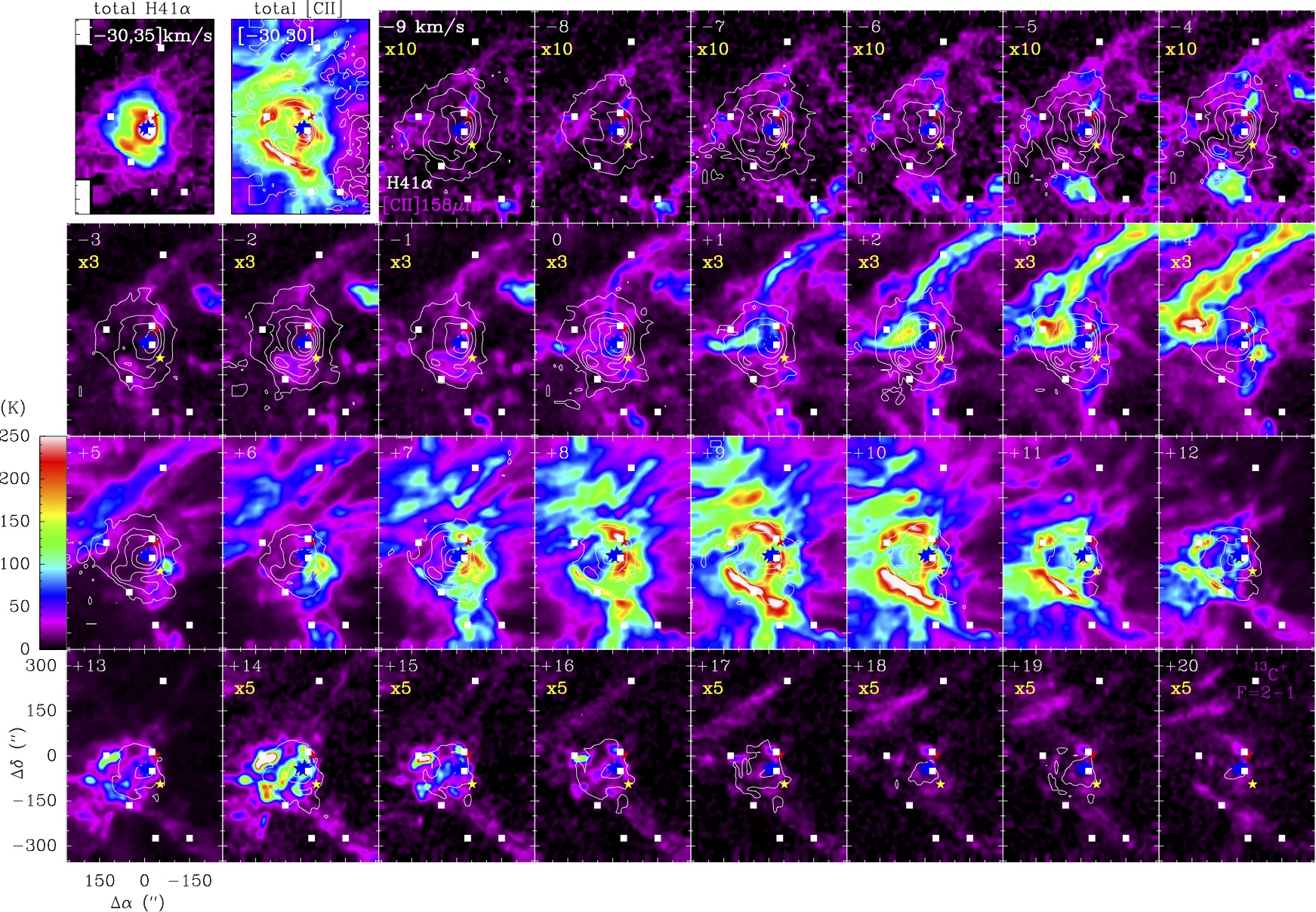}
\caption{Velocity channel maps in steps of 1\,km\,s$^{-1}$ from v$_{\rm LSR}=-9$ to $+20$\,km\,s$^{-1}$ 
([\CII] in colors and H41$\alpha$ line in contours). 
The [\CII]  main beam temperature scale in bins of 
1~km\,s$^{-1}$ is given in the color wedge (in some velocity intervals the [\CII] intensity per channel
 is multiplied by a factor that
is specified in each panel).  Offsets  in arcsec are given with respect to the HIFI map center.
The H41$\alpha$ contours go from 0.2 to 1.8~K\,km\,s$^{-1}$ in steps of 0.3~K\,km\,s$^{-1}$.
The [\CII] emission seen at \mbox{v$_{\rm LSR}=+20$\,km\,s$^{-1}$}
(last panel) is dominated by the \mbox{[$^{13}$\CII]\,$F$=2-1} line peak. 
Note that the first two upper-left panels
show total integrated line intensity maps for H41$\alpha$ (from 0 to 22 K\,km\,s$^{-1}$) and
[\CII]\,158\,$\mu$m (from 0 to 1300 K\,km\,s$^{-1}$). In the total [\CII] intensity map, H41$\alpha$ integrated line
intensities are shown as white contours (from  1 to 21~K\,km\,s$^{-1}$ in steps of 4~K\,km\,s$^{-1}$).
The Orion hot core, Orion~S condensation and  
Trapezium stars are  indicated by red, yellow and blue stars respectively. The positions where line spectra have been extracted
(shown in Figure~\ref{fig_spectra}) are  indicated by white squares.}
\vspace{0.25cm}
\label{fig_channels_H41}
\end{figure*}

\begin{figure*}[t]
\centering
\includegraphics[scale=0.605, angle=0]{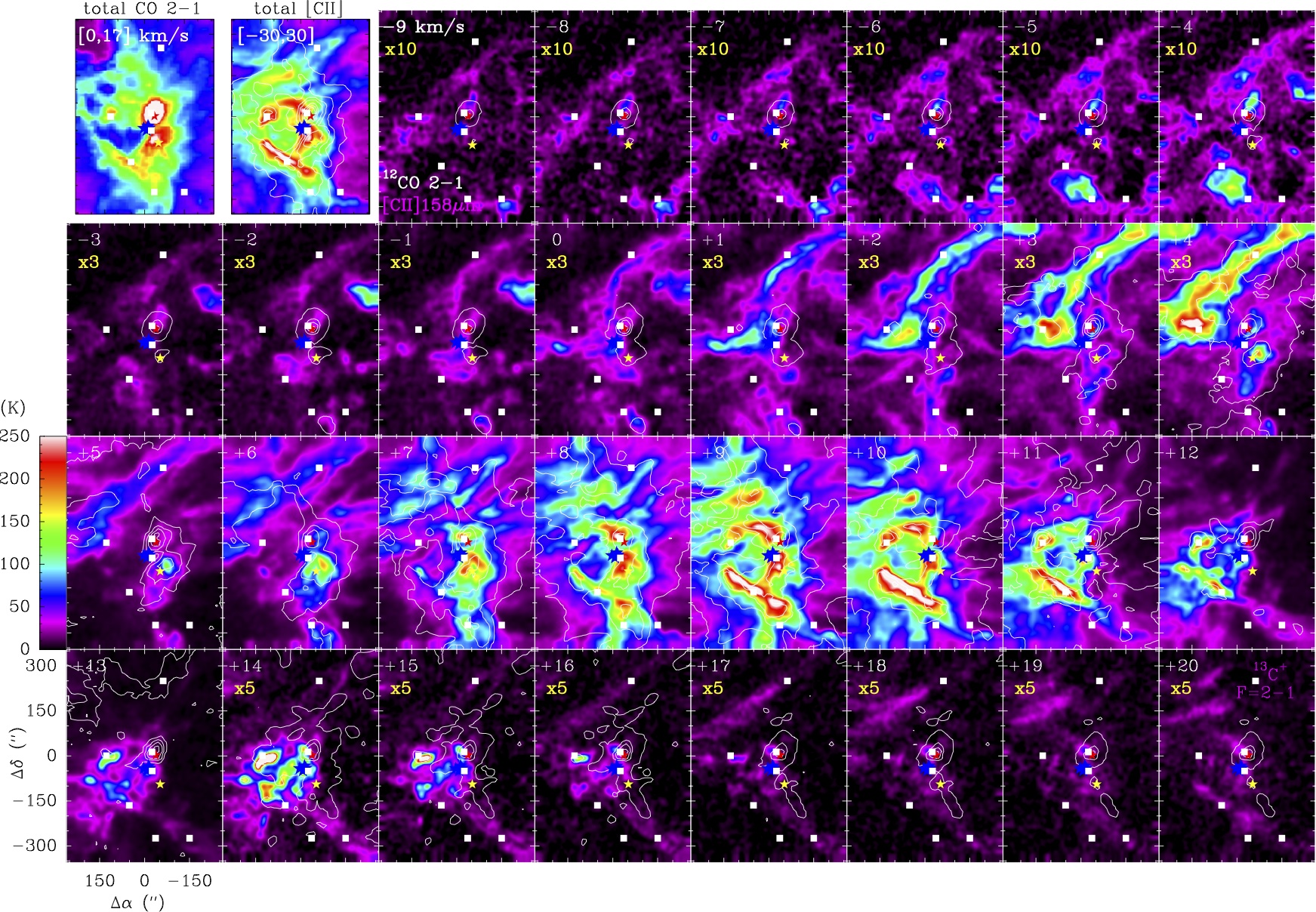}
\caption{Same as Figure~\ref{fig_channels_H41} but  the CO~2-1 line (in contours).
The \mbox{CO~2-1} contours go from 1.5 to 160~K\,km\,s$^{-1}$ in steps of 30~K\,km\,s$^{-1}$.
The high velocity \mbox{CO~2-1} emission 
($\rm{v}_{\rm LSR}$$<$0\,\,s$^{-1}$ and $\rm{v}_{\rm LSR}$$>$15\,\,s$^{-1}$)
mostly arises from molecular
outflows around Orion~BN/KL and Orion~S protostellar sources. At other positions, the observed [\CII] emission in these velocity ranges 
does not have CO emission counterpart (e.g., the blue-shifted [\CII] streamer at \mbox{v$_{\rm LSR}$=$-$2 to $+$5~km\,$^{-1}$}, 
or  the [\CII] emission in the southwest at \mbox{$\rm{v}_{\rm LSR} \gtrsim 12\,\rm{km\,s^{-1}}$} beyond the Orion Bar).
The first two upper-left panels
show total integrated line intensity maps of CO~2-1 (from 0 to 600 K\,km\,s$^{-1}$) and
[\CII]\,158\,$\mu$m (from 0 to 1300 K\,km\,s$^{-1}$). In the [\CII] total intensity map, CO~2-1 integrated line
intensities are shown as white contours, from  200 to 500~K\,km\,s$^{-1}$ in steps of 100~K\,km\,s$^{-1}$ (extended emission)
and from 800 to 1400~K\,km\,s$^{-1}$ in steps of 200~K\,km\,s$^{-1}$ (Orion~BN/KL outflows).
}
\vspace{0.25cm}
\label{fig_channels_CO}
\end{figure*}

\subsection{Kinematics: [\CII]\,158\,$\mu$m Velocity Channel Maps}
\label{subsect-channel}
 
Figures~\ref{fig_channels_H41} and \ref{fig_channels_CO} show continuum-subtracted [\CII]~158\,$\mu$m velocity channel maps 
compared to those of H41$\alpha$ and \mbox{CO~2-1} lines respectively. 
Both figures show the emission in 1\,km\,s$^{-1}$ velocity channels
from \mbox{v$_{\rm LSR}=-9$ to $+20$\,km\,s$^{-1}$} ([\CII]\,158\,$\mu$m in colored maps and 
H41$\alpha$ or CO~2-1 lines in white contours). They reveal the intricate small spatial-scale  kinematics of the 
ionized/PDR/molecular gas interfaces. 

The velocity-centroid of the  molecular cloud emission\footnote{The Orion~A molecular
 complex to which OMC\,1 belongs shows
a velocity gradient of 0.7\,km\,s$^{-1}$\,pc$^{-1}$ over $\sim$25\,pc \citep[e.g.,][]{Bal87}.}, 
traced by CO, lies at  \mbox{v$_{\rm LSR}\simeq$$+$(8-10)~km\,s$^{-1}$}. 
The [\CII]~158\,$\mu$m spectrum average over all mapped positions is characterized by a very bright component
peaking at \mbox{v$_{\rm LSR}\simeq +9.5$\,km\,s$^{-1}$}, with a line-width of $\sim$5\,km\,s$^{-1}$
(lowest panel in Figure~\ref{fig_spectra}),  in agreement with CO velocity and line widths.

This emission mostly arises from the  \HII/OMC\,1 interfaces close to the Trapezium
(inside the 200$''$ radius circle)
and also from the large-scale  FUV-illuminated face of the molecular cloud (outside). 
Therefore, \textit{the main [\CII] spectral component that coincides with CO 
(\mbox{{\rm v}$_{\rm LSR}$$\sim$5-17\,km\,s$^{-1}$})
is dominated by C$^+$ in dense PDR gas
and represents $\sim$85\,\%~of the total {\rm[\CII]} luminosity in the map}. 
The emission is so bright that the [$^{13}$\CII]\,$F$=2-1 line, the strongest  [$^{13}$\CII] hyperfine component
\citep{Oss13}, is also detected toward the highest column density 
peaks  (see the inset in Figure~\ref{fig_spectra} and the  channel maps between $+$17 and $+$20~km\,s$^{-1}$
in Figures~\ref{fig_channels_H41} and \ref{fig_channels_CO}).

Other specific regions such as the Orion Bar  PDR show [\CII]
line peak velocities at \mbox{v$_{\rm LSR}\simeq$$+$(10-12)~km\,s$^{-1}$} and thus 
 have a slightly differentiated kinematics.  
In fact, [\CII]  shows widespread emission in a broad range of velocities outside the main spectral component range.
On the other hand, the \mbox{CO~2-1} emission range is much more restricted, with the
high velocity CO emission seen  at  
$\rm{v}_{\rm LSR}$$<$0\,km\,s$^{-1}$ and $\rm{v}_{\rm LSR}$$>$15\,km\,s$^{-1}$ being very localized, and
mostly arising from the  molecular outflows surrounding Orion~BN/KL and S.

Despite the relatively flat  [\CII] intensity distribution
suggested by previous lower resolution observations,
the [\CII]  channel maps  show complicated structures and velocity patterns. 
This is true even away from  the Trapezium,
where filamentary structures and emission striations are seen at several velocity intervals. 
At  the \mbox{v$_{\rm LSR}\approx$$+$8-10~km\,s$^{-1}$} velocity of the neutral cloud, the 
 H41$\alpha$ emission  
delineates the far edges of the expanding \HII~blister in close interaction with the  cloud. This produces
a bright shell-shaped [\CII] emission  from compressed PDRs immediately beyond the ionization fronts. This is the
signature of the UV radiation \textit{eating into} the  cloud.

Also at  \mbox{v$_{\rm LSR}\approx$$+$8-10~km\,s$^{-1}$}, but away from the \HII~region, 
[\CII] shows narrow striations or 
sheet-like structures that are roughly
perpendicular to the main molecular ridge (north-to-south) and do not show much CO emission. 
As suggested by the overlap with the visible-light images of M42 (Figure~\ref{fig_nebula}), 
these photoevaporative [\CII] structures demonstrate that the entire field and velocity range is permeated by FUV photons.

The most negative velocities show H41$\alpha$  emission around the Trapezium. 
This is probably the  edge of the \HII~blister that is expanding toward the observer. The  analogous [\CII] emission
has some  spatial  correspondence but it also shows specific blue-shifted structures that do not follow the 
H41$\alpha$ emission  and  that must lie in the line of sight
toward OMC\,1.

\subsubsection{C$^+$ Gas with no CO Counterpart}

In certain velocity ranges the observed [\CII] emission  
does not have  any CO emission counterpart. 
The most outstanding  structure with no, or very little, CO emission is a blue-shifted  [\CII] emission \mbox{``streamer''}  
(\mbox{v$_{\rm LSR}$=$-$2 to $+$5~km\,$^{-1}$}) that goes north of Orion~BN/KL  and  also  a compact region that fills the
cavity inside the bright [\CII] shell-like structure (\mbox{Figure~\ref{fig_nebula}}). 
These structures are not coincident with H41$\alpha$ emission tracing \HII~gas,
 suggesting that they arise from warm neutral gas of low shielding ($A_V$$\approx$0-4\,mag),
in which very little or no CO exists (due to the elevated FUV field in the region and thus rapid photodissociation). 
Interestingly, they do coincide with foreground depressions of the visible-light emission known as the \textit{Dark Bay}  and the
\mbox{\textit{Northern Dark Lane}}  that separates M42 and M43 nebulae 
(blue contours in Figure~\ref{fig_nebula} $right$). 
 
The column density of material in these structures is sufficient to have significant hydrogen in molecular form
\citep[consistent with the detection of  radio OH lines in absorption toward the Dark Bay region, e.g.,][]{Bro05}.
Hence, they can be referred to \mbox{\textit{CO-dark H$_2$ gas}}  \citep[][]{Gre05,Wol10}, 
which can also be detected by dust extinction of visible light \citep{Ode00}.
In [\CII] emission, they represent large-scale structures of  C$^+$ gas  surrounding OMC\,1 and showing blue-shifted
velocities relative to the molecular cloud  (dark blue contours in \mbox{Figure~\ref{fig_nebula} right}).
In terms of  luminosity, 
the [\CII] emission from \mbox{v$_{\rm LSR}\simeq-2$ to $+$5${\rm \,km\,s^{-1}}$}, not 
 related to OMC\,1, 
accounts for $\sim$12\%~of the total [\CII] luminosity.

Surprisingly, the [\CII] line shows emission components at even more negative LSR velocities throughout the region,
in particular in the \mbox{v$_{\rm LSR}\simeq-10$ to $-$2\,km\,s$^{-1}$} range,
and a fainter blue emission wing at \mbox{v$_{\rm LSR}\simeq-25$ to $-$10\,km\,s$^{-1}$}
(see spectra in Figure~\ref{fig_spectra}).
Their combined contribution to the total [\CII] luminosity is however small, $\sim$3\%.

Some of the structures showing  [\CII] emission but no CO counterpart can be associated, 
both spatially and in   velocity space, with particular regions of
\HI~absorption observed against the radio continuum. 
They are produced by the ``Veil'', a collection of 
absorbing \HI~layers in the line of sight toward Orion  
 with little H$_2$, less than one part in 10$^6$  \citep[][]{Tro89,Abe06}.
For a detailed explanation of the different \HI~absorbing velocity components (\mbox{A, B, C, D ...})
see e.g., \citet{vdW13}.
In particular, components~A and B of the \HI~\textit{absorption} maps, produce
strong \HI~opacity in the \mbox{v$_{\rm LSR}$$\simeq$0 to +$6$\,km\,s$^{-1}$} range.
Components A and B are  interpreted as a foreground veil of nearly 
atomic gas  in front of the Orion nebula. 
Thus the [\CII] emission in the \mbox{v$_{\rm LSR}$$\simeq$0 to +$6$\,km\,s$^{-1}$} range, and outside
the Dark Bay and North Dark lane regions, likely arises from these neutral atomic layers of the Veil
 (cyan contours in Figure~\ref{fig_nebula} right).

Component~C of the \HI~absorption (from \mbox{v$_{\rm LSR}$$\simeq$$-$4 to $-2$\,km\,s$^{-1}$}) is only found
at the southwestern edge of OMC\,1 and we also detect [\CII] emission at those velocities and positions.
Finally, \HI~absorption components~D1 and D2 near the Bar  appear in the
\mbox{v$_{\rm LSR}$$\simeq$$-$20 to $-10$\,km\,s$^{-1}$} range and are
interpreted as an expanding shell of atomic gas \citep[][]{vdW13}. 
The most blue-shifted [\CII] emission (\mbox{v$_{\rm LSR}$$<$$-$10~km\,s$^{-1}$}) is detected toward the
southern regions of the map. Hence, part of this [\CII] emission  may  arise from D1 and D2.
 In addition, a fraction of this blue-shifted [\CII] emission may  arise from the foreground
 \HII~region that lies between
the Veil and the Trapezium, and that is not the main \HII~region interacting with  OMC\,1. 
[\NII]~$\lambda$6585\AA~emission line, and \PIII~$\lambda$1335\AA~and \SIII~$\lambda$1190\AA~absorption lines
 centered at \mbox{v$_{\rm LSR}$$<$$-$13~km\,s$^{-1}$} arise from this foreground
\HII~region \mbox{\citep[e.g.,][]{Abe06}}.
Unfortunately, the sensitivity of our [\CII] observations is insufficient to map these faint features and
 to determine a definite association.

\HI~\textit{emission} tracing denser 
gas associated with the molecular cloud
is detected \mbox{at v$_{\rm LSR}$$>+10$\,km\,s$^{-1}$}, 
outside the strong \HI~absorption range of the Veil. Indeed, we find  specific regions 
 that show coincident  [\CII] and 
\HI~red-shifted emission. The brightest one is the Orion Bar, where both  [\CII] and \HI~trace the most 
exposed PDR layers,
in which H$_2$ is photodissociated.  In addition,  
an elongated [\CII] and \HI~emission  feature without CO counterpart  
(\mbox{v$_{\rm LSR}$$\gtrsim$$+12$\,km\,s$^{-1}$}) extends in 
the southwest direction beyond the Bar and
along the Bar prolongation seen at visible wavelengths (Figure~\ref{fig_nebula}).
Another remarkable emission feature is the high-velocity [\CII] and \HI~line-wing 
detected toward Orion~BN/KL outflows. Its presence confirms that 
the  shocked material is illuminated by FUV radiation  \citep[][]{Che14,Goi15}.

All in all, our observations demonstrate the link between  [\CII] and both \HI~absorption (hydrogen mostly atomic gas)
and \HI~emission (hydrogen predominantly molecular). 
In particular, $\sim$15\% of the total [\CII] luminosity does not have a CO counterpart.
 Most  arises from cold \HI~gas and from \mbox{CO-dark H$_2$ gas} (e.g., the Dark Bay), but also from \HII~gas. The contribution
from ionized gas is expected to decrease with increasing electron density ($n_{\rm e}$) 
and   effective temperature ($T_{\rm eff}$)  of the ionizing stars 
(as carbon becomes doubly ionized). Both quantities are 
high toward the Trapezium, \mbox{$T_{\rm eff}$$\simeq$39,000\,K} for $\theta^1$\,Ori\,C star
\citep[][]{Sim06}, and
\mbox{$<n_{\rm e}>$$\simeq$(0.5-1)$\times$10$^4$~cm$^{-3}$} in the surrounding \HII~region \citep[][]{Zuc73}.
For these conditions,  models predict that the contribution of [\CII] from ionized gas
is small, of the order of $\lesssim$10\% of the {\mbox{PDR plus \HII} emission \citep[][]{Abe06a}. 
 This agrees with our general conclusion that
 most of the observed  [\CII]  luminosity  in Orion arises from dense PDR gas.

\begin{figure*}[ht]
\centering  
\includegraphics[scale=0.69, angle=0]{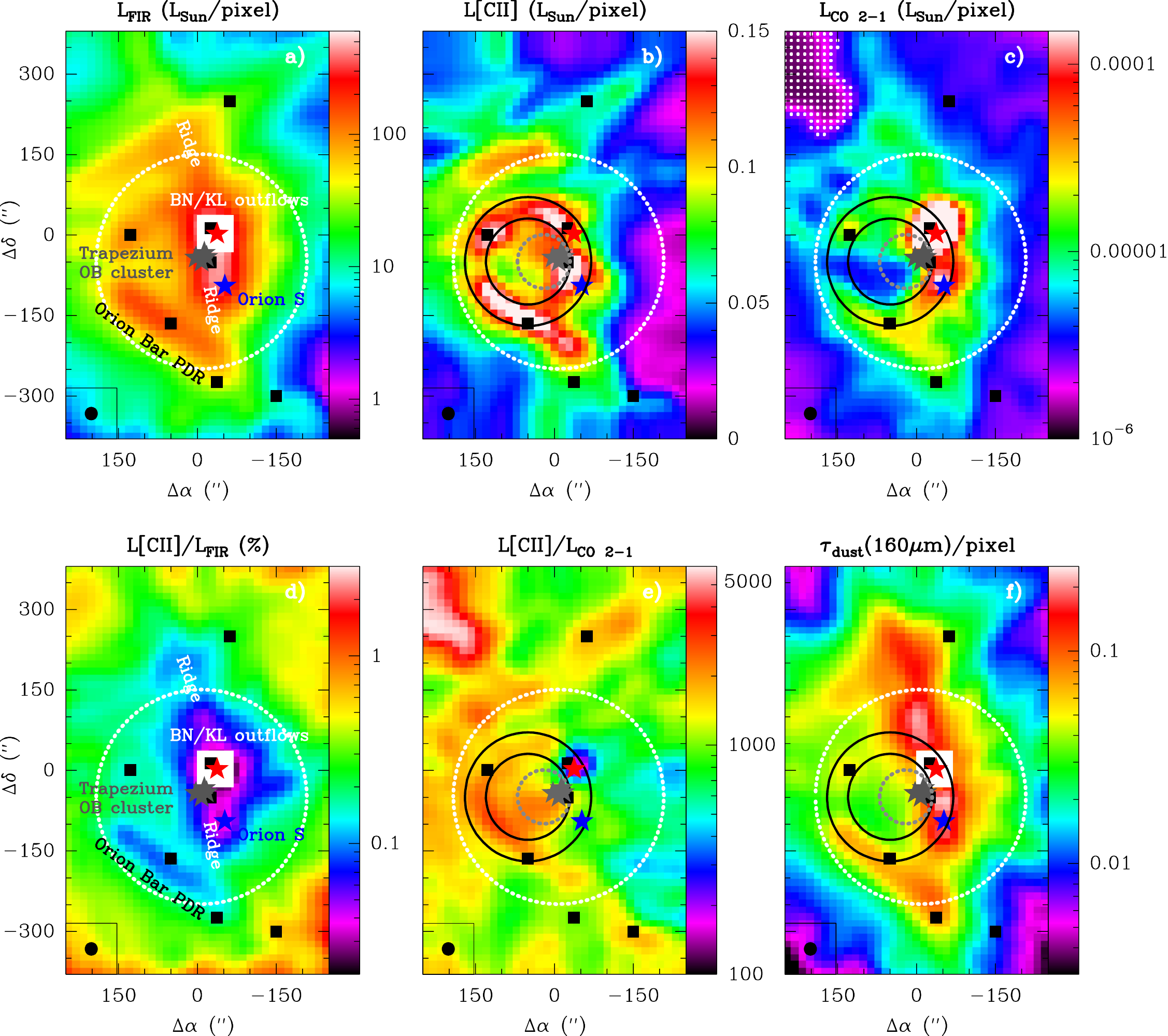}
\caption{Luminosity and dust opacity per pixel maps toward OMC\,1 convolved  to 25$''$ resolution.
(a)~\mbox{$L_{\rm FIR}$} map  ($G_0 \approx 200\,L_{\rm FIR}/L_{\odot}$ per pixel, see footnote 11),
(b)~\mbox{$L$[\CII]}, 
(c)~\mbox{$L_{\rm CO\,2-1}$} \citep[from][]{Ber14},
(d)~\mbox{$L$[\CII]/$L_{\rm FIR}$} in \%,
(e)~\mbox{$L$[\CII]/$L_{\rm CO\,2-1}$}, and 
(f)~dust opacity at 160\,$\mu$m from SED fits.
 The white squares in panel (c) represent positions of low \mbox{$L_{\rm CO\,2-1}$}
and high \mbox{$L$[\CII]/$L_{\rm CO\,2-1}$} ratios (up to $\sim$5000)  in panel 
(e) see also Figure~\ref{fig_corr}f.
The different circles and regions are defined in Section~\ref{sect-circles}. 
The black squares  indicate the positions of the spectra shown in Figure~\ref{fig_spectra}.
The white   square areas in the central BN/KL region  indicate positions where the \textit{Herschel} 
photometric observations are saturated.}
\vspace{0.25cm}
\label{fig_lums} 
\end{figure*}

\subsubsection{[\CII] Line Broadening and C$^+$ Gas Turbulence}
\label{subsub-turb}

The [\CII]~158\,$\mu$m line-width of  the main spectral component  varies between 
$\Delta$v$_{\rm FHWM}$$\sim$3.5~km\,s$^{-1}$ toward the extended ``face'' of  OMC\,1,
$\sim$4.0~km\,s$^{-1}$ toward the Orion Bar, and $\sim$5.5~km\,s$^{-1}$ toward the Trapezium.
The two latter values are likely affected by opacity broadening
(see Section~\ref{sub-excitation}
for a determination of [\CII] opacities).
In particular, for \mbox{$\tau_{\rm [CII]}$$\simeq$1-3},  opacity broadening increases
the line widths by \mbox{$\sim$10-30\%} \citep[e.g.,][]{Phi79, Oss13}.
We note that the line widths of the C91$\alpha$ radio recombination line toward the Orion Bar
are 2.0-2.5~km\,s$^{-1}$ \citep{Wyr97}.

In the dense PDR gas traced by the [\CII] main spectral component inside the 200$''$ radius circle, 
the  gas temperature is
 $T_{\rm k}$$\simeq$300-1000\,K (see  Figure~\ref{fig_PDR} and Section~\ref{sub-sub-rt}), 
 implying a thermal broadening of
 \mbox{$\sigma_{\rm th}$=($k$\,$T_{\rm k}$/$m_{\rm C^+}$)$^{1/2}$$\simeq$0.5-0.8~km\,s$^{-1}$}.
The nonthermal velocity dispersion $\sigma_{\rm turb}$ (from turbulence and macroscopic motions) is
\mbox{$\sigma_{\rm turb}$=($\sigma^2-\sigma_{\rm th}^2$)$^{1/2}$} where  \mbox{$\sigma$=$\Delta$v$_{\rm FHWM}$/2.355}.
For the above line-widths and gas temperatures, 
one obtains \mbox{$\sigma_{\rm turb}$=1.2-1.7~km\,s$^{-1}$} (after opacity broadening correction). 
This has to be compared with
the isothermal sound speed in the warm PDR gas,
\mbox{$c_{\rm PDR}$=($k$\,$T_{\rm k}$/$m$)$^{1/2}$$\simeq$1.4-3.0~km\,s$^{-1}$} 
(with $m$ the mean mass per atom). In other words,
the nonthermal velocity dispersion in the dense PDR gas  is subsonic or only weakly supersonic 
\mbox{(with Mach numbers $\sigma_{\rm turb}$/$c_{\rm PDR}$ of $\simeq$0.4-1.2)}.
 
In the intermediate \mbox{v$_{\rm LSR}$=$-$2 to $+$5~km\,$^{-1}$} components, 
 the [\CII] line is narrower, for example
$\Delta$v$_{\rm FHWM}$$\sim$2.5-3.0~km\,s$^{-1}$ toward the blue-shifted  streamer that coincides with
the North-Dark-Lane. 
On the other hand, the most blue-shifted [\CII]  component
with neither CO nor H41$\alpha$ counterpart, for example 
the features at \mbox{v$_{\rm LSR}$$\simeq$$-$7~km\,s$^{-1}$} toward \mbox{($-$150$''$,$-$300$''$)} or 
at \mbox{v$_{\rm LSR}$$\simeq$$-$5~km\,s$^{-1}$} toward \mbox{($-$37$''$,$-$274$''$)}, show again broader
line-widths of \mbox{$\Delta$v$_{\rm FHWM}$$\sim$4.7~km\,s$^{-1}$}  \mbox{(Figure~\ref{fig_spectra})}. 
Assuming optically thin [\CII] emission and $T_{\rm k}$$<$300~K, we find an  increased
$\sigma_{\rm turb}$$\simeq$2\,km\,s$^{-1}$ 
(or $\sigma_{\rm turb}$/$c_{\rm HI}$$>$2) in these CO-\textit{dark} components surrounding OMC\,1.
These results suggest that
the gas turbulence is more strongly dissipated in the dense \HII/OMC\,1 interfaces than in the   
surrounding, more turbulent, halo components.

\subsection{Surface Luminosity Maps}

Figure~\ref{fig_lums} shows maps of different luminosity\footnote{Line luminosity 
per pixel integrating 
all [\CII] spectral components.}
and luminosity ratios that are often discussed in the context 
of star formation and galaxy properties \citep[e.g.,][]{Car13}. 

\subsubsection{Dust Opacity, Mass and Surface Density}

Figure~\ref{fig_lums}(a) shows a map of $L_{\rm FIR}$ per pixel obtained by integrating the dust SED$^6$ fits  at each position.
Despite the relatively small area
covered ($\sim$1.25\,pc$^{2}$), the FIR intensity exhibits variations of $\sim$3 orders of magnitude 
(excluding Orion~BN/KL).
$L_{\rm FIR}$~is obviously high toward the high column 
density\footnote{Using \mbox{$N_{\rm H}\,\,({\rm cm^{-2}/pixel})= N({\rm H})+2\,N({\rm H_2}) = M_{\rm Gas}/(\mu\,m_{\rm H}\,D^2\,\Omega)$}, with 
$M_{\rm Gas}$ obtained from Equation~(\ref{eq-mass}).} 
regions in the molecular ridge \mbox{($N_{\rm H} \gtrsim 10^{23}$~cm$^{-2}$)}
and locally, it reaches even higher values toward the embedded star forming cores in Orion BN/KL.   
Figure~\ref{fig_lums}(f) shows a map of the dust opacity ($\tau_{\rm d,160}$) 
 from \mbox{Equation~(\ref{eq-flux})} for each line of sight.
The  observations reveal moderate FIR 
opacities toward Orion~S and along the molecular ridge (\mbox{$\tau_{\rm d,160} \approx 0.2$}).
The FIR opacity then decreases
from the Bar (\mbox{$\tau_{\rm d,160} \lesssim 0.1$}) 
to the  extended regions (\mbox{$\tau_{\rm d,160} \approx 0.01$}). 
Note that the 160\,$\mu$m dust emission toward Orion~BN/KL  is 
optically thick
\citep[$\tau_{\rm d,160}$$>$1,][and Sect.~\ref{subsub-dust}]{Goi15}.

The  total gas mass in the mapped region is computed from the optically thin dust emission at~250\,$\mu$m 
using
\mbox{Equation~(\ref{eq-mass})}. For the assumed grain properties (see Section~\ref{sect-circles}), we derive
\mbox{$M_{\rm Gas,Total}\simeq2600\,M_{\odot}$} in the $\sim$0.9\,pc\,$\times$1.4\,pc
 mapped area
\citep[excluding  Orion~BN/KL that  adds $\sim$100\,$M_{\odot}$;][]{Goi15}. This is equivalent to a  surface density of 
 \mbox{$\Sigma_{\rm Gas}$$\simeq$2000\,$M_{\odot}$\,pc$^{-2}$}. 
 We note that a $R_{\rm gd}$/$\kappa_{250}$ ratio higher than the assumed in this work will increase our gas mass estimate  
and the \mbox{$N_{\rm H}$-to-$A_V$} conversion value accordingly. 
 Unfortunately, the exact grain composition and  optical properties in molecular
clouds like OMC\,1 are not fully constrained \citep[][]{Gol97}, and
they likely change on small scales \citep{Ara12}.  Owing to the similar gas masses 
reported from large-scale $^{13}$CO and  submm continuum mapping \citep[][]{Bal87,Lis98,Ber14},
we  estimate that our   $M_{\rm Gas}$  and $N_{\rm H}$ values are uncertain by a factor of $\lesssim$2.

\begin{table*}[t]
\begin{center}
\caption{Average physical parameters toward  different regions of OMC\,1 \label{table-param}}
\begin{tabular}{lcccccc}
\tableline\tableline 
  & & & & & &  \\
  & $L$[\CII]/$L_{\rm FIR}$  & $L_{\rm CO\,2-1}$/$L_{\rm FIR}$ & $L$[\CII]/$L_{\rm CO\,2-1}$
  & $L_{\rm FIR}$/$M_{\rm Gas}$  & $G_0^b$  & $\tau_{\rm d,160}$  \\
  &  & & & ($L_\odot/M_\odot$) & (Habing field) & \\\tableline
  & & & & & & \\
OMC\,1 .......................................     & 3.8(5.4)$\times$10$^{-3}$ & 3.6(6.2)$\times$10$^{-6}$  & 1200(900) &  43(27) & 7.5$\times$10$^{3}$ & 0.03(0.03) \\
(full map\tablenotemark{a})&           &                           &      &    &                     &      \\
Extended Cloud .........................   & 4.7(6.0)$\times$10$^{-3}$ & 4.5(7.5)$\times$10$^{-6}$  & 1400(900) &  35(20)  & 3.2$\times$10$^{3}$& 0.02(0.02) \\
($R$$>$0.4\,pc)&                       &                      &      &     &                     &      \\
Dense PDR interface .................       & 1.1(0.5)$\times$10$^{-3}$ & 6.6(2.2)$\times$10$^{-7}$  & 1600(500) &  87(24) & 3.3$\times$10$^{4}$ & 0.08(0.05)\\
(0.16$<$$R$$<$0.24\,pc annulus\tablenotemark{a})&      &                      &      &    &                     &      \\
Toward \HII~around Trapezium ..      & 9.3(3.9)$\times$10$^{-4}$ & 5.4(1.8)$\times$10$^{-7}$  & 1700(400) &  126(18) & 3.3$\times$10$^{4}$ & 0.05(0.02)\\
($R$$<$0.1\,pc)						&      &                      &      &    &                     &      \\

& & & & & &  \\\tableline
\tableline
\end{tabular}
\tablecomments{Average values over the different \textit{template} cloud regions
defined in Section~\ref{sect-circles}.
 Numbers in parenthesis represent the standard deviation (1$\sigma$)
of each sample. 
$L_{\rm FIR}$ refers to the 40-500\,$\mu$m  range.
$^a$The central Orion~BN/KL region (50$''$$\times$50$''$) is not included in the computation of the average values due 
to saturation of the photometric measurements.
A low \mbox{$L$[\CII]/$L_{\rm FIR}$$\simeq$(2-8)$\times$10$^{-5}$} luminosity ratio and $\tau_{\rm d,160}$$\simeq$1-3 was 
inferred from \textit{Herschel}/PACS spectroscopic observations \citep{Goi15}. The mean
\mbox{$L$[\CII]/$L_{\rm CO\,2-1}$} ratio in this region is $\sim$270. 
$^b$Estimated from  $I_{\rm FIR}$ and Equation~\ref{eq-G0}.}
\end{center}
\end{table*}

\subsubsection{$L$[\CII]/$L_{\rm FIR}$ and $L$[\CII]/$L_{\rm CO\,2-1}$ Maps}

Figure~\ref{fig_lums}(d) shows the spatial distribution of the $L$[\CII]/$L_{\rm FIR}$ 
 luminosity ratio toward OMC\,1. $L$[\CII]/$L_{\rm FIR}$ varies from the more extended and translucent cloud component,
where \mbox{$N_{\rm H} \simeq 10^{21-22}$~cm$^{-2}$} and $L$[\CII] carries up to \mbox{$\sim$1-5\,\%~of} the FIR luminosity,
to the dense molecular ridge, where  \mbox{$N_{\rm H} \simeq 10^{24}$~cm$^{-2}$} and $L$[\CII] 
 only carries  $\sim$0.01\%.
The mean (median) value in the map is \mbox{$L$[\CII]/$L_{\rm FIR}=3.8\,(2.9) \times 10^{-3}$}. This is  similar to the value observed toward
 translucent clouds of the Galaxy  \mbox{\citep{Ing02}}. 
 Indeed, low angular resolution surveys of the Milky Way show that the typical $L$[\CII]/$L_{\rm FIR}$ ratio in regions of weak [\CII] emission 
is $\sim$4$\times$10$^{-3}$, and decreases to $\sim$10$^{-3}$ toward star-forming regions where the
[\CII] line is brighter \citep[ e.g.,][]{Nak98}.

The $L_{\rm FIR}$ map per pixel in Figure~\ref{fig_lums}(a) can be used to 
estimate, to first order, the spatial distribution of  $G_0$. Assuming that at 
large scales 
the entire column density of dust 
is  heated by absorption of FUV photons, dust grains
re-radiate in the FIR at a
characteristic temperature ($\approx T_{\rm d}$) and;
\begin{equation}
\label{eq-G0}
G_0 \approx \frac{1}{2}\,\frac{I_{\rm FIR}\,(\rm{erg\,s^{-1}\,cm^{-2}\,sr^{-1}})}{1.3 \times 10^{-4}},
\end{equation} 
\citep[see][]{Hol99} where $G_0$ is the FUV field in Habing units\footnote{At the distance of OMC\,1 and considering the pixel size in Fig.~\ref{fig_lums}(a),
the $L_{\rm FIR}$ scale per pixel is equivalent to $G_0 \approx 200\,L_{\rm FIR}/L_{\odot}$ (per pixel).
 This relation neglects
dust heating by embedded sources. Here we assume that owing to the high FUV field in the region, 
this extra heating does not dominate at large scales. For the same reason, and owing to the 
warm dust grain temperatures,
we neglect the dust emission at $>$500\,$\mu$m in the  $G_0$ estimation.}.  
The factor of 1/2 approximately takes into account the absorption of visible photons by dust.
 The maximum value near the Trapezium is \mbox{$G_0$$\simeq$10$^5$}, whereas the lowest values in the 
mapped area are \mbox{$G_0$$>$500}.
The mean  value inside the 200$''$ radius circle is \mbox{$G_0$$\simeq$2$\times$10$^4$}
(the $G_0$ used in the representative PDR models shown in Figure~\ref{fig_PDR}).
 These FUV fluxes are in good agreement with early estimates based
on lower angular resolution observations and PDR modelling \mbox{\citep[e.g.,][]{Tie85b,Sta93}}. 

The [\CII]  and \mbox{CO~2-1} luminosities in the observed area are
 $L$[\CII]=173\,$L_{\odot}$ and $L_{\rm CO\,2-1}$=0.15\,$L_{\odot}$.
 Figure~\ref{fig_lums}(e) shows the \mbox{$L$[\CII]/$L_{\rm CO\,2-1}$} map. The spatial distribution is markedly 
different from that of \mbox{$L$[\CII]/$L_{\rm FIR}$}. In particular, the $L$[\CII]/$L_{\rm CO\,2-1}$ ratio 
shows  differences locally, 
from a few hundred in the strong outflows around Orion~BN/KL, to $\sim$5000 in the peculiar northeastern positions
almost devoid of CO emission (see white marks in the $L_{\rm CO\,2-1}$ map shown in \mbox{Figure~\ref{fig_lums}(c))}.
At other positions of the map, the ratio reaches high values ($>$1000), either because the observations  trace 
regions of low extinction (i.e., low CO columns) or nearly edge-on PDRs for which
 the  C$^+$/CO transition is spatially resolved.
 \mbox{Table~\ref{table-param}} summarizes the mean
values extracted from the maps toward the different regions of the cloud.

\section{Analysis}
\label{sec-anal}

\subsection{[\CII]\,158\,$\mu$m Excitation and $C^+$ Column Densities}
\label{sub-excitation}

Collisions  with electrons, H atoms and H$_2$ molecules contribute to the excitation of  
the [\CII] $^2P_{3/2}$-$^2P_{1/2}$  fine structure levels \citep[see discussion by][]{Gol12}.  The critical densities ($n_{\rm cr}$)  
for collisions of C$^+$ with, $e$, H and H$_2$ are 
$\sim$44~$e$\,/\,cm$^{3}$ at \mbox{$T_{\rm e}$=8000~K} \citep{Wil02}, and 
$\sim$3000~(2400)~H\,/\,cm$^{3}$ \citep{Bar05} 
and $\sim$4500~(3400)~H$_2$\,/\,cm$^{3}$ at \mbox{$T_{\rm k} = 100$~(500)~K} \citep{Wie14}.
Owing to the similar collisional rates  with
H$_2$ and H, the [\CII] excitation in neutral gas is nearly independent of the molecular fraction.
As the gas  density increases above $n_{\rm cr}$, the excitation temperature
 $T_{\rm ex}$ of the $^2P_{3/2}$-$^2P_{1/2}$ transition tends to thermalize with $T_{\rm k}$. 
For optically thick lines, line-trapping  further reduces the critical densities, thus reaching thermalization
($T_{\rm ex}$=$T_{\rm k}$) at lower densities.
Also for optically thick lines, a strong FIR background with high $T_{\rm bg}$ values
increases $T_{\rm ex}$ but  reduces the [\CII] line intensity.

When $n_{\rm H} \lesssim n_{\rm cr}$, the [\CII] excitation becomes
sub-thermal \textit{($T_{\rm ex} < T_{\rm k}$)}. If 
significant C$^+$ columns of low-density gas exist in front of a  FIR continuum source, 
 the [\CII] line will be seen in absorption if  \mbox{$T_{\rm ex}<\frac{91.2}{ln\,(1 + 91.2/T_{\rm c})}$}
\mbox{\citep[e.g.,][]{Ger15}}, where 
\mbox{$h\nu/k=91.2$~K} is the equivalent temperature at 1900.537\,GHz,
and \mbox{$T_{\rm c}= J(T_{\rm bg})=\frac{91.2}{e^{91.2/T_{\rm bg}}-1}$} is the continuum brightness temperature
 at 158\,$\mu$m. 
[\CII] line self-absorptions can also be produced by excitation gradients
within the [\CII] emitting region \citep[e.g.,][]{Plu04,Gra12}.

One potential problem  in interpreting the [\CII]\,158\,$\mu$m emission 
is the difficulty inferring the line opacity. 
Toward many positions, our observations resolve the \mbox{[$^{13}$\CII] $F$=2-1} 
line\footnote{Corrected line strengths for the \mbox{[$^{13}$\CII] $F$=2-1}, 1-0 and 1-1 hyperfine lines
at 1900.466, 1900.950 and 1900.136\,GHz respectively have been published by \citet{Oss13}. The relative line strengths
are 0.625, 0.250 and 0.125. These are slightly different from those of \citet{Coo86}  that were used in the
analysis of \citet{Sta91b} and \citet{Bor96}, and resulted in $^{12}$C/$^{13}$C isotopic ratios lower than
the usual value of $\sim$67.} 
arising from the main spectral component at \mbox{v$_{\rm LSR}\simeq +9.5$\,km\,s$^{-1}$} 
(see spectra in Figure~\ref{fig_spectra} and maps in the highest-velocity 
panels of Figures~\ref{fig_channels_H41}~and~\ref{fig_channels_CO}).
The second brightest component, the \mbox{[$^{13}$\CII] $F$=1-0} line, is only detected toward a few positions of very
bright [\CII] emission  (e.g. the Orion Bar and BN/KL, see Figure~\ref{fig_spectra}).
Assuming that  the total integrated intensity of the three [$^{13}$\CII] lines is $I$[$^{13}$\CII] = $I_{F=2-1}$/0.625, one can derive the  
[\CII] line opacity from\footnote{Toward the Orion Bar and BN/KL, the \mbox{[$^{13}$\CII] $F$=2-1} line
is blended with the  red-shifted [\CII] emission. To avoid overestimating the [\CII]\,158\,$\mu$m line opacity toward  these positions,
we used the  [$^{13}$\CII] $F$=1-0 line (which is not blended with any spectral feature)  
and assumed \mbox{$I$[$^{13}$\CII] = $I_{F=1-0}/0.250$}.}:
\begin{equation}
\label{eq-opa}
\frac{1-e^{-\tau_{\rm [CII]}}}{\tau_{\rm  [CII]}} \simeq 
\frac{0.625\,T_{\rm P}({\rm [^{12}CII]})/T_{\rm P}({\rm [^{13}CII]},F=2-1)}{[\rm{^{12}C/^{13}C]}}
\end{equation}
\citep[e.g.,][]{Oss13} where $\tau_{\rm [CII]}$ is the \mbox{[\CII]\,158\,$\mu$m} opacity at line center,  $T_{\rm P}$ is the line peak 
 main beam temperature
(continuum-subtracted) and \mbox{[${\rm ^{12}C/^{13}C}$]} is the isotopic ratio  
\citep[$\sim$67 in Orion,][]{Lan90}. 
For optically thick [\CII]\,158\,$\mu$m  emission, the excitation temperature ($T_{\rm ex}$)
of the \mbox{$^{2}P_{3/2}-^{2}P_{1/2}$} transition  can be computed from:
\begin{equation}  
\label{eq-Tex}
T_{\rm ex}= \frac{91.2}{ln\,(1 + \frac{91.2}{T_{\rm P}({\rm ^{12}C^+}) + J(T_{\rm bg})})}\,\,\, {\rm (K)}.
\end{equation}
At these FIR wavelengths, the background is dominated by
dust emission  so that the continuum level adjacent to the [\CII] line,
for an equivalent background temperature of $20-35$\,K,
is \mbox{$T_{\rm c} = 1-7$\,K}, roughly the range of continuum levels
detected by HIFI. 
For the main  \mbox{[\CII]\,158\,$\mu$m} spectral component we find \mbox{$T_{\rm P}({\rm ^{12}C^+})\gg T_{\rm c}$}, 
thus the continuum emission  has a minor effect on $T_{\rm ex}$. However, the situation can be different in more extreme 
and luminous environments.

Once $T_{\rm ex}(^{2}P_{3/2}-^{2}P_{1/2})$ and $\tau_{\rm [CII]}$ have been determined, the C$^+$ column density 
can be computed from: 
\begin{equation}
\label{eq-N}
N({\rm C^+})  \simeq 1.5 \times 10^{17} \, \frac{1 + 2\,e^{-91.2/T_{\rm ex}}}{1 - e^{-91.2/T_{\rm ex}}}\,
\tau_{\rm [CII]}\cdot \Delta {\rm v_{\rm FWHM}} \,\,\,\,\,\, {\rm (cm^{-2})} 
\end{equation}
where $\Delta {\rm v_{\rm FWHM}}$ is the line width in km\,s$^{-1}$ of an assumed Gaussian line-profile.
From the  [\CII] and [$^{13}$\CII] spectra averaged over the mapped region (main spectral component) we
find \mbox{$<T_{\rm ex}> \simeq 100$~K}, \mbox{$<\tau_{\rm [CII]}> \simeq 1.3$} and
\mbox{$<N({\rm C^+})> \simeq 3\times 10^{18}$~cm$^{-2}$}. 
The highest C$^+$ column density peaks are found near the Trapezium, the Orion Bar, the East PDR  
and northeast of Orion~BN/KL.
In these emission peaks, the excitation temperatures and line opacities 
(\mbox{$T_{\rm ex} = 250-300$~K} and \mbox{$\tau_{\rm [CII]} \simeq 1-2$})
are higher than the average. 
The 25$''$~beam-averaged C$^+$ column density toward these peaks 
is \mbox{$N({\rm C^+}) \simeq 10^{19}$~cm$^{-2}$}.

\begin{figure*}[ht]
\centering
\includegraphics[scale=0.6, angle=0]{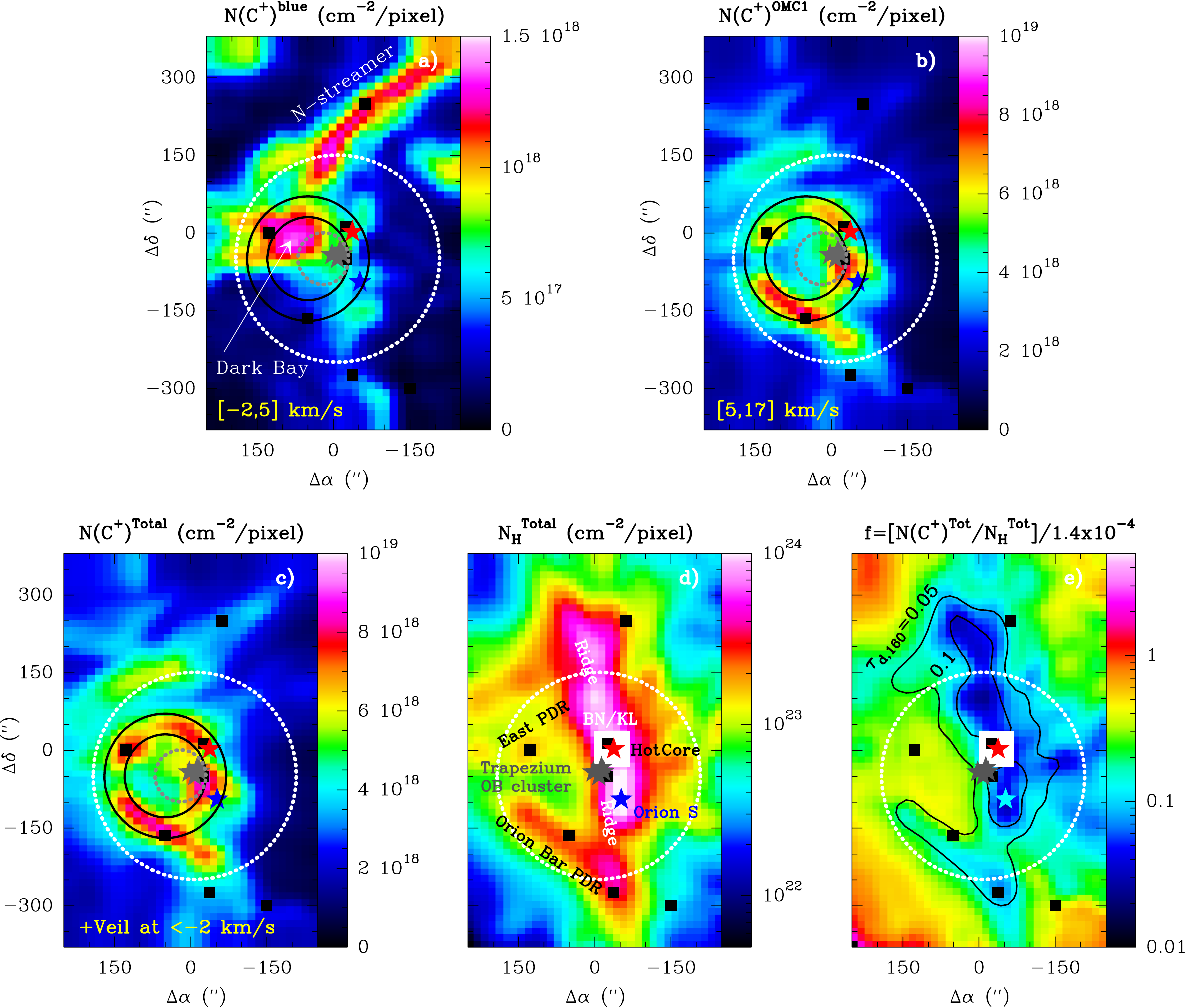}
\caption{Column density per pixel maps toward Orion. 
(a)~$N$(C$^+$)$^{\rm blue}$  extracted from the blue-shifted emission component from
\mbox{v$_{\rm LSR}$=$-$2 to $+$5~km\,$^{-1}$}.
(b)~$N$(C$^+$)$^{\rm OMC1}$ extracted from the main  emission component from
v$_{\rm LSR}$=$+$5 to $+$17~km\,$^{-1}$.
(c)~\mbox{$N$(C$^+$)$^{\rm Total}$=$N$(C$^+$)$^{\rm OMC1}$+$N$(C$^+$)$^{\rm blue}$+$N$(C$^+$)$^{\rm <-2\,km/s}$,} where $N$(C$^+$)$^{\rm <-2\,km/s}$=10$^{17}$~cm$^{-2}$ is adopted in all positions (see text).
(d)~$N_{\rm H}^{\rm Total}$=$N$(H)+2$N$(H$_2$) from the dust emission.
(e)~\mbox{[$N$(C$^+$)$^{\rm Total}$/$N_{\rm H}^{\rm Total}$]/1.4$\times$10$^{-4}$} color map, this is roughly the fraction ($f$) of the total column density traced by 
[\CII] toward Orion. This fraction scales with the total column density of dust
along each line of sight.  The 160\,$\mu$m dust opacity  
iso-contours at $\tau_{\rm d,160}$=0.1 (inner) and 0.05 (outer)
are shown in black.The different circles  are defined in Section~\ref{sect-circles}.}
\vspace{0.1cm}
\label{fig_columns_Cp_H2}
\end{figure*}

The inferred excitation temperatures can be used to further constrain
 \textit{the origin of the [\CII] emission
at the  \mbox{FUV-illuminated} surface of the molecular cloud}. In particular, the mean $T_{\rm ex}$$\gtrsim$100\,K values 
inferred from the map
require the fractional population of the $^2$$P_{3/2}$  level, 
\mbox{$n$($^2$$P_{3/2}$)/$n$($^2$$P_{1/2}$+$^2$$P_{3/2}$)}, 
to be $>$0.45.  The representative PDR model in Figure~\ref{fig_PDR} shows that the conditions for producing  
significant [\CII] emission are quite restrictive. For high FUV radiation fields, the C$^+$ layers
extend to $A_V$$\simeq$3-4\,mag but the gas temperature falls much faster for any reasonable density. 
Figure~\ref{fig_PDR} shows that a fractional population  $>$0.45 of the $^2$$P_{3/2}$  level 
requires high gas densities ($n_{\rm H}$$\simeq$5$\times$10$^3$ for even $T_{\rm k}$$\simeq$500\,K,
see Figure~\ref{fig_grid}).
Hence, the bright [\CII] emission from OMC\,1 traces narrow cloud surface layers ($A_V$$<$4\,mag) and not the entire line of sight.

 In the extended regions,  the 
[$^{13}$\CII] lines are not detected and the [\CII]\,158\,$\mu$m  line is weaker. This suggests that the  line is
optically thin or \textit{effectively} thin \citep[][]{Gol12}. The line opacity 
can be derived if one knows $T_{\rm ex}$.  Here we make the approximation
\mbox{$\tau_{\rm [CII]} \simeq T_{\rm P}$([\CII])/$J(T_{\rm ex})$}.
Figure~\ref{fig_columns_Cp_H2}(a) shows a column density 
map,  computed under these assumptions, of the blue-shifted component (v$_{\rm LSR}$$\simeq$$-$2 to $+$5~km\,s$^{-1}$) that includes
the Dark Bay and the Northern Dark Lane emission. We adopted a constant 
$T_{\rm ex}$$\simeq$80\,K (intermediate between $T_{\rm k}$
in the OMC\,1/\HII~interfaces and that in the atomic Veil \citep{Abe09}. The resulting $N$(C$^+$) toward the Dark Bay 
($\simeq$1.4$\times$10$^{18}$~cm$^{-2}$) 
is equivalent to
\mbox{$A_V$$\simeq$$N$(C$^+$)/4$\times$10$^{17}$$\simeq$3.5~mag} (i.e.,~significant H$_2$ must be present) 
 assuming that the bulk of the gas-phase carbon is in C$^+$, with an abundance of 1.4$\times$10$^{-4}$  \citep{Sof04}.
This value is consistent with the extinction peak determined toward the Dark Bay \citep{Ode00}. 
Other sightlines with $N$(C$^+$)$<$8$\times$10$^{17}$~cm$^{-2}$ ($A_V$$<$2\,mag)  are likely associated
with C$^+$ in nearly atomic \HI~gas (e.g., toward the Traprezium).
The integrated C$^+$ column density in the map 
(v$_{\rm LSR}$$\simeq$$-$2 to $+$5~km\,s$^{-1}$ component)
is  compatible with the observed [\CII] fractional  luminosity of $\sim$12\% (Section~\ref{subsect-channel}).

In addition, Figure~\ref{fig_columns_Cp_H2}(b) shows a $N$(C$^+$) map of the
 OMC\,1 component (v$_{\rm LSR}$$\simeq$5 to $+$17~km\,s$^{-1}$)
adopting $T_{\rm ex}$$\simeq$100\,K from the map averaged spectrum.
\mbox{Figure~\ref{fig_columns_Cp_H2}(c)} shows the resulting map of the \mbox{total C$^+$} column density
along each line of sight,  \mbox{$N$(C$^+$)$^{\rm Total}$=$N$(C$^+$)$^{\rm OMC1}$+$N$(C$^+$)$^{\rm blue}$+$N$(C$^+$)$^{\rm <-2\,km/s}$,} where we have
added, at each position, $N$(C$^+$)$^{\rm <-2\,km/s}$=10$^{17}$~cm$^{-2}$ from
additional foreground emission components at v$_{\rm LSR}$$<$$-$2~km\,s$^{-1}$.
The integrated column density in the $<$$-$2~km\,s$^{-1}$ component is consistent with the $\sim$3\%~contribution to the total [\CII] luminosity. This is equivalent to $A_V^{\rm <-2km/s}$$\simeq$0.25~mag
(i.e.,~fully atomic or ionized gas).
We finally note that along many sightlines, the C$^+$ column in the  Veil  is comparable to that computed in the OMC\,1 surface. In these approximate maps of the
C$^+$ components without CO counterpart, $\sim$30\,\%~of the C$^+$ mass is in 
\mbox{CO-dark H$_2$ gas}. 

Figure~\ref{fig_columns_Cp_H2}(d) shows the $N_{\rm H}^{\rm Total}$ column density map obtained from the dust emission$^{10}$.
Both $N$(C$^+$)$^{\rm Total}$ and $N_{\rm H}^{\rm Total}$ maps can be used to estimate 
the fraction ($f$) of the total column density traced by [\CII] along each line of sight.
In particular, Figure~\ref{fig_columns_Cp_H2}(e) shows the distribution of the
 \mbox{$f$=[$N$(C$^+$)/$N_{\rm H}$]/1.4$\times$10$^{-4}$}  ratio  at 25$''$ resolution. 
 Note that this map is similar to the \mbox{$L$[\CII]/$L_{\rm FIR}$}
 map shown in \mbox{Figure~\ref{fig_lums}(d)}.
The lowest  values $f\ll 1$ are inferred toward the high $N_{\rm H}$ column density ridge
(bluish regions with \mbox{$f\lesssim$5\,\%}).
These  fractions are consistent with the presence of a massive molecular cloud behind the [\CII] emitting  surfaces.
The typical fractions outside the ridge are}} \mbox{$f\simeq$20-30\,\%} (greenish regions).
Far from the Trapezium,  toward the extended cloud component, fractions higher than \mbox{$f$$\gtrsim$50\,\%} are only inferred toward the most translucent
lines of sight.
The [\CII] emission toward the  dense PDRs around the Trapezium is  slightly optically thick  and $T_{\rm ex}$ is higher than the average value assumed in the \mbox{$f$ map}. The corrected fraction  using the  column of
 \mbox{$N$(C$^+$)$\simeq$10$^{19}$\,cm$^{-2}$} computed from [$^{13}$\CII] is \mbox{$f$$\simeq$40\,\%}.
 
We also used the  C$^+$ column map to compute the ``PDR mass'' in OMC\,1. In particular,
we integrate the equivalent $N_{\rm H, PDR}$=$N$(C$^+$)$^{\rm OMC1}$/1.4$\times$10$^{-4}$ column traced by  [\CII] as
\mbox{$M_{\rm PDR} \simeq \mu\, m_{\rm H}\, \sum_{i} \, (N_{\rm H, PDR}^{i}\,A_i)$}, where $A_i$ is the area of pixel~$i$, $m_{\rm H}$ is the H atom mass and $\mu$ is the mean atomic weight. 
We derive \mbox{$M_{\rm PDR}$$\simeq$200\,$M_{\odot}$}, i.e., the PDR mass fraction traced by [\CII] 
is $\sim$8\,\% (within a factor of $\sim$2).

\subsubsection{[\CII] Radiative Transfer Models} 
\label{sub-sub-rt} 

In order to constrain the range of physical conditions that reproduce the observed [\CII]~158\,$\mu$m 
emission toward OMC\,1, 
we have run a grid of non-local and non-LTE radiative transfer models of the \mbox{$^{2}P_{3/2}-^{2}P_{1/2}$} doublet
that  take into account line trapping and opacity broadening \citep[see Appendix in][]{Goi06}.
For consistency, we included illumination by a FIR background continuum field and a combined treatment
of gas and dust emission/absorption. Hence, 
at a given wavelength the opacity is given by \mbox{$\tau_{\lambda} = \tau_{\rm [CII]} + \tau_{\rm d, 160}^{local}$}.
We used a constant column density of $N$(C$^+$)=3$\times$10$^{18}$\,cm$^{-2}$, the mean value
 we infer from the OMC\,1 map,
and a nonthermal velocity dispersion \mbox{$\sigma_{\rm nth} = 1.3$~km\,s$^{-1}$} (see Sec.~\ref{subsub-turb}).
The models use the latest available collisional rates of C$^+$ with $e$, H and H$_2$ 
(with a LTE H$_2$ ortho-to-para ratio at each $T_{\rm k}$) and assume  
\mbox{$n_{\rm e} = [\rm{C^+}/\rm {H}]\,n_{\rm H} = 1.4\times10^{-4}\,n_{\rm H}$} and $n$(H)=0.25\,$n$(H$_2$) 
(a molecular fraction,  2\,$n$(H$_2$)/[$n$(H)+2\,$n$(H$_2$)], of $\simeq$0.9). 
 For  gas densities \mbox{$n_{\rm H}$=2$\times$10$^{5}$~cm$^{-3}$} and
$T_{\rm k}$=500~K, this choice of parameters implies
that the [\CII] \mbox{$^2P_{3/2}$-$^2P_{1/2}$} collisional rate is due to collisions with H$_2$ molecules
($\sim$70\%), H atoms ($\sim$25\%) and electrons ($\sim$5\%). 
The FIR background radiation field is modelled as thermal emission at 30\,K (roughly
the mean dust temperature
we derive from SED fits) and the temperature of the dust grains mixed with the C$^+$ atoms is set to 50\,K
(typical of highly illuminated PDRs at $A_V \approx 2$\,mag).

Figure~\ref{fig_grid} shows model results in the form of \mbox{iso-$T_{\rm ex}$} contours.
The mean [\CII] excitation temperature in the \mbox{\textit{Herschel}/HIFI map} is \mbox{$T_{\rm ex}$$\simeq$100~K}.
This plot shows that the minimum gas density required to produce this  $T_{\rm ex}$=100\,K is
\mbox{$n_{\rm H}$$\simeq$5$\times$10$^3$~cm$^{-3}$}.
Toward the  \HII\,/OMC\,1 interfaces (the spherical shell of PDRs around the Trapezium) $T_{\rm ex}$ reaches higher
values ($\sim$250-300\,K). This sets a higher limit to the gas density  of  
$n_{\rm H}$$\geq$10$^5$\,cm$^{-3}$ (including the Orion Bar).
The more elevated temperatures in these dense PDRs (\mbox{$T_{\rm k}$$\geq$300\,K}) 
are  consistent with the \mbox{$T_{\rm k}$=400-700\,K} values
inferred from H$_2$ pure rotational lines \citep{All05} and with \mbox{$T_{\rm k}$$\simeq$540~K} derived from 
\HI~emission, both measurements toward the Orion Bar PDR \citep{vdW13}.\\

\begin{figure}[t]
\centering
\includegraphics[scale=0.41, angle=0]{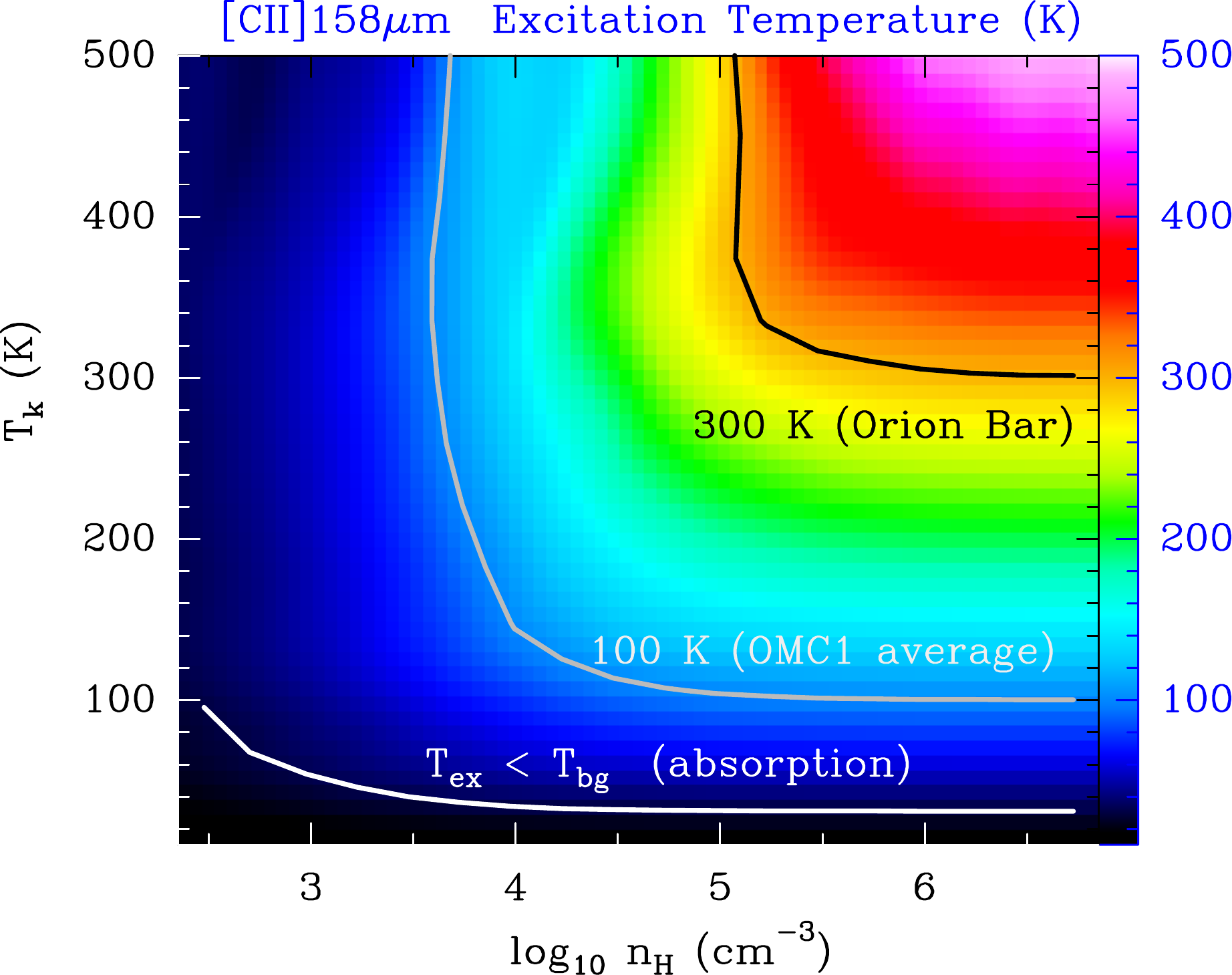}
\caption{Synthetic excitation temperatures for the [\CII]\,158\,$\mu$m transition obtained
from a grid of nonlocal non-LTE models (see text).}
\vspace{0.00cm}
\label{fig_grid}
\end{figure}

Finally,  the presence of a foreground halo of low-excitation material can produce [\CII]  
absorption.   Such a halo will also absorb or scatter the [\CII] line 
photons emitted inside the  cloud, if  the absorption is produced at similar velocities.   This may  be the case
of components~A  and B of the \HI~absorption in the Veil. These components peak at 
v$_{\rm LSR}$$\simeq$0-6~km\,s$^{-1}$ \mbox{\citep{vdW13}} and coincide with the  blue-wing of the
 [\CII] emission seen in the main spectral component.
The gas temperature and density in components~A and B of Orion's Veil are estimated from \HI~lines to 
 \mbox{$T_{\rm k}$$\simeq$50-80\,K} and 
\mbox{$n$(H)$\simeq$10$^{2.5-3.4}$~cm$^{-3}$} \citep{Tro89,Abe06}. 
For the FIR background in OMC\,1 and for
the parameters in our grid, we do predict [\CII] absorption for low excitation gas
(white line in Figure~\ref{fig_grid}).
Hence, the [\CII] emission dips seen in some velocity intervals may be produced by [\CII] absorption in the Veil
 (e.g. toward the Orion Bar PDR at 
v$_{\rm LSR}$$\simeq$$+1$\,km\,s$^{-1}$, toward the southwest \mbox{(-150$''$,-300$''$)} position at  
v$_{\rm LSR}$$\simeq$$+3$\,km\,s$^{-1}$ or  toward the \mbox{(-62$''$,250$''$)} position in the northern
streamer at v$_{\rm LSR}$$\simeq$$+6$\,km\,s$^{-1}$, see channel maps).

 In any case, our [\CII] excitation models confirm that the gas density in the FUV-illuminated face of OMC\,1 is high.
 The blue-shifted [\CII] emission  caused by the Veil also shows that the density must be relatively high in these 
 foreground components.
 This agrees with the general notion that the elevated stellar density in the
Orion  cluster is a consequence of hundreds of stars that formed from a remarkably dense condensation
\mbox{\citep[e.g.,][]{Gen89, Riv13}}.

\begin{figure*}[t]
\centering
\includegraphics[scale=0.47, angle=0]{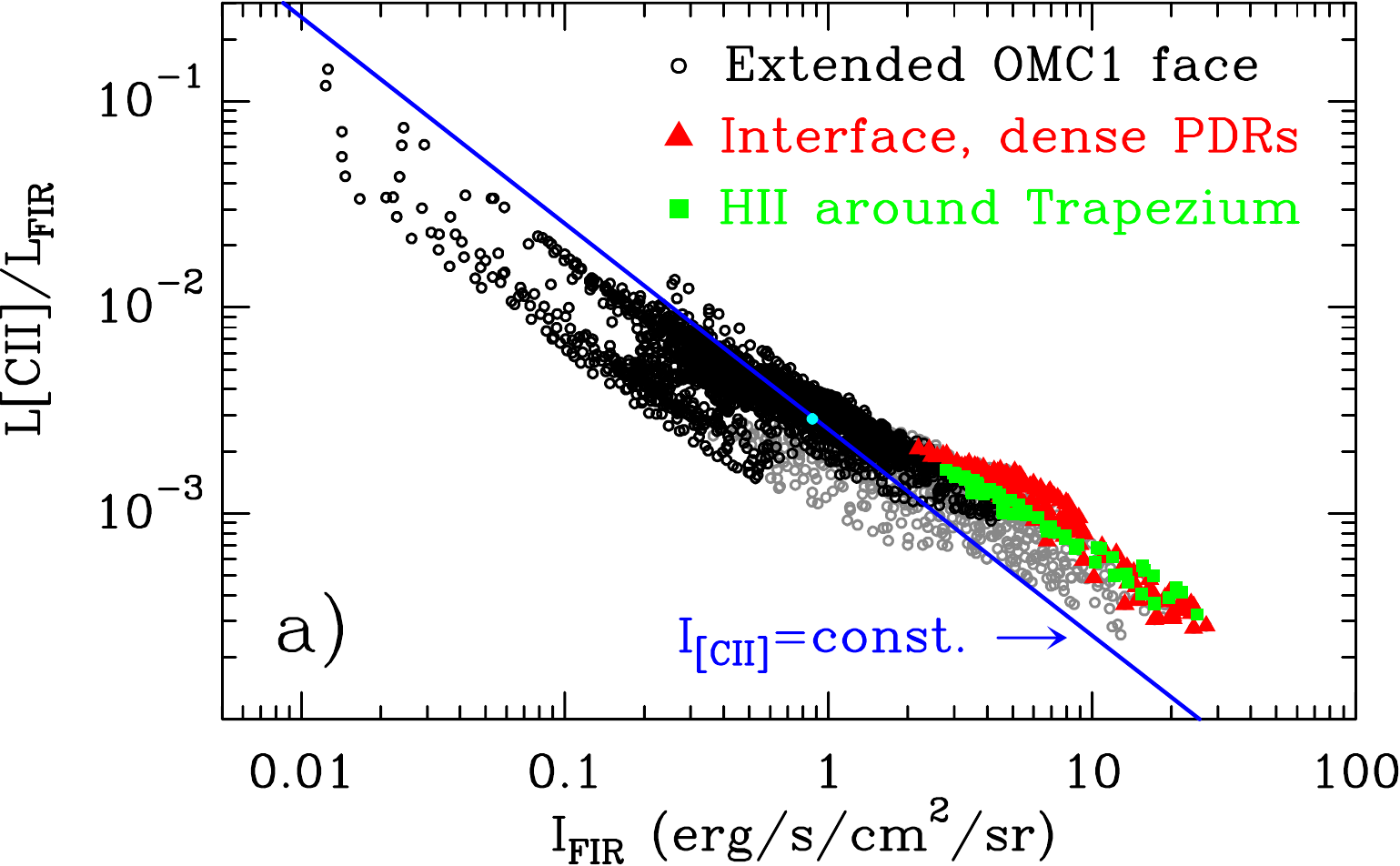} 
\hspace{0.2cm}
\includegraphics[scale=0.47, angle=0]{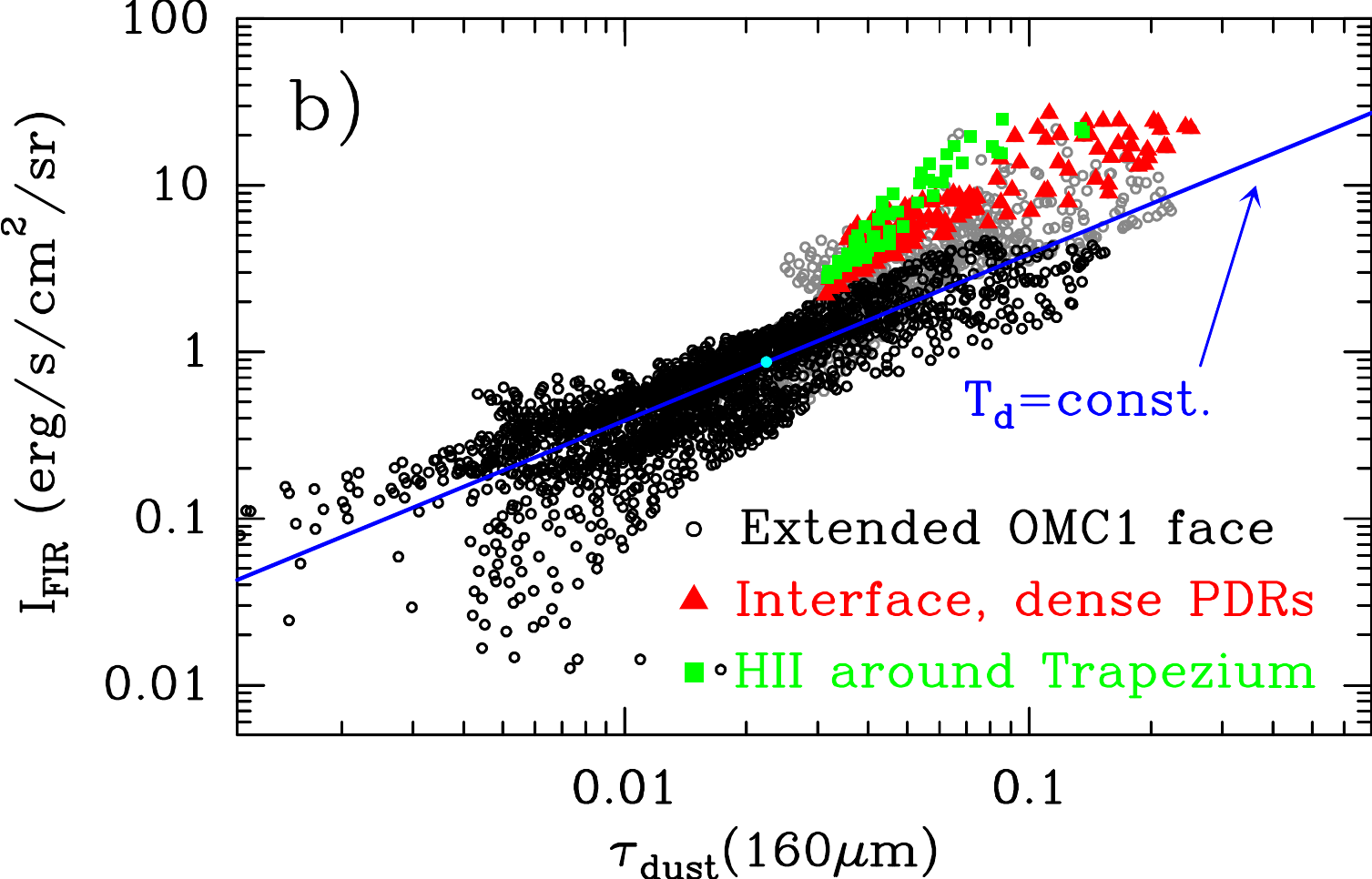}\\ 
\vspace{0.2cm}
\includegraphics[scale=0.47, angle=0]{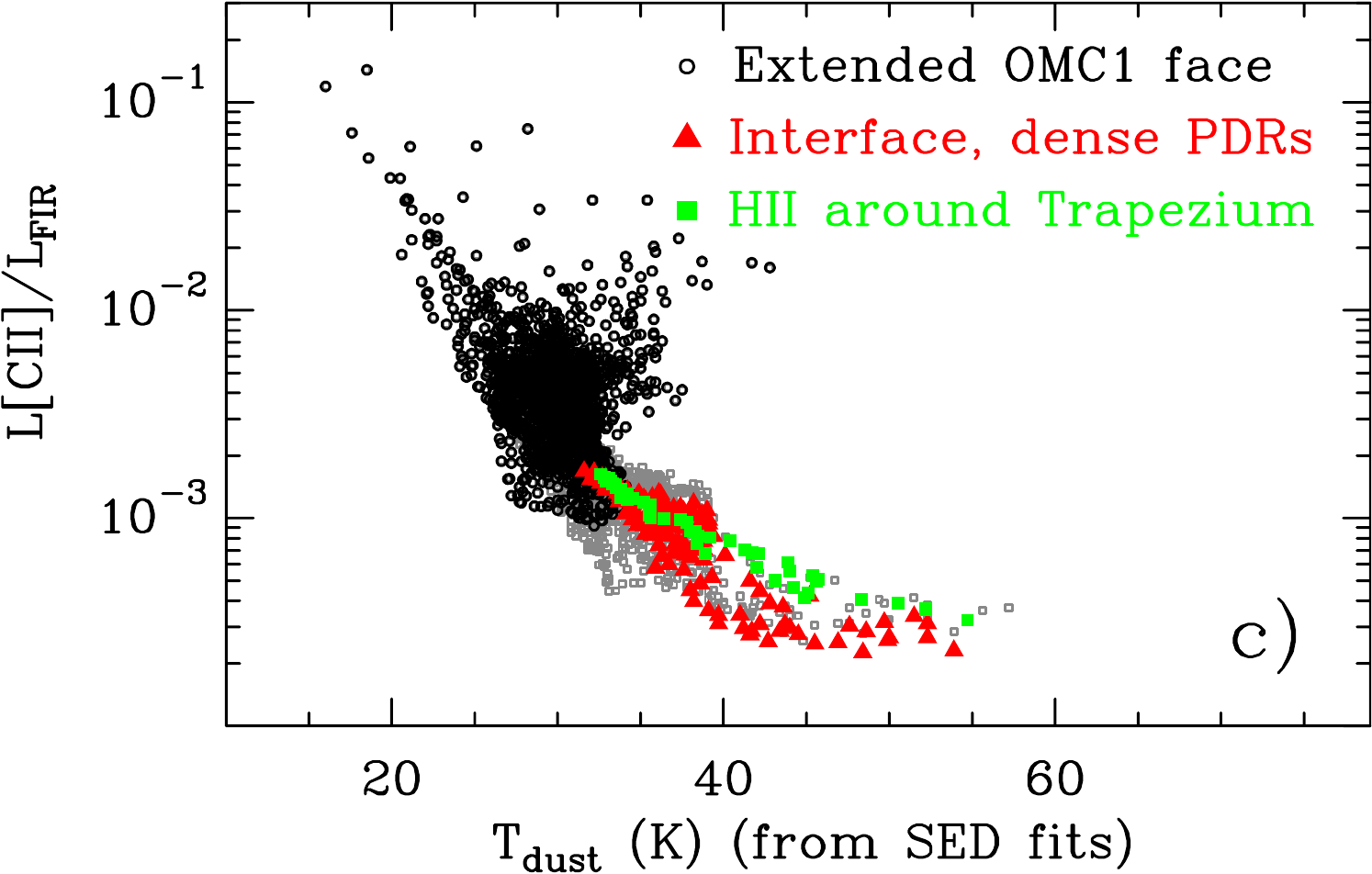} 
\hspace{0.2cm}
\includegraphics[scale=0.47, angle=0]{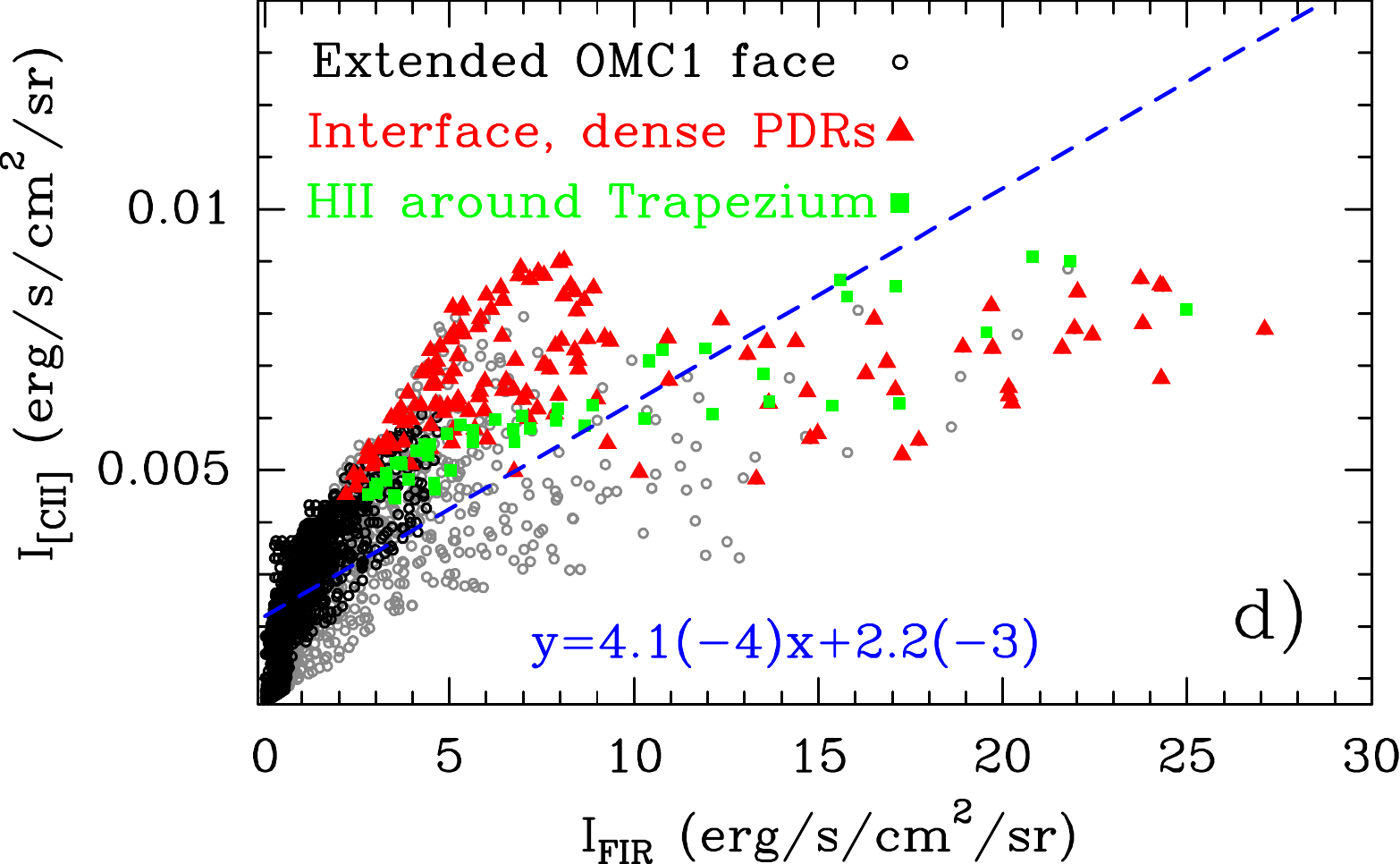}\\ 
\vspace{0.2cm}
\includegraphics[scale=0.47, angle=0]{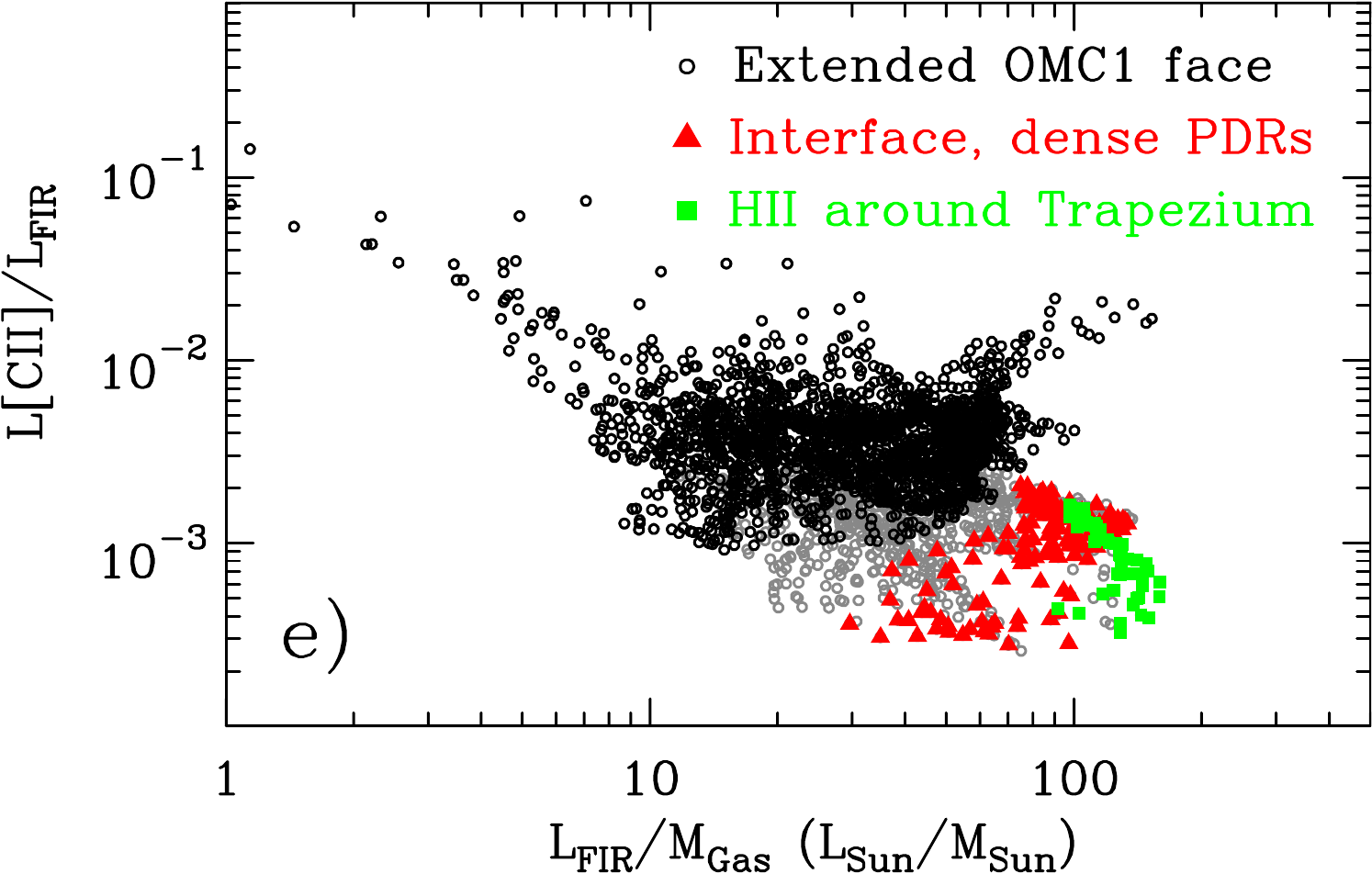} 
\hspace{0.2cm}
\includegraphics[scale=0.47, angle=0]{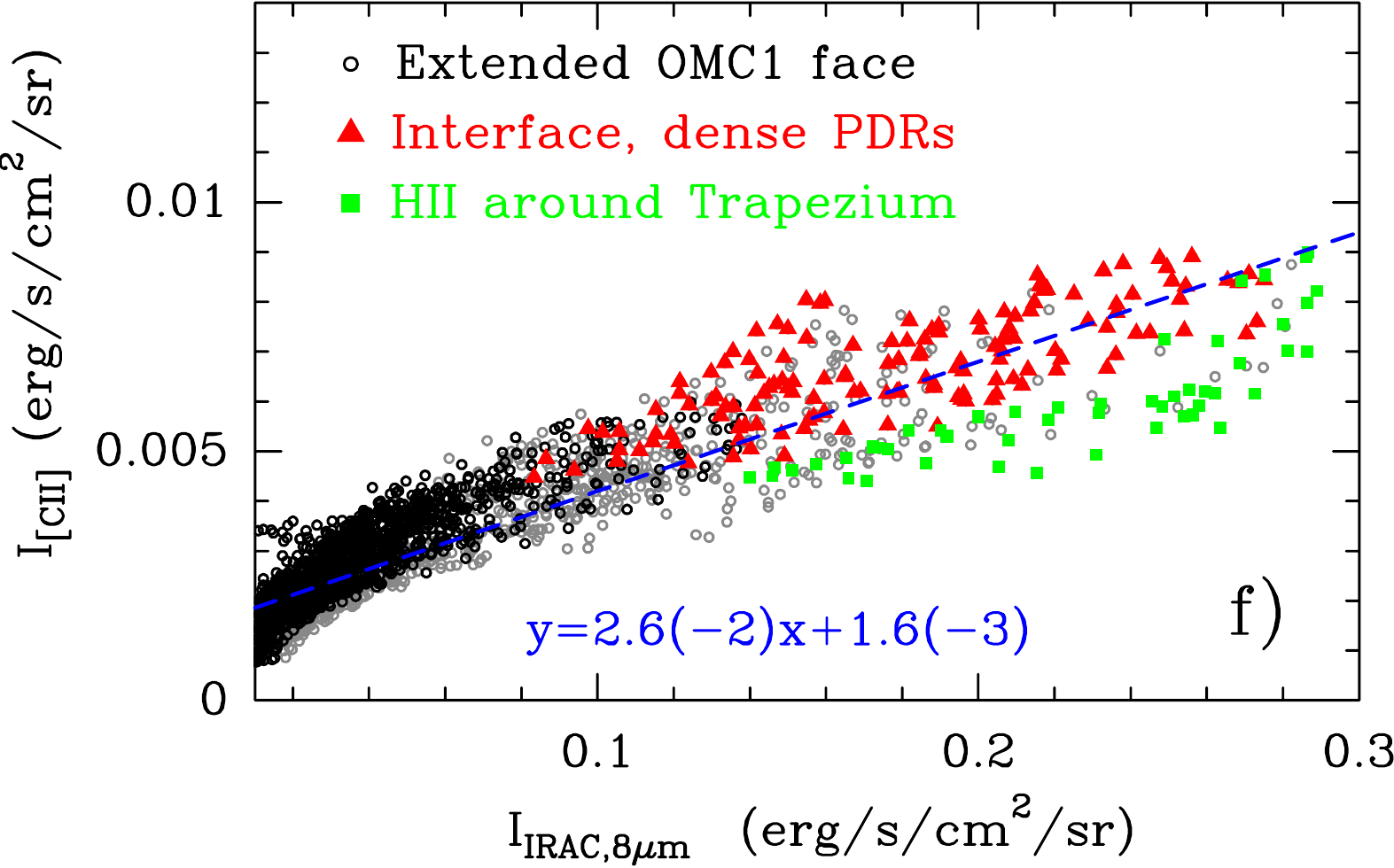}\\ 
\caption{Correlation plots extracted from the OMC\,1 maps and convolved  to a uniform  25$''$ resolution.
Black points represent positions outside the  \mbox{200$''$ (0.4\,pc)} radius circle  
(``extended cloud face'' component). Green squares represent positions toward the \HII~region component around the Trapezium.
Red triangles represent positions in the  ``dense PDR interface"  annulus (see e.g., Figure~\ref{fig_nebula}).
Gray points are other positions inside the \mbox{200$''$ (0.4\,pc)} radius circle that do not fall in the latter two categories.
Two 
 models (blue lines) are plotted in panels (a) and (b), one for a constant [\CII] intensity
cloud (a) and another for a cloud of
constant $T_{\rm d}$ (b). Both models are defined to intercept the median  $L$[\CII]/$L_{\rm FIR}$, $I_{\rm FIR}$ and $\tau_{d,160}$
of the map (cyan dots).}
\label{fig_corr}
\end{figure*}

\subsection{Correlation Diagrams}

Figure~\ref{fig_corr} shows correlation plots between several  quantities 
that are also used in the discussion of the  [\CII] extragalactic emission (one point, one galaxy).
Here we present their spatial correlation in a template massive star-forming region.
In these plots, the black points represent positions outside the  \mbox{200$''$} (0.4\,pc) radius circle, 
the ``extended cloud face'' component.
 Green squares represent lines of sight toward the 	\mbox{``\HII~region''} component around the Trapezium stars,
 and red triangles represent positions in the  \mbox{``dense PDR interface''}  annulus.
Gray points are other positions inside the \mbox{200$''$} circle that do not fall 
in the latter two categories.

Figure~\ref{fig_corr}(a) shows that the [\CII]/FIR ratio decreases with increasing FIR intensity. In OMC\,1
we find that \mbox{log\,([\CII]/FIR)$\simeq$$-$0.6$\cdot$log\,($I_{\rm FIR}$[erg\,s$^{-1}$\,cm$^{-2}$\,sr$^{-1}$])$-$2.6}, with a correlation
coefficient of $\rho$=$-$0.93.
This is reminiscent of the \mbox{``[\CII]} deficit'' observed in local ULIRGs
when compared to starburst and normal galaxies
\citep[e.g.,][]{Mal97,Luh98,Gra11}.
In a local complex like Orion, the decrease of $L$[\CII]/$L_{\rm FIR}$  
obviously occurs at orders of magnitude lower FIR luminosities and seems dominated
by the strong variations of $I_{\rm FIR}$ in the region. 
The blue line in Fig.~\ref{fig_corr}(a) shows a linear model for a constant  [\CII] intensity 
cloud (the median value of the map sample).

\mbox{Figure~\ref{fig_corr}(b)}  shows that $I_{\rm FIR}$ increases 
with $\tau_{\rm d,160}$. 
\mbox{Figure~\ref{fig_corr}(c)} shows a \mbox{$L$[\CII]/$L_{\rm FIR}$-$T_{\rm d}$} scatter plot.
In many studies, in which a complete SED cannot be constructed, $T_{\rm d}$ is replaced by the
$F_{60}/F_{100}$ or  $F_{70}/F_{160}$ flux ratio.
Theoretical models show that the photoelectic heating efficiency decreases for high $G_0/n_{\rm H}$ values (also producing higher
dust temperatures),
because grains and PAHs get more positively charged and it 
becomes more difficult to photo-eject additional electrons \citep{Tie85,Oka13}. 
As the grain charge increases for high $G_0$ fields,
the gas heating and cooling are reduced and a larger fraction of the FUV field is converted into dust FIR luminosity
(thus lower $L$[\CII]/$L_{\rm FIR}$ ratios can be expected).
Although such a trend is suggested by \mbox{Figure~\ref{fig_corr}(c)},
the dust temperatures obtained from the SED fits,  \mbox{$T_{\rm d}$=31$\pm$4\,K} (1$\sigma$),
do not change much over the mapped region and other effects seem to dominate the observed $L$[\CII]/$L_{\rm FIR}$
 variations at large spatial scales.

In particular, most of the dust emission in the extended cloud  is consistent with an roughly  constant temperature of \mbox{$T_{d}$$\simeq$30~K}. 
This leads to the approximately linear relation between dust opacity and FIR intensity
shown in \mbox{Figure~\ref{fig_corr}(b)}. 
However, the inferred dust temperatures  toward some lines of sight of small $\tau_{\rm d,160}$ are lower than 30~K, resulting in lower $L_{\rm FIR}$ values  
than those expected in a constant $T_{d}$ cloud (blue line \mbox{Figure~\ref{fig_corr}(b)}), as well as in high $L$[\CII]/$L_{\rm FIR}$  ratios in  Figure~\ref{fig_corr}(a).
On the other hand, the dust is warmer toward the dense PDRs and \HII~region, and  high
$I_{\rm FIR}$ intensities  are observed above the constant-$T_{d}$ blue line in \mbox{Figure~\ref{fig_corr}~(b)}.
The higher dust temperatures toward the central of OMC\,1  likely reflect
higher $G_0$ values and also contribution from
internal dust heating by embedded sources.
However, their associated $L$[\CII]/$L_{\rm FIR}$  ratios lie above the linear model that assumes constant
[\CII] emission in Figure~\ref{fig_corr}(a), revealing enhanced 
[\CII] intensity toward these regions. Still, the FIR intensity increases more than [\CII] (\mbox{Figure~\ref{fig_corr}(d)}),
and the resulting $L$[\CII]/$L_{\rm FIR}$ ratios are low toward lines of sight of high dust temperatures, see \mbox{Figure~\ref{fig_corr}(c)}
and Section~\ref{subsub-dust}.

Figure~\ref{fig_corr}(d) shows that, except for high FIR intensities, the [\CII]\,158\,$\mu$m intensities
are reasonably well correlated with the FIR emission ($\rho$=0.75), with
\mbox{$I_{\rm [CII]}$$\simeq$4.1$\times$10$^{-4}$$\cdot$$I_{\rm FIR}$$+$2.2$\times$10$^{-3}$} (in \mbox{erg\,s$^{-1}$\,cm$^{-2}$\,sr$^{-1}$} units).
The [\CII] intensities, however, are better correlated  with the PAH emission traced by the IRAC\,8\,$\mu$m image ($\rho$=0.91, 
\mbox{Figure~\ref{fig_corr}(f)}),
 with \mbox{$I_{\rm [CII]}$$\simeq$2.6$\times$10$^{-2}$$\cdot$$I_{\rm IRAC\,8\mu m}$$+$1.6$\times$10$^{-3}$} 
(also in \mbox{erg\,s$^{-1}$\,cm$^{-2}$\,sr$^{-1}$} units).
 As opposed to FIR, this correlation seems to
 hold for high [\CII] and PAH intensities.
Finally, Figure~\ref{fig_corr}(e) shows a correlation plot of $L$[\CII]/$L_{\rm FIR}$ with   
$L_{\rm FIR}/M_{\rm Gas}$. This plot shows much more scatter, but suggests
that  $L$[\CII]/$L_{\rm FIR}$  decreases from the extended cloud  
to the \mbox{``\HII~region''} component (highest $L_{\rm FIR}/M_{\rm Gas}$ values).

\section{Discussion}
\label{sec-disc}

\subsection{[\CII], CO and FIR Diagnostic Power}

Dense PDR models predict that the low-$J$ $^{12}$CO lines saturate (\mbox{$\tau$$\gg$1}) close to the CO 
abundance peak and thus the $L$[\CII]/$L_{\rm CO}$ ratio depends mostly on $N$(C$^+$) and $T_{k}$, 
i.e.,~on $G_0$ and $n_{\rm H}$ \citep[see e.g., PDR models by][]{Kau99}.
For densities below $n_{\rm cr}$$\sim$10$^4$\,cm$^{-3}$ and $G_0$$<$10$^3$, 
the [\CII]~158\,$\mu$m line dominates the gas cooling \citep{Kau99} and the $L$[\CII]/$L_{\rm FIR}$ ratio provides a  
good approximation to the grain photoelectric heating efficiency.
For higher densities or $G_0$ values (i.e., the OMC\,1/\HII~interfaces around the Trapezium),
[\CII] collisional de-excitation plays a role and [\OI] starts to dominate the warm gas cooling. 
[\OI]\,63,145\,$\mu$m lines are certainly useful \cite[][]{Her97} 
although they are also bright in dissociative shocks  where the [\CII] line is  faint \mbox{\citep[e.g.,][]{Goi15}}.
In addition, the opacity of the [\OI]\,63\,$\mu$m line is higher  
($\tau_{\rm [OI]}$=1.6$\tau_{\rm [CII]}$ for a [O/C]=2.2 abundance ratio)
and lies at  wavelengths  of higher dust opacity  (2.5-6.3 times higher for $\beta$=1-2).
The [\OI]\,63\,$\mu$m line opacity problem is difficult to circumvent, because $^{18}$O does not have nuclear spin 
(producing hyperfine splittings in $^{13}$C$^+$), and the isotopic shift with respect to  $^{16}$O is only $\sim$1.5\,km\,s$^{-1}$
\citep[$+$23.5\,MHz,][]{Bro93}.
Therefore, [\CII] is a robust tracer of the presence of FUV photons also in dense gas.

As noted in the literature, $L$[\CII]/$L_{\rm FIR}$  approximately follows 
 $L_{\rm CO\,2-1}/L_{\rm FIR}$  (Figure~\ref{fig_COcorr}).
For most of the mapped positions, the observed luminosities appear well correlated 
($\rho$=0.89 and slope of $\simeq$775). 
\citet{Sta91} suggested that, to a first approximation,
the $L$[\CII]/$L_{\rm FIR}$ and $L_{\rm CO}$/$L_{\rm FIR}$ luminosity ratios
can be compared with PDR models to constrain the FUV radiation field and the gas density.
Most of the points in \mbox{Figure~\ref{fig_COcorr}} roughly lie on a straight line. 
In the frame of PDR models \citep[e.g.,][]{Kau99} 
this line approximately follows the decrease 
of $G_0$ (from left to right) and the decrease of $n_{\rm H}$ (from bottom to top). 
Comparing our spatially resolved luminosity ratios\footnote{Scaling 
our data by $L_{\rm FIR}$(40-500\,$\mu$m)$\simeq$1.5\,$L_{\rm 42-122\,\mu m}$
and assuming $I_{\rm CO\,2-1}$$\simeq$8\,$I_{\rm CO\,1-0}$ (erg\,s$^{-1}$\,cm$^{-2}$\,sr$^{-1}$) 
for optically thick thermalized CO gas.}  
\mbox{(Figure~\ref{fig_COcorr})}
with the plot in Figure~5 of \citet{Sta10}, based on PDR models of \citet{Kau99} for a $A_V$=10\,mag slab,
one sees that  the lowest $L$[\CII]/$L_{\rm FIR}$$\simeq$3$\times$10$^{-4}$ and 
$L_{\rm CO\,2-1}$/$L_{\rm FIR}$$\simeq$3$\times$10$^{-6}$ values near the Trapezium 
are consistent with high $G_0$$\gtrsim$10$^{4}$   fields and  $n_{\rm H}$$\gtrsim$10$^5$~cm$^{-3}$.
On the other hand, the highest $L$[\CII]/$L_{\rm FIR}$$\gtrsim$10$^{-2}$ values, corresponding to
$L_{\rm CO\,2-1}$/$L_{\rm FIR}$$\lesssim$10$^{-4}$ ratios in the extended cloud component 
are consistent with lower FUV fields and gas densities, $G_0$$>$10$^{2}$ and $n_{\rm H}$$\approx$10$^{3-4}$~cm$^{-3}$ respectively.

\begin{figure}[t]
\centering
\includegraphics[scale=0.5, angle=0]{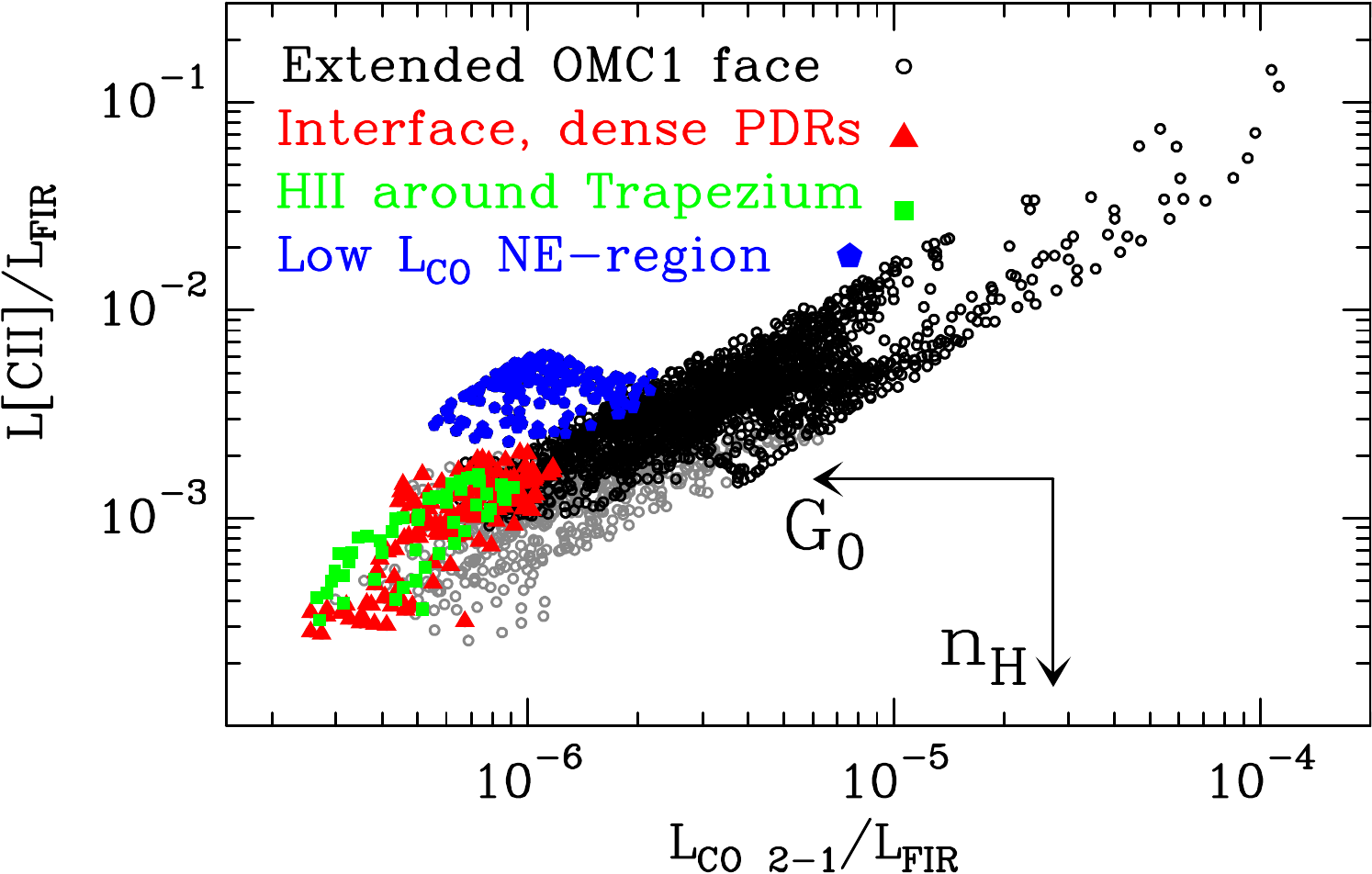}
\caption{Same as Fig.~\ref{fig_corr} but for $L$[\CII]/$L_{\rm FIR}$ versus $L_{\rm CO\,2-1}$/$L_{\rm FIR}$.
The blue pentagons represent lines of sight of low $L$(CO) and high $L$[\CII]/$L_{\rm FIR}$ ratios
(up to $\sim$5000) at the northeast of the map, see Figures~\ref{fig_lums}(c) and \ref{fig_lums}(e).
Arrows show the approximate directions of increasing $G_0$ and increasing $n_{\rm H}$ expected from PDR models
\citep[e.g.,][]{Kau99,Sta10}.}
\label{fig_COcorr}
\end{figure}

In addition, our  maps  reveal regions in which the luminosity ratios
lie outside the parameter space allowed 
by  \mbox{$A_V$=10\,mag} PDR models (blue pentagons in Figure~\ref{fig_COcorr}). 
These are specific positions with 
very high $L$[\CII]/$L_{\rm CO\,2-1}$$\simeq$5000 luminosities and low $\tau_{\rm d,160}$ 
opacities at the northeast of the map. 
Since they do not coincide with the \HII~region,
we conclude that they must be associated with lower density neutral gas at low $A_V$ (thus low CO abundances
for $G_0$$\gg$1 values).
Indeed, a much better match with PDR models is found by considering the $A_V$$\simeq$3\,mag 
cloud models of \citet{Kau99}.

\subsection{[\CII] and the Star Formation Rate in Orion}

\textit{Herschel}/HIFI has allowed the first velocity-resolved [\CII] survey of the Galactic 
plane, in which
the contribution of the different ISM phases to the [\CII] emission has been constrained \citep[e.g.,][]{Pin13}.
The combined [\CII] emission from all components added together
correlates with the SFR in similar manner as in nearby galaxies.
Including observations of the Milky Way, the LMC and several nearby galaxies dominated by star formation, 
\citet{Pin14} find: 
log\,(SFR\,[$M_{\odot}$\,yr$^{-1}$])$\simeq$0.89\,log\,($L$[\CII][erg\,s$^{-1}$])$-$36.3.
Using this correlation and the total [\CII] luminosity toward OMC\,1, 
we derive \mbox{SFR$\simeq$3.8$\times$10$^{-5}$~$M_{\sun}$\,yr$^{-1}$} in the mapped area. 
This is very similar to the rate obtained using
 \mbox{SFR($M_{\sun}$\,yr$^{-1}$)$\simeq L_{\rm FIR}/5.8\times10^9\,L_{\odot}$}
\citep[e.g.,][]{Ken98} and the observed  luminosity toward OMC\,1 
(\mbox{$L_{\rm FIR}$$\simeq$2$\times$10$^5$\,$L_{\odot}$}).

In addition, extinction maps of  galactic star-forming regions suggest
that the SFR correlates with the mass ($M'$) of the \mbox{UV-shielded} (\mbox{$A_V$$\gtrsim$7\,mag}) dense gas 
(\mbox{$n$(H$_2$)$>$10$^{4}$\,cm$^{-3}$}),
the reservoir  
that ultimately forms stars  
\mbox{\citep[e.g.,][]{And10,Lad10,Lad12}}.
These studies suggest   
\mbox{SFR\,[$M_{\odot}$\,yr$^{-1}$] $\simeq$ (1.2-1.8)$\times$10$^{-8}\,M'$\,($M_{\odot}$)}, and
this corrrelation holds from star-forming regions   to  galaxies \citep[][]{Gao04,Wu05}.
Taking the masses inferred from our observations, 
we estimate \mbox{$M'$$\simeq$$M_{\rm Gas,Total}$$-$$M_{\rm PDR}$$\simeq$2400\,$M_{\odot}$}
(using the total mass deduced from the dust emission and subtracting the PDR contribution inferred from [\CII]),
and use the above relation to derive 
\mbox{SFR$\simeq$(2.9-4.3)$\times$10$^{-5}$~$M_{\sun}$\,yr$^{-1}$}. 
Therefore, the different methods  provide consistent SFR values in Orion.

Taking into account the mapped area, our derived SFR  is 
equivalent to a  very high SFR surface density of 
\mbox{$\Sigma_{\rm SFR}$=(2.3-3.4)$\times$10$^{-5}$~$M_{\sun}$\,yr$^{-1}$\,pc$^{-2}$}. 
Together with the high surface density inferred from the mass estimates
(\mbox{$\Sigma_{\rm Gas}$$\simeq$2000~$M_{\sun}$\,pc$^{-2}$}), 
these values place the core of Orion at an extreme of a Kennicutt--Schmidt-type  surface density relation 
(\mbox{$\Sigma_{\rm SFR}\propto \Sigma_{\rm Gas}^\alpha$} with $\alpha$ close to 1). 
These,  and even higher  surface densities are also inferred toward luminous galaxy mergers hosting  intense star formation
\mbox{\citep[][]{Gen10}}. 

\subsection{$L$[\CII]/$L_{\rm FIR}$ Variations and Extragalactic Link}

The reduced $L$[\CII]/$L_{\rm FIR}$ luminosity observed  toward local ULIRGs
($L_{\rm FIR}$$\gtrsim$10$^{12}$~$L_{\sun}$) has been difficult to interpret.
Dust extinction, optically thick [\CII] emission, reduced photoelectric heating efficiency or
 soft UV fields from 
less massive stars have been invoked \mbox{\citep[e.g.,][]{Luh98}}.
In addition, extragalactic observations with \textit{Herschel} suggest that the 
``line deficit'' (relative to $L_{\rm FIR}$ or $L_{\rm FIR}$/$M_{\rm Gas}$) also applies to 
other FIR fine structure lines such as [\OI], [\NII] or [\OIII]
arising either from PDRs or \HII~regions \citep[][]{Gra11}.

\begin{figure}[t]
\centering 
\includegraphics[scale=0.87, angle=0]{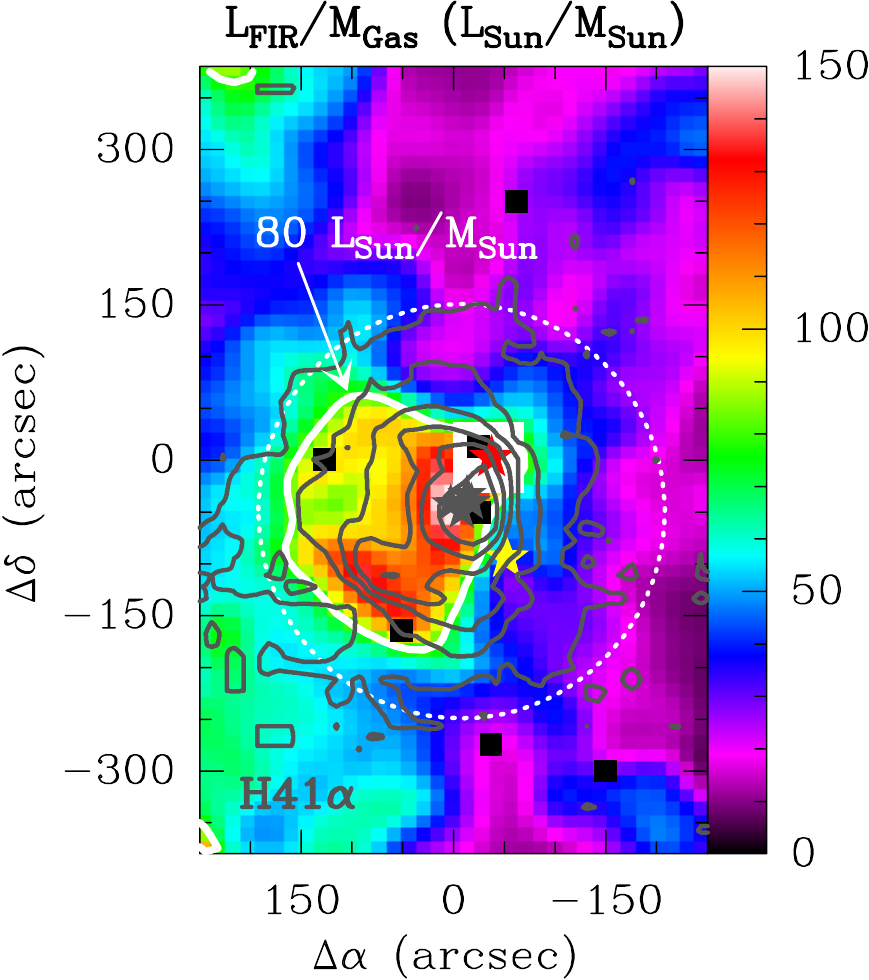}
\caption{Spatial distribution of the $L_{\rm FIR}/M_{\rm Gas}$  ratio in units of $L_{\odot}\,M_{\odot}^{-1}$
per pixel \mbox{(colored map)}
and H41$\alpha$ integrated intensity map in gray contours (from 2 to 22~K\,km\,s$^{-1}$ in steps of 4~K\,km\,s$^{-1}$). 
The  white thick curve shows the  \mbox{$L_{\rm FIR}/M_{\rm Gas}$=80~$L_{\odot}\,M_{\odot}^{-1}$}  iso-contour.
Note the good correspondence between the ionized gas emission and the regions of high $L_{\rm FIR}/M_{\rm Gas}$ ratios.
The red, yellow and the gray stars show the position of the Orion hot core, Orion S and the Trapezium stars respectively.
The black squares show the positions of the spectra shown in Figure~\ref{fig_spectra}.
}
\label{fig_L_M}
\end{figure}

\begin{figure*}[t]
\centering
\includegraphics[scale=0.72, angle=0]{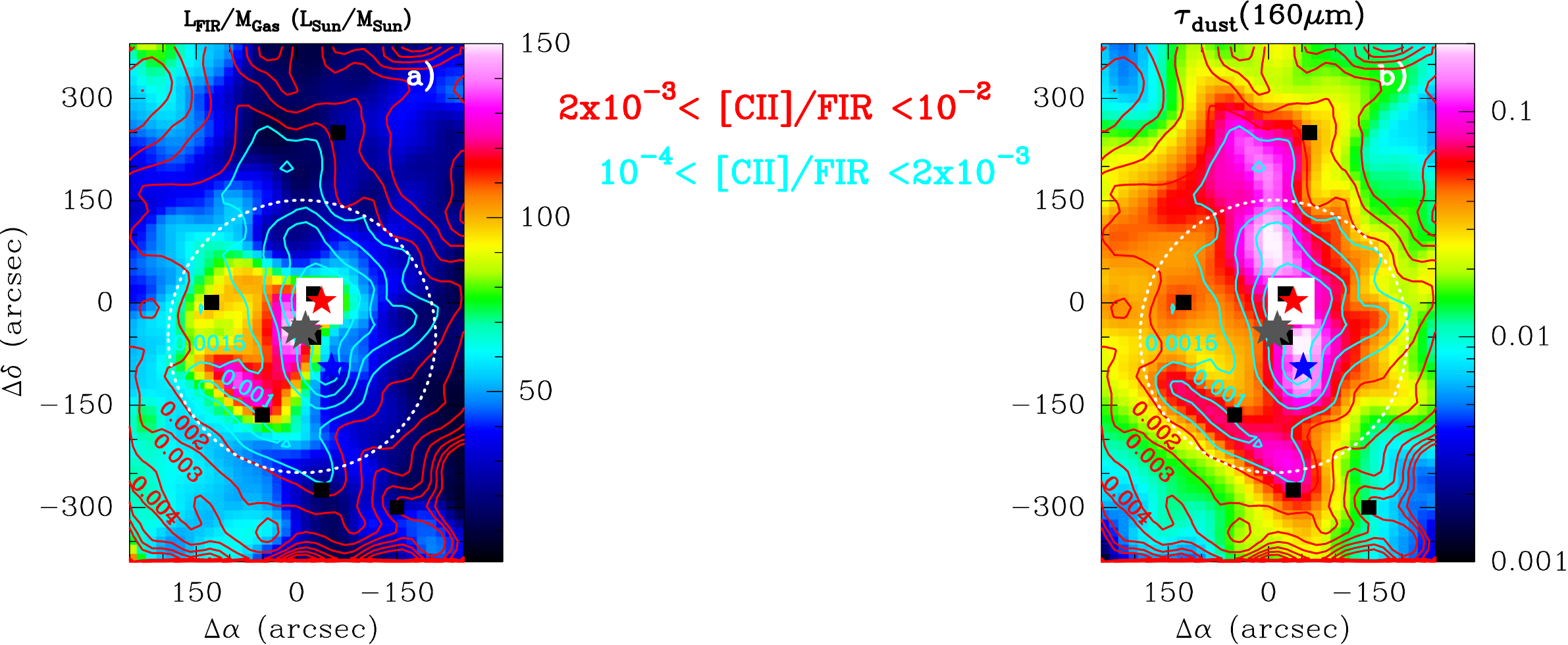}
\caption{Spatial distribution of two regimes found for 
 the $L$[\CII]/$L_{\rm FIR}$ luminosity ratio. Cyan contours for low
 $L$[\CII]/$L_{\rm FIR}$ ratios, 10$^{-4}$$-$1.5$\cdot$10$^{-3}$ toward the core of OMC\,1, 
 and red contours for high ratios, 2$\cdot$10$^{-3}$$-$10$^{-2}$ toward the extended cloud component
(the $L$[\CII]/$L_{\rm FIR}$ ratio increases from the center of the map and outwards).
$(a)$ The colored map shows the $L_{\rm FIR}/M_{\rm Gas}$ ratio, which to a first-approximation traces  ionized
 gas and the ionization parameter $U$. $(b)$~The colored map shows the spatial distribution of the 
 dust opacity at 160~$\mu$m.
The red, blue and the gray stars show the position of the Orion hot core, Orion S and the Trapezium stars respectively.
The black squares denote the positions of the spectra shown in Figure~\ref{fig_spectra}.}
\label{fig_final}
\end{figure*}

The $L_{\rm FIR}$/$M_{\rm Gas}$ ratio in particular, is expected to be proportional to the
ionization parameter $U$ (the number of H ionizing photons, $Q$(H), divided by the density $n_{\rm H}$ and
 by the distance to the ionizing stars).  
The  FIR line deficit  becomes more  apparent in galaxies with very high
\mbox{$L_{\rm FIR}$/$M_{\rm Gas}$$\gtrsim$80~$L_{\sun}$\,$M_{\sun}^{-1}$}  values \citep{Gra11,Gon15}. 
 This threshold has been suggested to separate normal galaxies from
 ultra-luminous mergers  characterized by very high SFRs \mbox{\citep[][]{Gen10}}.
 
These authors suggest that the ISM conditions and the star-formation efficiency in these
galaxies are different from those in normal and starburst galaxies. In this context, the FIR line deficits (including [\CII]) 
would be explained by the much higher ionization parameters and more extreme conditions of the \HII/PDR~regions they host 
\mbox{\citep[e.g.,][]{Gra11}}. 
Photoionization  models including dust grains  indeed predict reduced 
\mbox{$L$[\CII]/$L_{\rm FIR}$} values for high \mbox{log\,$U>-2$} parameters  \citep{Abe09}. 
One of the predicted effects is that grains absorb a higher fraction of the available UV photons relative to the gas, leaving
fewer photons to ionize and heat the gas.
Models of the radiative pressure on grains do
show that dusty \HII~regions have a different ionization balance and that they 
compress the gas and dust into a dense ionized shell \mbox{\citep{Dra11}}. This is very suggestive of
the shell-like structure revealed by the [\CII] and H41$\alpha$ emission around the Trapezium.
On the other hand, \citet{Dra11} concludes that owing to the accumulation of gas and dust in a dense shell, the
fraction of photons that ionize the gas actually increases (relative to a uniform \HII~region).
In this context, the effects of increasing $U$ may not be so severe.

Observational evidence for the presence of PAH and grains in \HII~regions is increasing
\mbox{\citep[][]{Och15}}. Their role in modulating the FUV field reaching the neutral cloud surface needs
to be taken into account.
The \HII~region around the Trapezium contains dust and is indeed characterized by a relatively high ionization parameter 
\citep[log\,$U$$\simeq$$-$1.48 and $Q$(H)$\sim$10$^{49}$~s$^{-1}$,][]{Bal91}. 
The presence of grains in the ionized nebula
can be inferred from the depleted abundances
of refractory elements like Mg, Si and Fe from the gas phase \mbox{\citep[e.g.,][]{Sim11}}.
  
In Orion, the sightlines with \mbox{$L_{\rm FIR}$/$M_{\rm Gas}$$>$80~$L_{\sun}$\,$M_{\sun}^{-1}$} show a very 
good spatial correspondence
with the dense \HII~ionized gas traced by the H41$\alpha$ line and so they do trace regions of high ionization parameter
(white thick contour in \mbox{Figure~\ref{fig_L_M})}
In particular, the  highest  \mbox{$L_{\rm FIR}$/$M_{\rm Gas}$} ratios 
are observed
close to the H41$\alpha$ emission peaks: the dense ionized gas toward the Trapezium cluster and also the ionization
front  that precedes the Orion Bar PDR.
These regions are  characterized by more extreme UV irradiation conditions ($G_0$$>$10$^4$).

We finally note that the luminosity of the large-scale vibrationally excited H$_2$ line emission toward OMC\,1 is
dominated by FUV-exited fluorescent emission in PDR gas \citep[$>$98\%, see][]{Luh94} even though the collisionally  
excited (shocked) H$_2$ dominates the small-scale emission of high surface brightness (e.g., toward BN/KL and other outflows).
Like the extended [\CII] emission we have mapped, this FUV-exited H$_2$ emission traces the FUV-illuminated face of Orion
and may also dominate the extragalactic near-IR H$_2$ emission.

\subsection{An Explanation for the ``[\CII] Deficit''}

Figure~\ref{fig_final} shows the spatial distribution of the two  $L$[\CII]/$L_{\rm FIR}$ regimes found
toward OMC\,1. 
Cyan contours for low \mbox{$\sim$10$^{-4}$$-$10$^{-3}$} ratios toward
the core of the cloud,
and red contours for high \mbox{$\sim$10$^{-3}$$-$10$^{-2}$}  ratios toward the extended cloud face component 
(increasing from  the center of OMC\,1 and outwards). 
In Figure~\ref{fig_final}~(\textit{left}), the $L$[\CII]/$L_{\rm FIR}$ contours are shown on top of
the \mbox{$L_{\rm FIR}$/$M_{\rm Gas}$} map. 
Figure~\ref{fig_final}~(\textit{right}) shows the same $L$[\CII]/$L_{\rm FIR}$
contours on top of the $\tau_{\rm d,160}$ map.

\subsubsection{Geometry  and $N$({\rm C$^+$}) Relative to Total $N_{\rm H}$ Column} 
\label{subsub-dust}

Since most of  the [\CII] emission arises from  \mbox{$A_V$$<$4\,mag} FUV-irradiated cloud surface 
layers, 
the  local opacity of the dust grains mixed with the C$^+$ atoms  ($\tau_{d,160}^{local}$) is generally small,
\mbox{$\tau_{d,160}^{local} \simeq N_{H}\cdot\mu\,m_{\rm H}\,\kappa_{160} / R_{\rm gd}  \approx 0.001$}.
Therefore, the  dust opacity does not seem  important locally, but it may be relevant if 
the  [\CII] sources are embedded in large columns of dust or if
many [\CII] emitting surfaces in a clumpy medium are included in the telescope beam.

The average 160\,$\mu$m dust opacity inside the 200$''$~circle region surrounding the Trapezium 
 is only moderate ($\tau_{\rm d,160}\simeq0.06$ or \mbox{$A_V$$>$50\,mag}). 
Therefore, the  160\,$\mu$m continuum emission traces
the column density of dust in the molecular cloud that lies behind
the ionized nebula, weighted by $T_{\rm d}$.   
The [\CII] emission, however, does not arise from the entire line of sight but 
from the FUV-illuminated face of the cloud.

\begin{figure}[b]
\centering
\vspace{0.1cm}
\includegraphics[scale=0.57, angle=0]{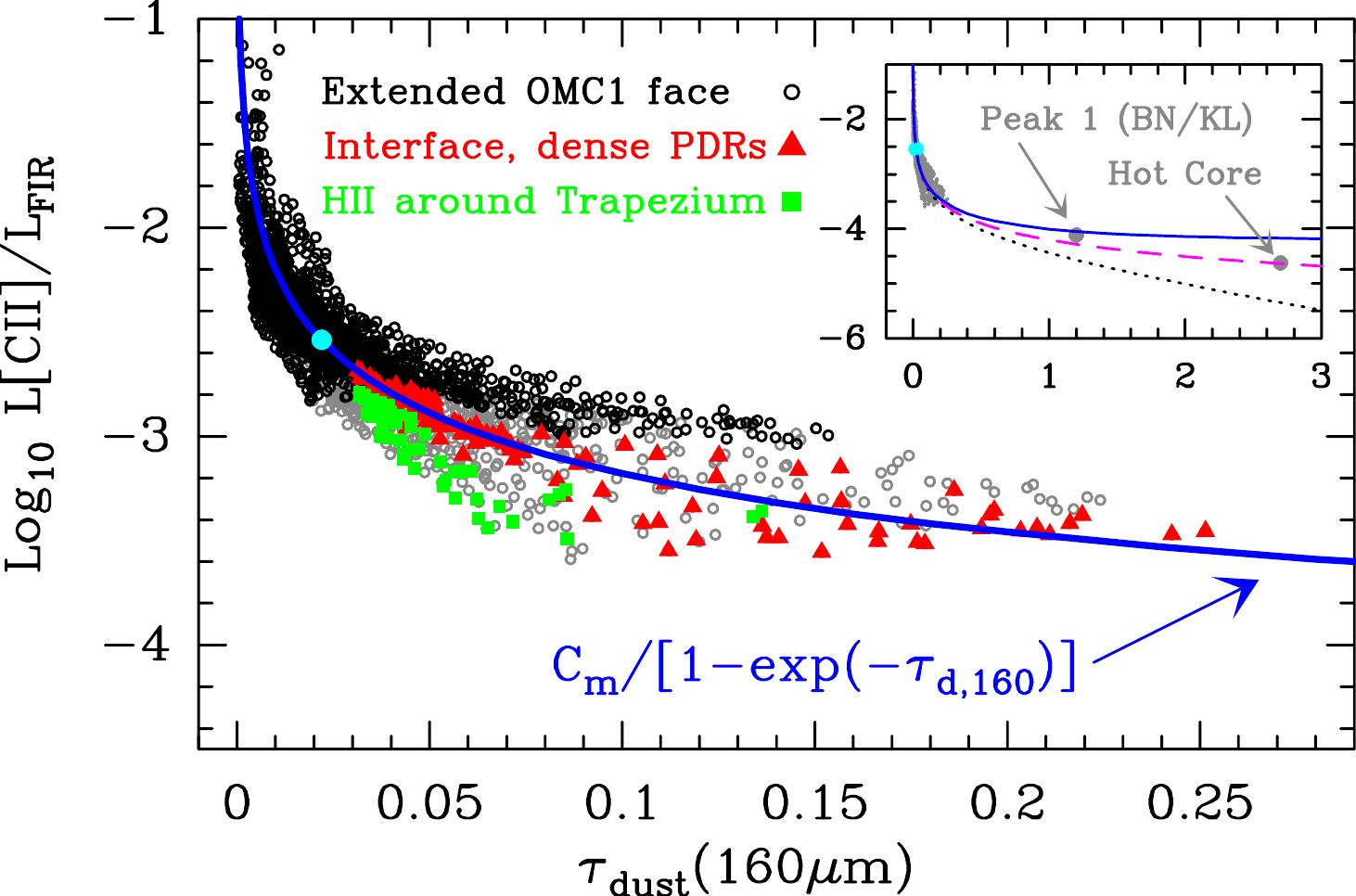}
\vspace{0.2cm}
\caption{Observed $L$[\CII]/$L_{\rm FIR}$  ratios as a function of the 160\,$\mu$m 
dust opacity. 
The different template regions are defined in Section~\ref{sect-circles}.
The inset panel shows an extension to higher dust opacity positions toward BN/KL.
The blue continuos, magenta dashed, and black dotted curves show  models of constant $I$[\CII]/$B$($T_{\rm d}$) emission
for different assumptions of the [\CII] emission location with respect to the cloud dust emission:
[\CII] in the foreground surface (as in OMC\,1), mixed C$^+$ gas and grains, and embedded [\CII] sources
 behind the cloud respectively (see sketch in Figure~\ref{fig_sketch}).
Models are defined to intercept the  median  $L$[\CII]/$L_{\rm FIR}$ and $\tau_{\rm d,160}$ values of the map (cyan dot).}  
\label{fig_dustcorr}
\end{figure}

\begin{figure*}[ht]
\centering
\vspace{0.1cm}
\includegraphics[scale=0.52, angle=0]{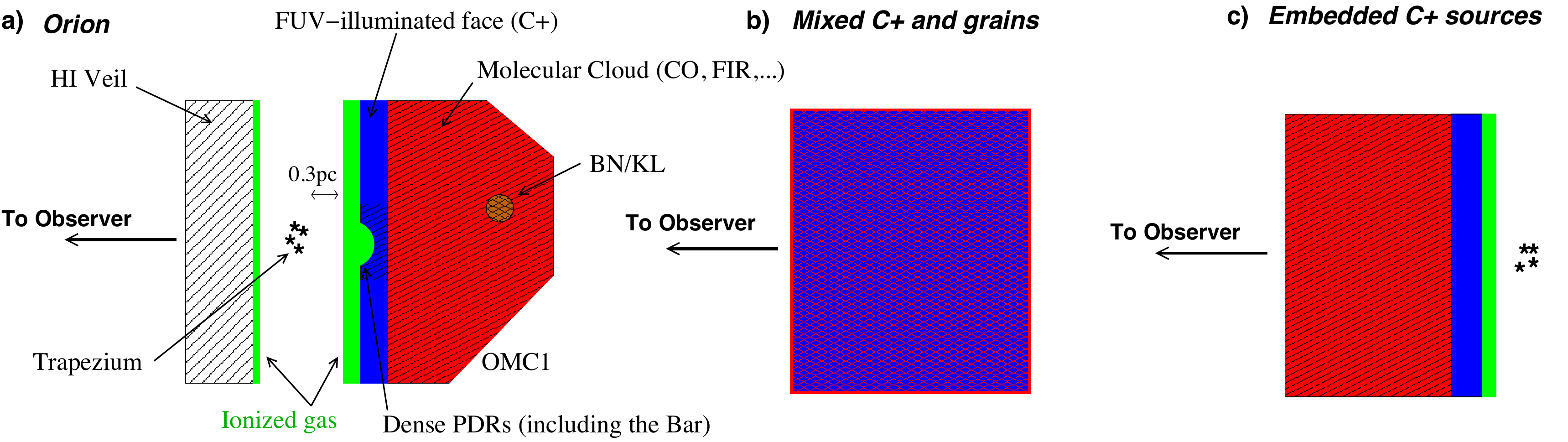}
\caption{$(a)$ Simplified geometry (not to scale) of OMC\,1 (red, molecular gas and dust), the FUV-illuminated cloud face (blue, 
dominating the [\CII] emission),  dense ionized gas in the \HII~region surrounding the Trapezium (green), and
 the atomic Veil (shaded, producing foreground  \HI~absorption). The observed [\CII] emission, $I_{\rm [CII]}$, reaches the observer (to the left) 
 almost unattenuated. At a given position, the observed  $\sim$160\,$\mu$m dust emission 
 is $B_{160}$($T_{\rm d})(1-e^{-\tau_{\rm d,160}})$, where $\tau_{\rm d,160}$ is the $\sim$160\,$\mu$m dust opacity along the
entire line of sight.
$(b)$ \textit{Mixed} C$^+$ and dust grains along the line of sight. In this case 
\mbox{$I_{\rm [CII]}$=$I_{\rm 0,[CII]}$$(1-e^{-\tau_{\rm d,160}})$/$\tau_{\rm d,160}$}, where $I_{\rm 0,[CII]}$ is
the unattenuated [\CII] emission.
$(c)$ [\CII] sources  \textit{embedded} behind large columns of  dust. In this case  
\mbox{$I_{\rm [CII]}$=$I_{\rm 0,[CII]}$\,$e^{-\tau_{\rm d,160}}$}.} 
\vspace{0.25cm}
\label{fig_sketch}
\end{figure*}

\mbox{Figure~\ref{fig_dustcorr}} shows the $L$[\CII]/$L_{\rm FIR}$  luminosity ratio 
as a function of  $\tau_{\rm d,160}$. Both quantities are clearly  tightly related. 
Therefore, the relative geometry  of the illuminating stars with respect to the molecular cloud is important 
to understand the variations of $L$[\CII]/$L_{\rm FIR}$. 
Toward OMC\,1, it can be idealized as a slab of constant [\CII] emission
 arising from the \mbox{\textit{foreground}} face of the molecular cloud
 (see a simplified sketch in Figure~\ref{fig_sketch}$a$).
In this case, and for  constant $T_{\rm d}$,   $L$[\CII]/$L_{\rm FIR}$ would vary as 
1/$(1-e^{-\tau_{\rm d,160}})$, with $\tau_{\rm d,160}$ being proportional to 
the dust column  toward each line of sight\footnote{In the most general case, 
$I_{\rm [CII]}$/$I_{\rm FIR}$$\simeq$$I_{\rm [CII]}$($T_{\rm k}$)/$[\int$$B_{\lambda}$($T_{\rm d})(1-e^{-\tau_{\rm d,\lambda}})d\lambda]$, 
with $T_{\rm k}$  and $T_{\rm d}$  only coupled inside clouds at high densities $n_{\rm}$$>$10$^6$\,cm$^{-3}$.
For the 1$\sigma$ range of $T_{\rm d}$ and $\tau_{\rm d,160}$ values toward OMC\,1, $I$[\CII]/$I_{\rm FIR}$ can be 
approximated by
\mbox{$I_{\rm [CII]}$/$I_{\rm FIR}$$\approx$$C\cdot I$[\CII]/$[B_{160}$($T_{\rm d})(1-e^{-\tau_{\rm d,160}})]$}, with the $C$ constant varying
 a factor of $\sim$2 depending on the exact $T_{\rm d}$ and $\tau_{\rm d,160}$ values. Thus, if $T_{\rm d}$  does
not vary much, the FIR intensity 
 raises with increasing total column density through the cloud, as shown in Figure~\ref{fig_corr}(b).}. 
 When this  slab model is defined to intercept the median $L$[\CII]/$L_{\rm FIR}$
and $\tau_{\rm d,160}$ values of the sample, the resulting curve 
satisfactorily reproduces the $\sim$3 orders of magnitude $L$[\CII]/$L_{\rm FIR}$  variations  observed 
throughout the map (blue  curve in Figure~\ref{fig_dustcorr}). 
The observed  scatter with respect to the model is likely produced by small variations of $T_{\rm d}$, grain charge, and 
of $I$[\CII] in lines of sight with the same $\tau_{\rm d,160}$ but different $G_0$.

\citet{Ger15} have carried out small maps of the [\CII]~158\,$\mu$m line toward
a limited sample\footnote{At the average distance of  $\sim$4~kpc of these massive 
star-forming regions, the $\sim$50$''$$\times$50$''$ mapped areas  are equivalent to $\sim$1$\times$1~pc$^2$. The typical
spatial resolution toward these more distant SFR regions with \textit{Herschel} is $\sim$0.2~pc.} of
 more distant massive star-forming regions in the Milky Way
(W31C, W49N, W51, DR21(OH), ...). 
These observations show that the [\CII] absorption produced by  diffuse clouds on 
the line of sight  can reduce the $L$[\CII]/$L_{\rm FIR}$ ratio 
by  $\sim$50~\%, if the v$_{\rm LSR}$ range of the
absorbing clouds  is not  spectrally-resolved. Velocity-\textit{unresolved} observations toward bright FIR continuum sources 
with foreground diffuse clouds thus result in apparently weaker [\CII] luminosities.
However, the most important  effect revealed
by the velocity-resolved [\CII] emission observations is the decrease of $L$[\CII]/$L_{\rm FIR}$ by large factors 
as one moves from the extended cloud component  to the FIR continuum peak.
Just like we have demonstrated in Orion
(the $L$[\CII]/$L_{\rm FIR}$ and the $f$=[$N$(C$^+$)/$N_{\rm H}$]/1.4$\times$10$^{-4}$ maps shown in
Figures \ref{fig_lums}(d)  and \ref{fig_columns_Cp_H2}(e) are similar),
 we believe that the decrease of $L$[\CII]/$L_{\rm FIR}$ 
is due to the increase in the
total column density throughout the  cloud relative to the (surface) [\CII] emitting column.

In addition, local low-metallicity galaxies such as the nearby LMC  \citep[e.g.,][]{Leb12}, but also early
galaxies at $z$$\sim$6 \citep{Cap15}, show enhanced [\CII] emission relative to their FIR emission.
These values  are interpreted as the consequence of  higher penetration of FUV photons due to the lower dust content
(lower extinction). This agrees with our view of galaxies having clouds of larger C$^+$ 
column densities relative to their total dust column.

\subsubsection{Dust Attenuation} 
\label{subsub-dust2}

The Trapezium cluster and associated \HII~region are only half enveloped 
by the molecular cloud. The [\CII] emitting region is thus facing us and is
mostly unattenuated by foreground dust. Other massive star-forming regions or galaxies may be  
\mbox{\textit{embedded}} in
 large amounts of dust (buried starbursts in galactic nuclei in particular). 
 In such cases, the [\CII] emission itself will be attenuated
by $e^{-\tau_{\rm d,160}}$ (a dust screen), 
where $\tau_{\rm d,160}$ now represents the dust opacity between the observer
and the [\CII] emitting layers (Figure~\ref{fig_sketch}$c$). For $N$(H$_2$)$\gtrsim$10$^{24}$\,cm$^{-2}$ ($\tau_{\rm d,160}$$\gtrsim$1), the expected
$L$[\CII]/$L_{\rm FIR}$ values can  be very low, lower than 5$\times$10$^{-5}$ 
(black dotted curve in the inset of \mbox{Figure~\ref{fig_dustcorr}}).
The prototypical example in the Milky Way  is Sgr~B2, the most massive starburst in the Galactic Center.
Sgr~B2 also shows a pronounced decrease of $L$[\CII]/$L_{\rm FIR}$  as the dust opacity increases from 
the extended cloud to the (FIR) optically thick  cores and embedded \HII~regions 
\mbox{\citep[e.g.,][]{Goi04,Etx13}}.
This reasoning agrees with recent observations of  local LIRGs in which the 
$L$[\CII]/$L_{\rm FIR}$  deficit  is restricted to their nuclei and not their extended disks \citep{Die14}.

An intermediate case may exist in which the C$^+$ gas and the FIR emitting dust grains are well \textit{mixed}
through the entire line of sight (Figure~\ref{fig_sketch}$b$).
In this case, the intrinsic  [\CII] emission will be attenuated by \mbox{$(1-e^{-\tau_{\rm d,160}})$/$\tau_{\rm d,160}$} 
\citep[e.g.,][]{Thr90}. This model provides less extreme corrections at large extinctions 
(magenta dashed curve in Figure~\ref{fig_dustcorr}). Given the surface origin of the [\CII] emission,
this model is however less realistic for OMC\,1, but it may apply to low metallicity galaxies (lower
$A_V$/$N_H$ and dust-to-gas mass ratios)
in which larger columns of C$^+$ relative to the total dust column exist due to the much higher penetration
of  the FUV radiation field.

Protostellar sources in Orion~BN/KL  are  embedded in large column densities of material
and thus very low $L$[\CII]/$L_{\rm FIR}$ ratios can be expected. 
Using the \textit{Herschel}/PACS spectroscopic data \citep[][]{Goi15} we derive
$L$[\CII]/$L_{\rm FIR}$$\simeq$2.4$\times$10$^{-5}$  and  $\tau_{\rm d,160}$$\simeq$2.7 toward the 
hot~core and IRc sources,  and $L$[\CII]/$L_{\rm FIR}$$\simeq$7.7$\times$10$^{-5}$ and  $\tau_{\rm d,160}$$\simeq$1.2 
toward the adjacent H$_2$~Peak\,1 outflow region.
Both positions are shown in the inset panel of Figure~\ref{fig_dustcorr}. In the frame of the simple geometrical models
discussed above, the $L$[\CII]/$L_{\rm FIR}$ ratio toward \mbox{H$_2$~Peak\,1} is still compatible with the
\textit{foreground} surface [\CII] emission model. 
The $L$[\CII]/$L_{\rm FIR}$ ratio toward the hot core is  even lower, but this is likely a consequence of 
the higher FIR luminosity due to internal dust heating from the protostellar sources.
In any case, the \textit{embedded} [\CII] emission model provides extinction corrections that are too large.

\subsubsection{Inhomogeneities and FUV radiation penetration} 

The structure of molecular clouds is not perfectly homogeneous \citep[e.g.,][]{Stu90,Fal91} and  FUV radiation
from massive stars can  penetrate to deeper cloud depths than in homogeneous clouds.
Our representative (homogeneous) PDR models for $n_{\rm H}$=2$\times$10$^{5-4}$\,cm$^{-3}$,  and dust grain properties 
appropriate to Orion,
predict \mbox{$N$(C$^+$)$\simeq$(1.2-1.6)$\times$10$^{18}$\,cm$^{-2}$}.
These C$^+$ column densities are a factor of $\simeq$2 larger than the predictions of PDR models using standard
ISM  grain properties
(Section~\ref{sect-circles}).
However, they are still lower, by a factor of $\sim$3, than the $N$(C$^+$) columns inferred toward many positions
of OMC\,1  (see Section~\ref{sec-anal}).
These  higher values suggest a degree of inhomogeneity at $\sim$25$''$ scales ($\sim$0.05\,pc).
As noted by \citet{Sta93}, an inhomogeneous structure facilitates the propagation
of FUV photons far from the Trapezium and contributes to the increase  in the fraction of the volume
that is effectively producing [\CII]  (more surfaces along the line of sight).

\section{Summary and Conclusions}

We have presented the first  $\sim$7.5$'$$\times$11.5$'$~($\sim$0.9\,pc\,$\times$\,1.4\,pc) velocity-resolved   map of the [\CII]\,158\,$\mu$m line toward OMC\,1.
In combination with FIR photometry and 
 H41$\alpha$  and \mbox{CO~$J$=2-1} line maps, we obtained the following results:

\begin{enumerate} 

\item  The main contribution  to the [\CII] luminosity ($\sim$85~\%) is from the extended, FUV-illuminated
face of the cloud ($G_0$$>$500, $n_{\rm H}$$>$5$\times$10$^3$\,cm$^{-3}$) and from  dense PDRs 
($G_0$$\gtrsim$10$^4$, $n_{\rm H}$$\gtrsim$10$^5$\,cm$^{-3}$)  at the interface
between OMC\,1 and the \HII~region surrounding the \mbox{Trapezium} cluster. In addition,
$\sim$15\,\%~of the [\CII] emission arises from a different gas component without CO counterpart.
Part of this emission  can be associated with  filamentary structures that 
 photoevaporate from the cloud and with depressions  of the visible-light emission previously  seen 
 in the images of the  Orion nebula (the Dark Bay, the Northern Dark Lane and  
 other components in the atomic \HI~gas Veil).
 An additional but minor contribution can
be associated with  ionized gas.

\item  The highest C$^+$ column density peaks are found toward the Trapezium, the Orion Bar,
and specific PDRs to the east of BN/KL. 
These bright [\CII] emission regions (\mbox{$T_{\rm P} \gtrsim 150$~K}) follow a spherical shell structure 
(in projection on the sky)
surrounding
the back edge of the blister \HII~region. 
Several of these OMC\,1/\HII~interfaces have nearly edge-on orientations. 
The excitation temperatures 
and line opacities derived from the [\CII] and [$^{13}$\CII] lines 
are \mbox{$T_{\rm ex} = 250-300$~K} and  \mbox{$\tau_{\rm [CII]} \simeq 1.5-3$}.
Hence, the [\CII] line toward these peaks is  slightly optically thick, and traces high excitation dense PDR gas 
(\mbox{$T_{\rm k}$$\gtrsim$300\,K} and \mbox{$n_{\rm H}$$\gtrsim$10$^5$~cm$^{-3}$}).

\item  $L$[\CII]/$L_{\rm FIR}$ decreases from the extended cloud component 
(\mbox{$\sim$10$^{-2}$$-$10$^{-3}$}) to the cloud center 
(\mbox{$\sim$10$^{-3}$$-$10$^{-4}$}). The latter values are reminiscent
of the ``[\CII] line deficit'' seen toward local ULIRGs.
The $L$[\CII]/$L_{\rm FIR}$ variations 
can be explained 
by the approximately  face-on  geometry  of the  molecular cloud relative to the externally illuminating stars.
This creates a [\CII] emission layer at the FUV-illuminated  surface of the cloud. The FIR emission, however, approximately 
traces the bulk of material along each line of sight.
Indeed, $L$[\CII]/$L_{\rm FIR}$ is spatially correlated with the dust column density in a way that 
agrees with a simple face-on slab model.
Trying to link the local and the extragalactic emission, we conclude that
the   [\CII] emitting column relative to the total dust  column density along each line of sight  
determines  the observed  $L$[\CII]/$L_{\rm FIR}$ variations, with the lowest ratios observed toward the highest column density peaks.
This also implies that the specific grain properties and dust-to-gas mass ratio are important parameters
 in determining the $L$[\CII]/$L_{\rm FIR}$ ratio.

\item  We 
show that the $L_{\rm FIR}$/$M_{\rm Gas}$ ratio follows the hydrogen ionizing photons traced by the H41$\alpha$ line, reaching 
 $L_{\rm FIR}$/$M_{\rm Gas}$$>$80~$L_{\odot}$\,$M_{\odot}^{-1}$ toward the
\HII~region around the Trapezium. These high $L_{\rm FIR}$/$M_{\rm Gas}$ values are interesting proxies of the ionization parameter $U$,
and are similar to the threshold where luminous galaxy mergers start to show ``[\CII] emission deficits''.
Indeed, the surface densities   we infer toward OMC\,1 (\mbox{$\Sigma_{\rm Gas}$$\simeq$2000~$M_{\sun}$\,pc$^{-2}$} and
\mbox{$\Sigma_{\rm SFR}$=(2.3-3.4)$\times$10$^{-5}$~$M_{\sun}$\,yr$^{-1}$\,pc$^{-2}$}) 
are comparable to those inferred in  galaxies  hosting vigorous star-formation. 
The interesting point of course is that these conditions, those of a young massive cluster
still surrounded by its dense parental molecular cloud,  can dominate the FIR emission at kpc galactic scales.

\end{enumerate}

The [\CII]~158~$\mu$m line is
undoubtedly a bright and powerful diagnostic of the FUV-illuminated ISM. 
In combination with other
tracers ([\OI], CO, dust SEDs, PAHs  and \HI), a  very detailed picture of the dominant environment 
and  prevailing physical conditions can be extracted. In addition, observations of  
molecular ions (CO$^+$, HOC$^+$, CF$^+$ or CH$^+$) directly related to C$^+$,  are becoming interesting tools to
 constrain the properties of the \mbox{``C$^+$ layers''} in 
FUV-illuminated H$_2$ gas \citep[e.g.,][]{Sto95,Fue03,Neu06,Guz12,Nag13}. 
Even more \textit{exotic} ions such as OH$^+$ or ArH$^+$ are more specific tracers of 
 \mbox{low H$_2$ fraction} (nearly atomic) gas  where [\CII] is also expected \citep[][]{Neu10,vdT13,Sch14}.
Most of them can be observed with ALMA.

\acknowledgments

We acknowledge helpful comments and suggestions from our referee.
We thank Spanish MINECO for funding support under grants CSD2009-00038, AYA2009-07304 and \mbox{AYA2012-32032}.
We  also thank the ERC for support under grant \mbox{ERC-2013-Syg-610256-NANOCOSMOS}.
Part of this  research was carried out at the Jet Propulsion Laboratory (JPL), California Institute of Technology (Caltech), 
under a contract with the National Aeronautics and Space Administration (NASA).
Support for this work was provided by NASA through an award issued by JPL/Caltech.
This work was supported by the German
\emph{Deut\-sche For\-schungs\-ge\-mein\-schaft, DFG\/} project
number SFB 956, C1.
This work was in part supported by the CNRS program ``Physique et Chimie du
Milieu Interstellaire” (PCMI)''.\\

{\it Facilities:} \facility{Herschel Space Observatory}, \facility{IRAM-30m}\\

{\it \textbf{Notes Added after Proofs:}} Three relevant papers were discovered after proof reading, below 
we include their most pertinent information related to this work:
\begin{itemize}

\item O'Dell \& Harris (2010, AJ, 140, 985) presented a more detailed geometrical model 
of the Orion molecular cloud, Orion nebula and Veil (see their Figure 13).

\item D\'{\i}az-Santos et al. (2013, ApJ, 774, 68) determined the  $L$[\CII]/$L_{\rm FIR}$ ratio in 
a sample of 202 local LIRG and ULIRG galaxies. The $L$[\CII]/$L_{\rm FIR}$ ratio was found to scale
with the strength of the the 9.7\,$\mu$m silicate absorption feature, except for ``optically thick'' galaxies
showing very large 9.7\,$\mu$m strengths. Unlike the MIR dust emission/absorption, the FIR/submm  emission
is more sensitive to the \text{total} column density of dust in dense clouds. Thus, their observed trend is consistent
with our detailed interpretation of the $L$[\CII]/$L_{\rm FIR}$ variations.

\item Compared to local ULIRGs, Rigopoulou et al. (2014, ApJL, 781, L15) found enhanced 
$L$[\CII]/$L_{\rm FIR}$ luminosity ratios in
a sample of \textit{intermediate} redshift (0.2$<$z$<$0.8) ULIRGs observed with \textit{Herschel}/SPIRE.
This result is more similar to observations of \textit{high} redshift galaxies with ALMA, and can also be understood
as variations of the [\CII] emitting column relative to the total dust  column density toward each galaxy.

\end{itemize}

\clearpage

\appendix

\section{Appendix: SED fitting examples}

\begin{figure}[h]
\centering
\includegraphics[scale=0.5, angle=0]{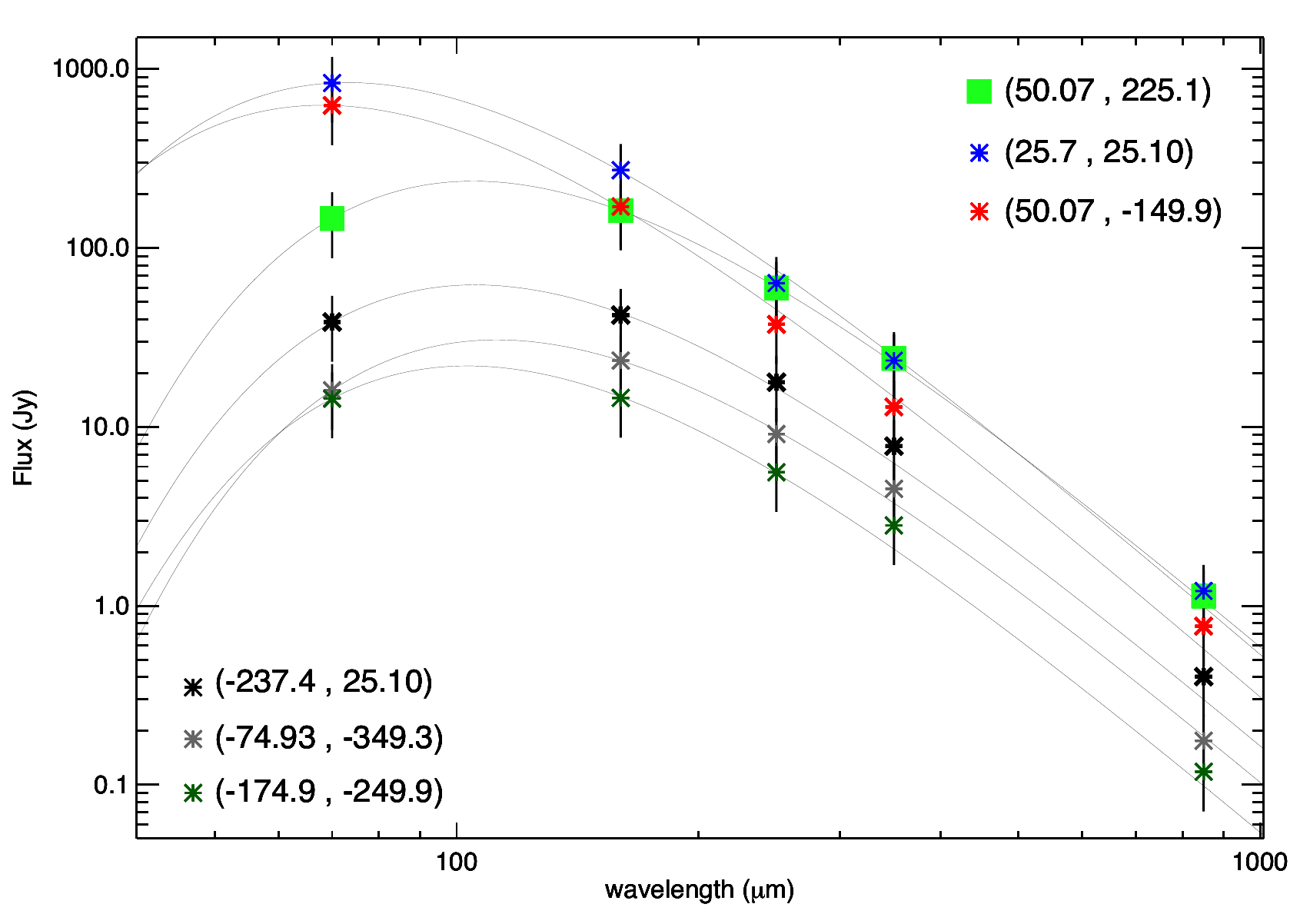}
\caption{Appendix: Example of SED fits to  \textit{Herschel}/PACS-SPIRE and \textit{JCMT}/SCUBA2 photometric data
 in Jy/pixel units. All convolved
at a uniform angular resolution of 25$''$. Selected positions are indicated by their offsets with respect to the
map center in arcsec.}
\label{fig_SED_fits}
\end{figure}


\begin{thebibliography}{}



\bibitem[Abel(2006a)]{Abe06a} 
Abel, N.~P.\ 2006, \mnras, 368, 1949 

\bibitem[Abel et al.(2006b)]{Abe06} 
Abel, N.~P., Ferland, G.~J., O'Dell, C.~R., Shaw, G., \& Troland, T.~H.\ 2006b, \apj, 644, 344 

\bibitem[Abel et al.(2009)]{Abe09}
Abel, N.~P., Dudley, C., Fischer, J., Satyapal, S., \& van Hoof, P.~A.~M.\ 2009, \apj, 701, 1147 

\bibitem[Allamandola et al.(1989)]{All89} 
Allamandola, L.~J., Tielens, A.~G.~G.~M., \& Barker, J.~R.\ 1989, \apjs, 71, 733 

\bibitem[Allers et al.(2005)]{All05} 
Allers, K.~N., Jaffe, D.~T., Lacy, J.~H., Draine, B.~T., \& Richter, M.~J.\ 2005, \apj, 630, 368 

\bibitem[Andr{\'e} et al.(2010)]{And10} 
Andr{\'e}, P., Men'shchikov, A., Bontemps, S., et al.\ 2010, \aap, 518, LL102 

\bibitem[Audit \& Hennebelle(2005)]{Aud05} 
Audit, E., \& Hennebelle, P.\ 2005, \aap, 433, 1 

\bibitem[Arab et al.(2012)]{Ara12} 
Arab, H., Abergel, A., Habart, E., et al.\ 2012, \aap, 541, A19

\bibitem[Baldwin et al.(1991)]{Bal91} 
Baldwin, J.~A., Ferland, G.~J., Martin, P.~G., et al.\ 1991, \apj, 374, 580 

\bibitem[Bally et al.(1987)]{Bal87} 
Bally, J., Langer, W.~D., Stark, A.~A., \& Wilson, R.~W.\ 1987, \apjl, 312, L45 

\bibitem[Bally(2008)]{Bal08} 
Bally, J.\ 2008, hsf1, 459 

\bibitem[Barinovs et al.(2005)]{Bar05} 
Barinovs, {\u G}., van Hemert, M.~C., Krems, R., \& Dalgarno, A.\ 2005, \apj, 620, 537 

\bibitem[Bernard-Salas et al.(2012)]{Ber12} 
Bernard-Salas, J., Habart, E., Arab, H., et al.\ 2012, \aap, 538, AA37 

\bibitem[Bern{\'e} et al.(2014)]{Ber14} 
Bern{\'e}, O., Marcelino, N., \& Cernicharo, J.\ 2014, \apj, 795, 13 

\bibitem[Bertoldi \& Draine(1996)]{Ber96} 
Bertoldi, F., \& Draine, B.~T.\ 1996, \apj, 458, 222 

\bibitem[Blake et al.(1987)]{Bla87} 
Blake, G.~A., Sutton, E.~C., Masson, C.~R., \& Phillips, T.~G.\ 1987, \apj, 315, 621 

\bibitem[Boreiko \& Betz(1996)]{Bor96} 
Boreiko, R.~T., \& Betz, A.~L.\ 1996, \apjl, 467, L113

\bibitem[Brisbin et al.(2015)]{Bri15} 
Brisbin, D., Ferkinhoff, C., Nikola, T., et al.\ 2015, \apj, 799, 13 

\bibitem[Brogan et al.(2005)]{Bro05} 
Brogan, C.~L., Troland, T.~H., Abel, N.~P., Goss, W.~M., 
\& Crutcher, R.~M.\ 2005, ASPC, 343, 183

\bibitem[Brown et al.(1993)]{Bro93} 
Brown, J.~M., Evenson, K.~M., \& Zink, L.~R.\ 1993, \pra, 48, 3761 

\bibitem[Capak et al.(2015)]{Cap15} 
Capak, P.~L., Carilli, C., Jones, G., et al.\ 2015, \nat, 522, 455 

\bibitem[Cardelli et al.(1989)]{Car89} 
Cardelli, J.~A., Clayton, G.~C., \& Mathis, J.~S.\ 1989, \apj, 345, 245

\bibitem[Carilli \& Walter(2013)]{Car13} 
Carilli, C.~L., \& Walter, F.\ 2013, \araa, 51, 105 

\bibitem[Chen et al.(2014)]{Che14} 
Chen, J.-H., Goldsmith, P.~F., Viti, S., et al.\ 2014, \apj, 793, 111 

\bibitem[Cooksy et al.(1986)]{Coo86} 
Cooksy, A.~L., Blake,  G.~A., \& Saykally, R.~J.\ 1986, \apjl, 305, L89 

\bibitem[Crawford et al.(1985)]{Cra85} 
Crawford, M.~K., Genzel, R., Townes, C.~H., \& Watson, D.~M.\ 1985, \apj, 291, 755 

\bibitem[Cuadrado et al.(2015)]{Cua15} 
Cuadrado, S., Goicoechea, J.~R., Pilleri, P., et al.\ 2015, \aap, 575, A82 

\bibitem[Cubick et al.(2008)]{Cub08} 
Cubick, M., Stutzki, J., Ossenkopf, V., Kramer, C., Rollig, M.\ 2008, \aap, 488, 623 

\bibitem[Dalgarno \& McCray(1972)]{Dal72} 
Dalgarno, A., \& McCray, R.~A.\ 1972, \araa, 10, 375 

\bibitem[de Graauw et al.(2010)]{dG10} 
de Graauw, T., Helmich, F.~P., Phillips, T.~G., et al.\ 2010, \aap, 518, LL6 

\bibitem[D{\'{\i}}az-Santos et al.(2014)]{Die14} 
D{\'{\i}}az-Santos, T., Armus, L., Charmandaris, V., et al.\ 2014, \apjl, 
788, LL17 

\bibitem[Draine(2011)]{Dra11} 
Draine, B.~T.\ 2011, \apj, 732, 100 

\bibitem[Etxaluze et al.(2013)]{Etx13} 
Etxaluze, M., Goicoechea, J.~R., Cernicharo, J., et al.\ 2013, \aap, 556, AA137 

\bibitem[Falgarone et al.(1991)]{Fal91} 
Falgarone, E., Phillips, T.~G., \& Walker, C.~K.\ 1991, \apj, 378, 186 

\bibitem[Fuente et al.(2003)]{Fue03} 
Fuente, A., Rodr{\i}guez-Franco, A., Garc{\i}a-Burillo, S., Mart{\i}n-Pintado, J., \& Black, J.~H.\ 2003, \aap, 406, 899 

\bibitem[Gao \& Solomon(2004)]{Gao04} 
Gao, Y., \& Solomon, P.~M.\ 2004, \apj, 606, 271 

\bibitem[Genzel \& Stutzki(1989)]{Gen89} 
Genzel, R., \& Stutzki, J.\ 1989, \araa, 27, 41 

\bibitem[Genzel et al.(2010)]{Gen10} 
Genzel, R., Tacconi, L.~J., Gracia-Carpio, J., et al.\ 2010, \mnras, 407, 2091 

\bibitem[Gerin et al.(2015)]{Ger15} 
Gerin, M., Ruaud, M., Goicoechea, J.~R., et al.\ 2015, \aap, 573, AA30 

\bibitem[Goicoechea et al.(2004)]{Goi04} 
Goicoechea, J.~R., Rodr{\'{\i}}guez-Fern{\'a}ndez, N.~J., \& Cernicharo, J.\ 2004, \apj, 600, 214 

\bibitem[Goicoechea et al.(2006)]{Goi06} 
Goicoechea, J.~R., Pety, J., Gerin, M., et al.\ 2006, \aap, 456, 565 

\bibitem[Goicoechea \& Le Bourlot(2007)]{Goi07} 
Goicoechea, J.~R., \& Le Bourlot, J.\ 2007, \aap, 467, 1 

\bibitem[Goicoechea et al.(2015)]{Goi15}
Goicoechea, J.~R., Chavarr{\'{\i}}a, L., Cernicharo, J., et al.\ 2015, \apj, 799, 102 

\bibitem[Goldsmith et al.(1997)]{Gol97} 
Goldsmith, P.~F., Bergin, E.~A., \& Lis, D.~C.\ 1997, \apj, 491, 615 

\bibitem[Goldsmith et al.(2012)]{Gol12}
Goldsmith, P.~F., Langer, W.~D., Pineda, J.~L., \& Velusamy, T.\ 2012, \apjs, 203, 13 

\bibitem[Gonz{\'a}lez-Alfonso et al.(2015)]{Gon15} 
Gonz{\'a}lez-Alfonso, E., Fischer, J., Sturm, E., et al.\ 2015, \apj, 800, 69

\bibitem[Graci{\'a}-Carpio et al.(2011)]{Gra11} 
Graci{\'a}-Carpio, J., Sturm, E., Hailey-Dunsheath, S., et al.\ 2011,  \apjl, 728, LL7 

\bibitem[Graf et al.(2012)]{Gra12} 
Graf, U.~U., Simon, R., Stutzki, J., et al.\ 2012, \aap, 542, L16 

\bibitem[Grenier et al.(2005)]{Gre05} 
Grenier, I.~A., Casandjian, J.-M., \& Terrier, R.\ 2005, Sci, 307, 1292 

\bibitem[Griffin et al.(2010)]{Gri10} 
Griffin, M.~J., Abergel, A., Abreu, A., et al.\ 2010, \aap, 518, L3 

\bibitem[Guzm{\'a}n et al.(2012)]{Guz12} 
Guzm{\'a}n, V., Pety, J., Gratier, P., et al.\ 2012, \aap, 543, LL1 

\bibitem[Habing(1968)]{Hab68} 
Habing, H.~J.\ 1968, \bain, 19, 421 

\bibitem[Herrera-Camus et al.(2015)]{Her15} 
Herrera-Camus, R., Bolatto, A.~D., Wolfire, M.~G., et al.\ 2015, \apj, 800, 1

\bibitem[Herrmann et al.(1997)]{Her97} 
Herrmann, F., Madden, S.~C., Nikola, T., et al.\ 1997, \apj, 481, 343 

\bibitem[Hildebrand(1983)]{Hil83} 
Hildebrand, R.~H.\ 1983, \qjras, 24, 267 

\bibitem[Hogerheijde et al.(1995)]{Hog95} 
Hogerheijde, M.~R., Jansen, D.~J., \& van Dishoeck, E.~F.\ 1995, \aap, 294, 792 

\bibitem[Holland et al.(2013)]{Hol2013} 
Holland, W.~S., Bintley, D., Chapin, E.~L., et al.\ 2013, \mnras, 430, 2513 

\bibitem[Hollenbach \& Tielens(1999)]{Hol99} 
Hollenbach, D.~J., \& Tielens, A.~G.~G.~M.\ 1999, RvMP, 71, 173 

\bibitem[Ingalls et al.(2002)]{Ing02} 
Ingalls, J.~G., Reach, W.~T., \& Bania, T.~M.\ 2002, \apj, 579, 289 

\bibitem[Kapala et al.(2015)]{Kap15} 
Kapala, M.~J., Sandstrom, K., Groves, B., et al.\ 2015, \apj, 798, 24 

\bibitem[Kaufman et al.(1999)]{Kau99} 
Kaufman, M.~J., Wolfire, M.~G., Hollenbach, D.~J., \& Luhman, M.~L.\ 1999, \apj, 527, 795 

\bibitem[Kennicutt(1998)]{Ken98} 
Kennicutt, R.~C., Jr.\ 1998, \apj, 498, 541 

\bibitem[Kester et al.(2014)]{Kes14} 
Kester, D., Avruch, I., \& Teyssier, D.\ 2014, AIPC, 1636, 62 

\bibitem[Lada et al.(2010)]{Lad10} 
Lada, C.~J., Lombardi, M., \& Alves, J.~F.\ 2010, \apj, 724, 687 

\bibitem[(2012)]{Lad12} 
Lada, C.~J., Forbrich, J., Lombardi, M., \& Alves, J.~F.\ 2012, \apj, 745, 190 

\bibitem[Langer \& Penzias(1990)]{Lan90} 
Langer, W.~D., \& Penzias, A.~A.\ 1990, \apj, 357, 477 

\bibitem[Lebouteiller et al.(2012)]{Leb12} 
Lebouteiller, V., Cormier, D., Madden, S.~C., et al.\ 2012, \aap, 548, A91 

\bibitem[Le Petit et al.(2006)]{LP06} 
Le Petit, F., Nehm{\'e}, C., Le Bourlot, J., \& Roueff, E.\ 2006, \apjs, 164, 506 

\bibitem[Lee(1968)]{Lee68} 
Lee, T.~A.\ 1968, \apj, 152, 913 

\bibitem[Li \& Draine(2001)]{Li01} 
Li, A., \& Draine, B.~T.\ 2001, \apj, 554, 778 

\bibitem[Lilley \& Palmer(1968)]{Lil68} 
Lilley, A.~E., \& Palmer, P.\ 1968, \apjs, 16, 143 

\bibitem[Lis et al.(1998)]{Lis98} 
Lis, D.~C., Serabyn, E., Keene, J., et al.\ 1998, \apj, 509, 299 

\bibitem[Luhman et al.(1994)]{Luh94} 
Luhman, M.~L., Jaffe, D.~T., Keller, L.~D., \& Pak, S.\ 1994, \apjl, 436, L185 

\bibitem[Luhman et al.(1998)]{Luh98} 
Luhman, M.~L., Satyapal, S., Fischer, J., et al.\ 1998, \apjl, 504, L11 

\bibitem[Madden et al.(1993)]{Mad93} 
Madden, S.~C., Geis, N., Genzel, R., et al.\ 1993, \apj, 407, 579 

\bibitem[Madden et al.(1997)]{Mad97} 
 Madden, S.~C., Poglitsch, A., Geis, N., Stacey, G.~J., \& Townes, C.~H.\ 1997, \apj, 483, 200

\bibitem[Malhotra et al.(1997)]{Mal97} 
Malhotra, S., Helou, G., Stacey, G., et al.\ 1997, \apjl, 491, L27 

\bibitem[Malhotra et al.(2001)]{Mal01} 
Malhotra, S., Kaufman, M.~J., Hollenbach, D., et al.\ 2001, \apj, 561, 766 

\bibitem[Megeath et al.(2012)]{Meg11} 
Megeath, S.~T., Gutermuth, R., Muzerolle, J., et al.\ 2012, \aj, 144, 192 

\bibitem[Mookerjea et al.(2003)]{Moo03} 
Mookerjea, B., Ghosh, S.~K., Kaneda, H., et al.\ 2003, \aap, 404, 569 

\bibitem[Nagy et al.(2013)]{Nag13} 
Nagy, Z., Van der Tak, F.~F.~S., Ossenkopf, V., et al.\ 2013, \aap, 550, AA96 

\bibitem[Nakagawa et al.(1998)]{Nak98} 
Nakagawa, T., Yui, Y.~Y., Doi, Y., et al.\ 1998, \apjs, 115, 259 

\bibitem[Neufeld et al.(2006)]{Neu06} 
Neufeld, D.~A., Schilke, P., Menten, K.~M., et al.\ 2006, \aap, 454, L37 

\bibitem[Neufeld et al.(2010)]{Neu10} 
Neufeld, D.~A., Goicoechea, J.~R., Sonnentrucker, P., et al.\ 2010, \aap, 521, LL10 

\bibitem[Ochsendorf \& Tielens(2015)]{Och15} 
Ochsendorf, B.~B., \& Tielens, A.~G.~G.~M.\ 2015, \aap, 576, A2 

\bibitem[O'Dell(2001)]{Ode01} 
O'Dell, C.~R.\ 2001, \araa, 39, 99 

\bibitem[O'Dell \& Yusef-Zadeh(2000)]{Ode00} 
O'Dell, C.~R., \& Yusef-Zadeh, F.\ 2000, \aj, 120, 382

\bibitem[O'Dell et al.(2009)]{Ode09} 
O'Dell, C.~R., Henney, 
W.~J., Abel, N.~P., Ferland, G.~J., \& Arthur, S.~J.\ 2009, \aj, 137, 367

\bibitem[Okada et al.(2013)]{Oka13} 
Okada, Y., Pilleri, P., Bern{\'e}, O., et al.\ 2013, \aap, 553, A2 

\bibitem[Ossenkopf et al.(2013)]{Oss13} 
Ossenkopf, V., R{\"o}llig, M., Neufeld, D.~A., et al.\ 2013, \aap, 550, AA57 

\bibitem[Phillips et al.(1979)]{Phi79} 
Phillips, T.~G., Huggins, P.~J., Wannier, P.~G., \& Scoville, N.~Z.\ 1979, \apj, 231, 720 

\bibitem[Pilbratt et al.(2010)]{Pil10} 
Pilbratt, G.~L., Riedinger, J.~R., Passvogel, T., et al.\ 2010, \aap, 518, LL1 

\bibitem[Pineda et al.(2013)]{Pin13} 
Pineda, J.~L., Langer, W.~D., Velusamy, T., \& Goldsmith, P.~F.\ 2013, \aap, 554, AA103 

\bibitem[Pineda et al.(2014)]{Pin14} 
Pineda, J.~L., Langer, W.~D., \& Goldsmith, P.~F.\ 2014, \aap, 570, AA121 

\bibitem[Plume et al.(2004)]{Plu04} 
Plume, R., Kaufman, M.~J., Neufeld, D.~A., et al.\ 2004, \apj, 605, 247 

\bibitem[Poglitsch et al.(2010)]{Pog10}
Poglitsch, A., Waelkens, C., Geis, N., et al.\ 2010, \aap, 518, L2 

\bibitem[Riechers et al.(2014)]{Rie14}
Riechers, D.~A., Carilli, C.~L., Capak, P.~L., et al.\ 2014, \apj, 796, 84 

\bibitem[Rivilla et al.(2013)]{Riv13} 
Rivilla, V.~M., Mart{\'{\i}}n-Pintado, J., Jim{\'e}nez-Serra, I., \& Rodr{\'{\i}}guez-Franco, A.\ 2013, \aap, 554, AA48 

\bibitem[Robberto et al.(2013)]{Rob13} 
Robberto, M., Soderblom, D.~R., Bergeron, E., et al.\ 2013, \apjs, 207, 10 

\bibitem[Roelfsema et al.(2012)]{Roe12} 
Roelfsema, P.~R., Helmich, F.~P., Teyssier, D., et al.\ 2012, \aap, 537, AA17 

\bibitem[Rodriguez-Fernandez et al.(2006)]{Rod06} 
Rodriguez-Fernandez, N.J., Braine, J., Brouillet, N., \& Combes, F.\ 2006, \aap, 453, 77 

\bibitem[Rodriguez-Franco et al.(1998)]{Rod98} 
Rodriguez-Franco, A., Martin-Pintado, J., \& Fuente, A.\ 1998, \aap, 329, 1097 

\bibitem[Rosenthal et al.(2000)]{Ros00} 
Rosenthal, D., Bertoldi, F., \& Drapatz, S.\ 2000, \aap, 356, 705 

\bibitem[Russell et al.(1980)]{Rus80} 
Russell, R.~W., Melnick, G., Gull, G.~E., \& Harwit, M.\ 1980, \apjl, 240, L99 

\bibitem[Sanders \& Mirabel(1996)]{San96} 
Sanders, D.~B., \& Mirabel, I.~F.\ 1996, \araa, 34, 749 

\bibitem[Schilke et al.(2014)]{Sch14} 
Schilke, P., Neufeld, D.~A., M{\"u}ller, H.~S.~P., et al.\ 2014, \aap, 566, AA29 

\bibitem[Sim{\'o}n-D{\'{\i}}az et al.(2006)]{Sim06} 
Sim{\'o}n-D{\'{\i}}az, S., Herrero, A., Esteban, C., \& Najarro, F.\ 2006, \aap, 448, 351 

\bibitem[Sim{\'o}n-D{\'{\i}}az \& Stasi{\'n}ska(2011)]{Sim11} 
Sim{\'o}n-D{\'{\i}}az, S., \& Stasi{\'n}ska, G.\ 2011, \aap, 526, AA48 

\bibitem[Sofia et al.(2004)]{Sof04} 
Sofia, U.~J., Lauroesch, J.~T., Meyer, D.~M., \& Cartledge, S.~I.~B.\ 2004, \apj, 605, 272 

\bibitem[Stacey et al.(1991)]{Sta91}
Stacey, G.~J., Geis, N., Genzel, R., et al.\ 1991, \apj, 373, 423 
 
\bibitem[Stacey et al.(1991b)]{Sta91b} 
Stacey, G.~J., Townes, C.~H., Geis, N., et al.\ 1991b, \apjl, 382, L37

\bibitem[Stacey et al.(1993)]{Sta93} 
Stacey, G.~J., Jaffe, D.~T., Geis, N., et al.\ 1993, \apj, 404, 219 

\bibitem[Stacey et al.(2010)]{Sta10} 
Stacey, G.~J., Hailey-Dunsheath, S., Ferkinhoff, C., et al.\ 2010, \apj, 724, 957 

\bibitem[St{\"o}rzer et al.(1995)]{Sto95} 
St{\"o}rzer, H., Stutzki, J., \& Sternberg, A.\ 1995, \aap, 296, L9 

\bibitem[St{\"o}rzer \& Hollenbach(1998)]{Sto98} 
St{\"o}rzer, H., \& Hollenbach, D.\ 1998, \apj, 495, 853 

\bibitem[Stutzki \& Guesten(1990)]{Stu90} 
Stutzki, J., \& Guesten, R.\ 1990, \apj, 356, 513 

\bibitem[Tercero et al.(2010)]{Ter10} 
Tercero, B., Cernicharo, J., Pardo, J.~R., \& Goicoechea, J.~R.\ 2010, \aap, 517, AA96 

\bibitem[Thronson et al.(1990)]{Thr90} 
Thronson, H.~A., Jr., Majewski, S., Descartes, L., \& Hereld, M.\ 1990, \apj, 364, 456 

\bibitem[Tielens \& Hollenbach(1985)]{Tie85} 
Tielens, A.~G.~G.~M., \& Hollenbach, D.\ 1985, \apj, 291, 722 

\bibitem[Tielens \& Hollenbach(1985b)]{Tie85b} 
Tielens, A.~G.~G.~M., \& Hollenbach, D.\ 1985b, \apj, 291, 747

\bibitem[Troland et al.(1989)]{Tro89} 
Troland, T.~H., Heiles, C., \& Goss, W.~M.\ 1989, \apj, 337, 342 

\bibitem[van der Tak et al.(2013)]{vdT13} 
van der Tak, F.~F.~S., Nagy, Z., Ossenkopf, V., et al.\ 2013, \aap, 560, AA95 

\bibitem[van der Werf et al.(2013)]{vdW13} 
van der Werf, P.~P., Goss, W.~M., \& O'Dell, C.~R.\ 2013, \apj, 762, 101 

\bibitem[Wiesenfeld \& Goldsmith(2014)]{Wie14} 
Wiesenfeld, L., \& Goldsmith, P.~F.\ 2014, \apj, 780, 183 

\bibitem[Wilson \& Bell(2002)]{Wil02} 
Wilson, N.~J., \& Bell, K.~L.\ 2002, \mnras, 337, 1027 

\bibitem[Wolfire et al.(2010)]{Wol10} 
Wolfire, M.~G., Hollenbach, D., \& McKee, C.~F.\ 2010, \apj, 716, 1191 

\bibitem[Wu et al.(2005)]{Wu05} 
Wu, J., Evans, N.~J., II, Gao, Y., et al.\ 2005, \apjl, 635, L173 

\bibitem[Wyrowski et al.(1997)]{Wyr97} 
Wyrowski, F., Schilke, P., Hofner, P., \& Walmsley, C.~M.\ 1997, \apjl, 487, L171

\bibitem[Zuckerman(1973)]{Zuc73} 
Zuckerman, B.\ 1973, \apj, 183, 863 

\end{thebibliography}
\end{document}